\documentclass[12pt]{article}
\usepackage{latexsym}
\usepackage{epsfig,amssymb,euscript,slashed}
\usepackage[linktocpage]{hyperref}
\usepackage{amsmath}
\usepackage{cite} 
\usepackage{array,calc,epsfig}
\usepackage{bbm}
\usepackage{fancybox}

\numberwithin{equation}{section} 

\usepackage{amsmath}                                      
\usepackage{amsthm}
\usepackage{amssymb}
\usepackage{cancel}
\usepackage{graphics}
\usepackage{multind}
\usepackage{wrapfig}
\usepackage{amssymb}
\usepackage{comment}
\usepackage{latexsym}
\usepackage{mathtools}
\usepackage{dsfont}
\usepackage{mathrsfs}
\usepackage{hyperref}
\usepackage{pdfpages}
\usepackage{float}
\usepackage{graphicx}
\usepackage{framed}
\usepackage{amsfonts}
\usepackage{bm}
\usepackage{simplewick}

\usepackage{xcolor}
\usepackage[framemethod=TikZ]{mdframed}
\allowdisplaybreaks[4]

\usepackage{subcaption}

\DeclarePairedDelimiter{\abs}{\lvert}{\rvert}

\newcommand{\CO}{\mathcal{O}}
\newcommand{\CN}{\mathcal{N}}

\newcommand{\CK}{\mathcal{K}}

\newcommand{\CV}{\mathcal{V}}
\newcommand{\CS}{\mathcal{S}}

\newcommand{\CW}{\mathcal{W}}

\newcommand{\pd}{\partial}
\newcommand{\half}{\frac{1}{2}}
\newcommand{\ep}{\epsilon}

\def\be{\begin{equation}}
\def\ee{\end{equation}}
\def\bea{\begin{eqnarray}}
\def\eea{\end{eqnarray}}

\newcommand{\nn}{\nonumber}

\newcommand{\diff}{\mathrm{d}}

\newcommand{\air}{\alpha}
\newcommand{\hir}{\eta}
\newcommand{\kir}{\iota}
\newcommand{\HL}{\tilde{H}}
\newcommand{\KL}{\tilde{K}}

\begin{document}
\font\cmss=cmss10 \font\cmsss=cmss10 at 7pt

\begin{flushright}{
\scriptsize DFPD-19-TH-01} \\  
\end{flushright}
\hfill
\vspace{18pt}
\begin{center}
{\Large 
\textbf{ General Supersymmetric AdS$_5$ Black Holes  \\ \vspace{0.2cm} with Squashed Boundary
}}

\end{center}

\vspace{8pt}
\begin{center}
{\textsl{Alessandro Bombini$^{\,a, b, c}$ and Lorenzo Papini$^{\,a, b}$}}

\vspace{1cm}

\textit{\small ${}^a$ Dipartimento di Fisica ed Astronomia ``Galileo Galilei",\\
 Universit\`a di Padova,\\
 Via Marzolo 8, 35131 Padova, Italy} \\  \vspace{6pt}

\textit{\small ${}^b$ I.N.F.N. Sezione di Padova, \\
Via Marzolo 8, 35131 Padova, Italy}\\ \vspace{6pt}

\textit{\small ${}^c$ Institut de Physique Th\'eorique, \\
Universit\'e Paris Saclay, CEA, CNRS, \\
Orme des Merisiers, F-91191 Gif sur Yvette, France}\\ 

\vspace{6pt}

\end{center}

\vspace{12pt}

\begin{center}
\textbf{Abstract}
\end{center}

\vspace{4pt} {\small
  \noindent

We present a new family of asymptotically locally AdS$_5$ squashed supersymmetric black hole solutions of Fayet-Iliopoulos gauged ${\cal N}=2$, $D=5$ supergravity with two vector multiplets that have a natural uplift to type IIB supergravity. Our new family of black holes is characterized by three parameters, of which two control the horizon geometry while the latter regulates the squashing at the boundary. We evaluate the main physical properties of the family of solutions using holographic renormalization and find that the entropy is independent on the squashing and it is reproduced by using the angular momentum and the Page charges. In previously known solutions Page and holographic charges are equal, due to the vanishing of the Chern-Simons term that here, instead, is relevant. This result suggests that for asymptotically locally AdS$_5$ solutions we should refer to the Page charges to describe the thermodynamics of the system.

}

\vspace{1cm}

\thispagestyle{empty}

\vfill
\vskip 5.mm
\hrule width 5.cm
\vskip 2.mm
{
\noindent  {\scriptsize e-mails:  {\tt alessandro.bombini@pd.infn.it , lorenzo.papini@pd.infn.it} }
}

\setcounter{footnote}{0}
\setcounter{page}{0}

\newpage
\tableofcontents


\section{Introduction} \label{sec:Introduction}
One of the main outstanding achievements of string theory is the description of asymptotically flat black holes in terms of microscopic constituents such as strings and branes~\cite{Polchinski:1995mt, Breckenridge:1996is, Callan:1996dv, Maldacena:1996ky, Polchinski:1996fm, Polchinski:1996na}, by which is possible to compute their entropy by microstate counting, reproducing correctly the Bekenstein-Hawking entropy as a statistical Boltzmann entropy~\cite{Strominger:1996sh,Sen:2008vm}.

When the AdS/CFT correspondence was conjectured~\cite{Maldacena:1997re,Gubser:1998bc,Witten:1998qj} it seemed very natural to use it to provide a description of asymptotically Anti-de-Sitter (AdS) black holes in terms of microscopical states in a dual quantum field theory, thus extending the results obtained for asymptotically flat black holes. In particular, since the original correspondence is AdS$_5$/CFT$_4$, it appeared natural to attack the problem starting with supersymmetric asymptotically AdS$_5$ black holes. However, although such black holes were constructed fifteen years ago~\cite{Gutowski:2004ez} and many generalization immediately followed~\cite{Gutowski:2004yv,Chong:2005da,Chong:2005hr,Kunduri:2006ek}, all the first attempts to provide an interpretation of their entropy in terms of a quantum field theory computation were basically unsuccessful~\cite{Kinney:2005ej} and this problem has remained a puzzle for many years. 

The same problem in one lower dimension has been solved four years ago starting in~\cite{Benini:2015eyy}.  There, a certain class of supersymmetric asymptotically locally AdS$_4$ black holes is analyzed and their entropy is obtained by a localization computation in the dual ABJM field theory. More precisely, the authors managed to reproduce the black hole entropy by Legendre transforming the large $N$ contribution of the topologically twisted index of the ABJM theory, introduced in~\cite{Benini:2015noa}. The field theoretic computation performed in~\cite{Benini:2015eyy} and related papers, as well as the corresponding interpretation on the gravity side discussed for example in~\cite{Halmagyi:2017hmw,Cabo-Bizet:2017jsl}, shade light on many non-trivial steps one must follow in order to reproduce the entropy; it was then reasonable to expect that these results could be inspiring to solve the problem also in five-dimensions.

A solution to the five-dimensional enigma has finally been proposed in~\cite{Cabo-Bizet:2018ehj,Choi:2018hmj,Benini:2018ywd,Honda:2019cio,ArabiArdehali:2019tdm} (see also~\cite{Hosseini:2017mds} for previous progress) by reconsidering the field theory partition function dual to the black hole allowing for complex values of the fugacities. The approaches used in the above-mentioned papers are quite different and the relation between them remains as an interesting puzzle. 


In each of these field theoretic computations it is crucial to understand which are the field theory states that contribute to the entropy. Further information on this may be collected by studying black holes which are not globally asymptotically AdS, but just locally. Such black holes may be obtained by deforming the conformal boundary and some solutions with these characteristics have been constructed in minimal five-dimensional gauged supergravity~\cite{Blazquez-Salcedo:2017kig,Blazquez-Salcedo:2017ghg} and in five-dimensional Fayet-Iliopoulos gauged supergravity~\cite{Cassani:2018mlh}. In all these papers, the authors considered a cohomogeneity-one
ansatz with local SU(2) $\times$ U(1) $\times$ U(1) symmetry and managed to obtain supersymmetric black holes with a non conformally-flat boundary geometry containing a squashed three-sphere. These solutions are thus asymptotically locally AdS$_5$ (AlAdS$_5$) rather than asymptotically AdS$_5$. 
In the squashed solution of minimal gauged supergravity~\cite{Blazquez-Salcedo:2017kig,Blazquez-Salcedo:2017ghg} the geometry of the event horizon results completely frozen so that the entropy is uniquely fixed; instead the squashing at the boundary can assume any value. Therefore this AlAdS$_5$ solution behaves differently from the general asymptotically AdS$_5$ one of~\cite{Gutowski:2004ez}, since in the latter the entropy is controlled by the same parameter regulating the horizon geometry and can therefore vary. The minimal gauged supergravity solution has been generalized in~\cite{Cassani:2018mlh}, where the authors constructed a two-parameters family of squashsed black holes in Fayet-Iliopoulos five-dimensional gauged supergravity with an arbitrary number of vector multiplets. Of the two parameters on which the solution depends, one controls the horizon geometry, which is thus not completely frozen anymore, while the other one determines the squashing at the boundary. The entropy of the solution is regulated by only one parameter and behaves again differently with respect to the general solution of~\cite{Gutowski:2004yv}, where the entropy is controlled by three parameters. Although quite general and valid for any number of vector multiplets, the solution of~\cite{Cassani:2018mlh} is obtained by imposing a particular ansatz on the scalar fields which constraints all their components orthogonal to the scalar vacuum expectation values in the supersymmetric AdS$_5$ vacuum to be the same. In principle each of such components is controlled by a different function; the authors impose their ansatz by requiring that all those functions are the same.

The main aim of the present paper is to construct general squashed solutions in five-dimensional Fayet-Iliopoulos gauged supergravity with two vector multiplets without imposing any ansatz. This means that we will abandon the restrictive conditions imposed in~\cite{Cassani:2018mlh} and let the scalar fields unconstrained. We will look for solutions with two vector multiplets only because this is the case for which the solutions can be uplifted to be solutions of type IIB supergravity in ten dimensions, and are thus particularly relevant from a string-theoretical perspective\footnote{It might be interesting to notice that the bosonic sector of the five-dimensional supergravity under consideration is a consistent truncation of eleven-dimensional supergravity on a space with boundary~\cite{Colgain:2014pha}.}. From an holographic point of view, the Fayet-Iliopoulos five-dimensional supergravity coupled to $n_V$ vector multiplets should be dual to a subsector of an $\CN = 1$ SCFT composed by an $\CN = 1$ energy-momentum tensor and U$(1)^{n_V}$ flavour current multiplets. This description is made more precise in the particular $n_V = 2$ case we are considering, where the supergravity theory can be regarded as a consistent truncation of type IIB supergravity and it is therefore holographically dual to a deformation of $\CN = 4$ super-Yang-Mills\footnote{Note that the supergravity theory with $n_V = 2$ vector multiplets has gauge group U$(1)^3$;  for this reason this particular case is sometimes dubbed as U$(1)^3$ theory~\cite{Gutowski:2004yv,Kunduri:2005zg}. We shall occasionally use this name to refer to the theory throughout the paper.}.

In~\cite{Cassani:2018mlh} the conditions, originally given in~\cite{Gutowski:2004yv}, to be imposed in order to have a supersymmetric solution to Fayet-Iliopoulos gauged supergravity are rearranged and partially solved. This process results into $n_V + 1$ coupled ordinary differential equations. We explicitly examine the $n_V = 2$ case we are interested in and we obtain three coupled ordinary differential equations which we simplify as much as we can using the particular form of the various functions in the U$(1)^3$ theory. However the equations remain very complicated and we could not find any new analytic solution. We therefore pass to a perturbative approach and we try to construct a near-horizon family of candidate black hole solutions and  a near-boundary family of candidate AlAdS$_5$ solutions. Remarkably enough, we find numerically that the two families such obtained match in the bulk and we are able to construct the whole solution for a large part of the parameter space. We thus obtain a new three-parameter family of supersymmetric black holes, which posseses both non trivial and unconstrained gauge and scalar fields. Our solution generalizes the one-parameter one of~\cite{Blazquez-Salcedo:2017ghg,Blazquez-Salcedo:2017kig} found in minimal gauged supergravity and also the two-parameters one of~\cite{Cassani:2018mlh} since our scalar fields are unconstrained. From a technical point of view, the approach we adopt here to perform the perturbative and numerical analysis is similar in spirit to the one of~\cite{Cassani:2018mlh}; however, the results and outcomes we find are more general since we abandon the simplifying ansatz imposed there on the scalar fields, so that we obtain an infinite set of new solutions, enlarging the family found there.

The horizon geometry and the horizon properties of our solution are controlled by two of the three total parameters, which are also responsible for the Page charges and the angular momentum. The last parameter regulates instead the squashing of the boundary geometry, but does not influence the horizon, whose geometry is completely independent on the squashing. 

The paper is organized as follows. In sec.~\ref{sec:Gauged_Sugra} we give an essential presentation of Fayet-Iliopoulos gauged supergravity and discuss its main features, including its uplift to string theory in the case $n_V = 2$. We also write the U$(1)^3$ theory supersymmetry equations we need to solve and present the related first integrals. In sec.~\ref{sec:Solution} we construct our family of solutions, first by analyzing the near-boundary and near-horizon regimes separately and then matching them numerically. In sec.~\ref{sec:Properties} we present the main physical properties of our solution. To evaluate some of them we employ the technique of holographic renormalization. We explicitly compute the entropy and find that, as was also obtained in~\cite{Cassani:2018mlh}, it is remarkably reproduced by a simple formula containing the Page charges, instead of the holographic charges. We conclude in sec.~\ref{sec:Conclusion}, where we discuss our results.
In app.~\ref{app_more_perturbative} we report more details about the perturbative analysis of sec.~\ref{sec:Solution}, turning also the near-boundary solution in Fefferman-Graham coordinates. Finally in app.~\ref{app_HR} we briefly review the main features of holographic renormalization and how this is used to compute the physical properties described in sec.~\ref{sec:Properties}.


\section{$\mathbf{\CN = 2, D=5}$ Gauged Supergravity and its uplift}
\label{sec:Gauged_Sugra}

In this section we briefly introduce the $\CN=2$ five-dimensional Fayet-Iliopoulos gauged supergravity and we discuss which conditions should be imposed to its various bosonic fields in order to obtain supersymmetric solutions. For the purposes of the present paper, we are interested only in cohomogeneity one supersymmetric solutions which possess a local SU(2)$\times$U(1)$\times$U(1) symmetry, which implies that the supersymmetry equations are just ordinary differential equations (ODE's) rather than partial differential equations. These supersymmetry equations were originally given in~\cite{Gutowski:2004yv} and were recast into a simpler form in~\cite{Cassani:2018mlh}. We recall that the general case is characterized by an arbitrary number of scalar fields $n_V$ while, in the case of interest for the present paper, we will consider $n_V = 2$ since this is the case in which the uplift to type IIB supergravity is possible. The authors of~\cite{Cassani:2018mlh} specialize to the particular case where all the components of the scalar fields orthogonal to the scalar vacuum expectation values in the supersymmetric AdS$_5$ vacuum are controlled by the same $H$ function (up to some constants $q_I$); here, we will instead specialize to the U$(1)^3$ theory case with $n_V = 2$ without imposing any simplifying ansatz. The same orthogonal components will thus be treated in full generality and let be free to assume any value; they will be controlled by three different functions $H_1, \, H_2$ and $H_3$ linked by a unique constraint. Solutions to this particular theory are interesting because they have a natural uplift to type IIB supergravity, as we will discuss in sec.~\ref{sec:uplit}.

\subsection{The theory} \label{sec:the_theory}
We now briefly review the main features of the five-dimensional $\CN=2$ Fayet-Iliopoulos gauged supergravity we consider in this paper. Here we give some essential information about the theory and we refer to~\cite{Gutowski:2004yv, Gunaydin:1983bi, Gunaydin:1984ak} for further details.

The bosonic sector of the theory is composed by the metric $g_{\mu \nu}$, by $n_V + 1$ Abelian gauge fields $A^I_\mu$ and by $n_V$ real scalar fields $\Phi^I$. For convenience it is customary to parametrize the latter using $n_V + 1$ real fields $X^I$ which fulfil the constraint
\begin{equation}
\label{constraint}
\frac{1}{6} C_{IJK} X^I X^J X^K = 1\,,
\end{equation}
where $C_{IJK}$ is a constant, symmetric tensor satisfying
\be\label{3Cinto1C}
C_{IJK} C_{J'(LM} C_{PQ)K'} \,\delta^{JJ'}\delta^{KK'}   \,=\, \frac{4}{3}\delta_{I(L} C_{MPQ)}\,.
\ee
The constraint~\eqref{constraint} can be more easily written by introducing lower-index scalars $X_I$ defined as
\begin{equation}
\label{Def_XI_Down}
X_I=\frac{1}{6} C_{IJK} X^J X^K \, ,
\end{equation}
so that now~\eqref{constraint} becomes
\begin{equation}
X_I X^I =1 \, .
\end{equation}
We can also obtain the inverse relation 
\begin{equation}
\label{Upper_Scalars}
X^I = \frac{9}{2} C^{IJK} X_J X_K \, ,
\end{equation}
by defining the tensor $C^{IJK}$ such that $C^{IJK} = \delta^{I I^\prime}  \delta^{J J^\prime}  \delta^{K K^\prime} \, C_{IJK}$.
The Fayet-Iliopoulos gauging procedure introduces $n_V + 1$ parameters $V_I$ in order to gauge a U(1) subgroup of the R-symmetry SU(2). In the bosonic sector the main consequence of this gauging is the introduction of the following scalar potential
\begin{equation}
\label{scalarpot}
\mathcal{V} = - 27 C^{IJK}V_IV_JX_K \,,
\end{equation}
which is fully consistent with supersymmetry. 

The bosonic action of the theory in mostly-plus signature is: 
{\small
\begin{equation}
\label{Bulk_action}
S = \frac{1}{2 \, \kappa^2} \int{ \left[ \sqrt{\abs{g}} \left(R - 2 \,  \CV \right) - Q_{IJ} F^I \wedge \star F^J - Q_{IJ} \, \diff  X^I \wedge \star \diff X^J - \frac{1}{6} C_{IJK} F^I \wedge F^J \wedge A^K \right] }  ,
\end{equation}}%
where $g = \det g_{\mu \nu}$ is the determinant of the metric, $F^I = \diff A^I$ are the $n_V + 1$ field strengths and $\kappa^2 = 8 \, \pi \, G_N$. The scalars appear in the action as contracted with the kinetic matrix $Q_{IJ}$, which reads:
\be\label{Qmatrix}
Q_{IJ} = \frac{9}{2}X_IX_J - \frac{1}{2}C_{IJK}X^K\, ,
\ee
and satisfies 
\be
Q_{IJ}X^J = \frac{3}{2}X_I \, .
\ee
From the action~\eqref{Bulk_action} it is possible to derive the Einstein, Maxwell and scalar equations~\cite{Gutowski:2004yv}
\begin{subequations}
\begin{align}
R_{\mu \nu} - Q_{IJ} F^I_{\mu \rho} F^{J \, \rho}_\nu - Q_{IJ} \nabla_\mu X^I \nabla_\nu X^J + \frac{1}{6} \, g_{\mu \nu} \left(- 4 \CV + Q_{IJ} \, F^I_{\rho \sigma} \, F^{J \, \rho \sigma} \right) = 0 \, , \label{Einstein_equations} \\
\diff \left( Q_{IJ} \star F^J \right) + \frac{1}{4}\, C_{IJK} F^J \wedge F^K =0  \, ,  \label{Maxwell_general} \\
\diff(\star \diff X_I) - \left( \frac{1}{6}\, C_{MNI} - \half \, X_I C_{MNJ} X^J   \right) \diff X^M \wedge \star \diff X^N \qquad\; \notag \\
+ \left( X_M X^P C_{NPI} - \frac{1}{6}\, C_{MNI} - 6 X_I X_M X_N + \frac{1}{6} \, X_I C_{MNJ} X^J  \right) F^M \wedge \star F^N \qquad\; \notag\\
+6 \left( 6 X_I C^{MPQ} V_M V_P X_Q - C^{MPQ} V_M V_P C_{QIJ} X^J   \right) \star 1 = 0 \,.
\end{align}
\end{subequations}

Assuming that $C^{IJK}  V_I  V_J V_K > 0 $, there is a supersymmetric AdS$_5$ vacuum allowed by the theory which is characterized by a radius $\ell$ and by the constant values of the scalars $\bar{X}^I$. These are furthermore determined by the Fayet-Iliopoulos parameters $V_I$ as
\begin{equation}
\bar{X}_I = \ell \, V_I \, .
\end{equation}
We will always use this relation to trade the parameters $V_I$ with the $\bar{X}_I$ in the rest of the paper. Using~\eqref{Upper_Scalars} we can therefore rewrite the scalar potential $\CV$ as 
\begin{equation}
\CV = - 6 \, \ell^{-2}  \bar{X}^I  X_I \, .
\end{equation}

We are interested in AlAdS solutions to five-dimensional Fayet-Iliopoulos gauged supergravity. These have a holographic interpretation in terms of a dual four-dimensional $\CN=1$ SCFT, including cases where the latter SCFT is not in a trivial state such as the conformal vacuum. However the holographic interpretation is actually under control in the cases where a consistent uplift of the theory to string theory or M-theory does exist. In section~\ref{sec:uplit} we will briefly review how it is possible to embed the five-dimensional $\CN=2$ Fayet-Iliopoulos supegravity we study in this paper as a consistent truncation of type IIB supergravity.

\subsection{Supersymmetry equations for the U(1)$^3$ theory} \label{sec:susyU1}

\subsubsection{The ansatz for the solution}
We introduce the SU(2) left-invariant one forms
\begin{align}
\label{One_forms_hatted}
\hat\sigma_1 \,&=\, \cos{\hat\psi}\, \diff \theta + \sin{\hat\psi} \sin\theta \,\diff \phi\ , \nn\\
\hat\sigma_2 \,&=\, -\sin{\hat\psi} \,\diff \theta + \cos{\hat\psi} \sin\theta \,\diff \phi \ , \nn\\
\hat\sigma_3 \,&=\,  \diff {\hat\psi} + \cos\theta \,\diff  \phi \  ,
\end{align}
where $\hat \psi$ denotes a coordinate different than $\psi$ to be introduced later. These one-forms satisfy $\diff  \hat{\sigma}_i = - \frac{1}{2} \epsilon_{ijk} \, \hat{\sigma}_j \wedge \hat{\sigma}_k$. We choose the set of coordinates $(y, \, \rho, \,  \theta, \, \phi, \, \hat{\psi})$ to describe our solution and we assume the following ansatz for the five-dimensional metric
\begin{equation}
\label{metric}
\diff s^2= -f^2(\diff y + w\, \hat\sigma_3)^2 + f^{-1} \left[\,\diff \rho^2 + a^2 (\hat\sigma^2_1+\hat\sigma^2_2) + (2 a a^\prime)^2 \,\hat\sigma^2_3\, \right]\, ,
\end{equation} 
where all the unknown functions $f$, $w$, $a$ are dependent on the $\rho$ coordinate only and for the rest of the paper the prime symbol will denote differentiation with respect to this coordinate. In the minimal theory, the functions $f$ and $w$ can be rewritten in terms of $a$ only; in particular $f$ assumes the form
\begin{equation}
f_{\rm min} = \frac{12 \, a^2  a^\prime}{\ell^2 (a^2 a^{\prime\prime\prime} - a^\prime + 7 a a^\prime a^{\prime \prime} + 4(a^\prime)^3)} \, .
\end{equation}
The gauge fields are given by 
\begin{equation}
\label{gauge_field}
A^I=X^I f \left(\diff y + w \,\hat\sigma_3\right) + U^I \hat\sigma_3\ ,
\end{equation}
where $U^I (\rho)$ are $n_V +1$ unknown functions to be further determined. The field strengths following from the above gauge fields are:
\begin{equation}
\label{Field_Strength}
F^I = - \left(f \, X^I \right)^\prime (\diff y+w\, \hat\sigma_3)\wedge\diff\rho + \left(f  w^\prime  X^I + \big(U^I\big)' \right) \diff\rho\wedge\hat\sigma_3 -  \left(f  w  X^I + U^I \right) \hat\sigma_1\wedge\hat\sigma_2\ ,
\end{equation}
so that their Hodge dual are\footnote{We choose $\diff y \wedge \diff \rho \wedge \hat \sigma_1 \wedge \hat \sigma_2 \wedge \hat \sigma_3$ to be our positive orientation, as it was done in~\cite{Gutowski:2004yv}.}:
\begin{align}
\label{Field_strength_star}
\star F^I &= 2a^3a'f^{-2} \left(f X^I \right)^\prime \hat\sigma_{123} + \frac{af}{2a^\prime} \left(f w^\prime X^I + (U^I)' \right) (\diff y+w\hat\sigma_3)\wedge\hat\sigma_1\wedge\hat\sigma_2 \nn\\[1mm]
&\quad \, - \frac{2a'}{a}f \left(f  w  X^I + U^I \right) \diff y \wedge\diff\rho\wedge \hat{\sigma}_3 \ .
\end{align}
Finally the scalar fields are only functions of the radial coordinate $\rho$, so that $X^I = X^I (\rho)$.

\subsubsection{The supersymmetry equations}
Rearranging the supersymmetry equations originally given in~\cite{Gutowski:2004yv}, the authors of~\cite{Cassani:2018mlh} showed that all the supersymmetric solutions of the $\CN=2, \, D=5$ Fayet-Iliopolous supegravity under consideration can be obtained by solving the following set of equations:
\be\label{Maxwell_orthog}
\left[ H_I'' - \left(\frac{3a'}{a}+\frac{a''}{a'}\right) H_I' + \frac{2p}{3a^2}H_I + \frac{24}{\ell^2a^4}\left( \bar{Q}_{IJ} - \frac{3}{2}\bar{X}_I \bar{X}_J \right) (CHH)^J   \right]'=0\, , 
\ee
\begin{align}
\label{eqfora_runningX}
&\bigg( \nabla^2 f_\text{min}^{-1} + \frac{8}{\ell^2} f_\text{min}^{-2} - \frac{\ell^2 g^2}{18} + f_\text{min}^{-1}\, g \bigg)' + \frac{4 a' g}{a f_\text{min}}\nn\\[1mm]
+& \bar{X}_IC^{IJK}\!\left\{\frac{36}{\ell^2a^3a'}\left[\left(\! \frac{H_JH_K}{a^4}\!\right)' - \frac{3a}{2a'} H_J'\left(\!\frac{H_K}{a^4}\!\right)'\; \right] \right\}' -\frac{216}{\ell^2}\bar{X}_IC^{IJK} \frac{H_J'}{a^3a'}\!\left(\!\frac{H_K}{a^4}\!\right)'  = 0\,,
\end{align}
where $H_I = H_I (\rho)$ are $n_V + 1$ functions which are defined such that
\begin{equation}
f^{-1} \, X_I = f_{\text{min}}^{-1} \, \bar{X}_I + \frac{H_I^\prime}{a^3 \, a^\prime} \, ,
\end{equation} 
and satisfy the constraint
\begin{equation}
\label{Gen_Constraint}
\bar{X}^I \, H_I = 0 \, .
\end{equation}
Moreover, we have defined
\begin{subequations}
\begin{align}
\label{P}
p &= -1 + 2 a a^{\prime\prime} + 4 (a^\prime)^2 \,, \\
g &= -\frac{a'''}{a'}-3 \frac{a''}{a}-\frac{1}{a^2} + 4 \frac{a'^2}{a^2} \, .
\end{align}
\end{subequations}
Our aim in this section is to explicitly rewrite the supersymmetry equations~\eqref{Maxwell_orthog}, \eqref{eqfora_runningX} and all the various objects defined in~\cite{Cassani:2018mlh} for the case $n_V = 2$.  We have the indices $I$, $J$, $K$ running from 1 to 3 and
\begin{equation}
\label{CIJKinU1}
C_{IJK} = C^{IJK} = 
\begin{cases}
1 \qquad \text{if $(IJK)$ is a permutation of $(123)$} \,,\\
0 \qquad \text{otherwise} \,.
\end{cases} 
\end{equation}
We have also that the constraint on the scalars~\eqref{constraint} becomes
\begin{equation}
X^1 \, X^2 \, X^3 = 1 \, ,
\end{equation} 
and the kinetic matrix \eqref{Qmatrix} for the scalars is given by
\begin{equation}
Q_{IJ} = \frac{9}{2} \, \text{diag} \, \bigg( \left(X_1 \right)^2 , \, \left(X_2 \right)^2 , \, \left(X_3 \right)^2 \bigg) .
\end{equation}
The corresponding value $\bar{X}^I$ of the scalars in the AdS$_5$ vacuum is
\begin{equation}
\bar{X}^I = 1 \,, \qquad \Rightarrow \qquad \bar{X}_I = \frac{1}{3}  \,,
\end{equation}
so the corresponding kinetic matrix in the same vacuum is just
\begin{equation}
\bar{Q}_{IJ} = \frac{1}{2} \, \, \mathcal{I}_{3 \times 3} \,.
\end{equation}
Using these relations we can explicitly write the supersymmetry equations in the context of the U$(1)^3$ theory we are considering. The equations will depend only on $a(\rho)$ and on three functions $H_1 (\rho)$, $H_2 (\rho)$,  $H_3 (\rho)$ which control the scalars. However from eq.~\eqref{Gen_Constraint} we have the constraint
\begin{equation}
\label{H_constraint}
H_1 + H_2 + H_3 = 0 \,.
\end{equation}
This implies that we can eliminate one of the $H_I$ functions. For example we choose to use this constraint to replace $H_3$ with
\begin{equation}
\label{H3_sub}
H_3 = - H_1 - H_2 \,,
\end{equation}
so that $H_3$ will never appear anymore throughout the paper.
We define two particular combinations of $H_1$ and $H_2$  which will appear in the supersymmetry equations:
\begin{subequations}
\begin{align}
\Sigma(H_1 , H_2) &= - \left(H_1^2 + H_2^2 + H_1 \, H_2 \right) , \label{Sigma_Constrained} \\
\Lambda(H_1 , H_2) &= - \left[ 2 \, H_1^\prime \, \left(\frac{H_1}{a^4} \right)^\prime + 2 \, H_2^\prime \, \left(\frac{H_2}{a^4} \right)^\prime + \, H_1^\prime \, \left(\frac{H_2}{a^4} \right)^\prime + H_2^\prime \, \left(\frac{H_1}{a^4} \right)^\prime \, \right] . \label{Lambda_Constrained}
\end{align}
\end{subequations}
We now proceed to rewrite the supersymmetry equations. In order to do this we let the index $I$ run from 1 to 3, we use~\eqref{H3_sub} to eliminate $H_3$ whenever it appears and we perform all the necessary contractions recalling~\eqref{CIJKinU1}. Doing so we obtain the following three equations
\begin{equation}
\label{eqforH}
\left[ H_1'' - \left(\frac{3a'}{a}+\frac{a''}{a'}\right) H_1' + \frac{2p}{3a^2}H_1 + \frac{8}{\ell^2 a^4}\left( H_1^2 - 2 H_2^2 - 2 H_1 H_2 \right) \right]'= 0 \, ,
\end{equation}
\begin{equation}
\label{eqforK}
\left[ H_2'' - \left(\frac{3a'}{a}+\frac{a''}{a'}\right) H_2' + \frac{2p}{3a^2}H_2 + \frac{8}{\ell^2 a^4}\left(- 2 H_1^2 + H_2^2 - 2 H_1 H_2 \right) \right]'= 0 \,,
\end{equation}
\begin{equation}
\label{eqforaU1}
\bigg( \nabla^2 f_\text{min}^{-1} + \frac{8}{\ell^2} f_\text{min}^{-2} - \frac{\ell^2 g^2}{18} + f_\text{min}^{-1} g \bigg)' + \frac{4 a' g}{a f_\text{min}} 
+ \left\{\frac{12}{\ell^2 \, a^3 \, a'}\left[2 \, \left(\frac{\Sigma}{a^4}\right)' - \frac{3 \, a}{2 \, a'} \, \Lambda \; \right] \right\}' -\frac{72 \, \Lambda }{\ell^2 \, a^3 \, a^\prime} = 0\,.
\end{equation}
We have to solve these three equations~\eqref{eqforH},~\eqref{eqforK} and~\eqref{eqforaU1} in order to find new solutions in the U$(1)^3$ theory. 

Once $a$, $H_1$ and $H_2$ are determined, all the other functions are fixed in terms of these. We now report the explicit expressions of them in the U$(1)^3$ case. These are straightforwardly obtained by the general relations reported in~\cite{Cassani:2018mlh} by setting $n_V = 2$ and performing the necessary contractions. 
We start with the function $f$ that is determined as
\begin{equation}
\label{f_from_a_U1}
f = \left[ f_{\text{min}}^{-3} - 9 \, f^{-1}_{\text{min}} \, \left(h_1^2 + h_2^2 + h_1 \, h_2 \right) - 27 \left(h_1^2 \, h_2 + h_1 \, h_2^2 \right) \right]^{-1/3} \ ,
\end{equation}
where $h_I$ are given in terms of $H_I$ as 
\be\label{h_from_H}
h_I = \frac{H_I'}{a^3a'}\,,
\ee
so that
\be\label{scalaransatz}
f^{-1}X_I = f_{\rm min}^{-1}\bar X_I + h_I\, .
\ee
They obviously satisfy the constraint $h_1 + h_2 + h_3 = 0$ as the $H_I$ do. We proceed with the $w$ function
\begin{equation}
\begin{split}
\label{w_from_a_U1}
w = - \frac{\ell \, a^2}{4} & \Bigg\{ \nabla^2 (f_{\rm min}^{-1}) + \frac{8}{\ell^2}f_{\rm min}^{-2}  - \frac{\ell^2g^2}{18} + f_{\rm min}^{-1}\,g  \\
&  \qquad + \frac{12}{\ell \, a^3 \, a^\prime} \left[ 2 \left( \frac{\Sigma \left(H_1, H_2 \right) }{a^4} \right)^\prime - \frac{3 \, a}{2 \, a^\prime} \, \Lambda \left(H_1 , H_2 \right) \right] \! \Bigg\} .
\end{split}
\end{equation}
The functions $U^1$, $U^2$ and $U^3$ are given in terms of $H_I$  by 
\begin{align}
\label{UI_function_U1}
& U^1 = \frac{\ell}{3} \, p - \frac{12} {\ell \, a^2} \, H_1 \,, \\
& U^2 = \frac{\ell}{3} \, p - \frac{12} {\ell \, a^2} \, H_2 \,, \label{UI_function_U2}\\
& U^3 = \frac{\ell}{3} \, p + \frac{12} {\ell \, a^2} \, \left( H_1 + H_2 \right)  ,\label{UI_function_U3}
\end{align}
and it is immediate to see that they satisfy
\begin{equation}
\label{UI_functions_constraint}
U^1 + U^2 + U^3 = \ell \, p \,,
\end{equation}
so that only $U^1$ and $U^2$ are independent. 
Finally we consider the scalars; for the lower-index scalars $X_I$ we have from equation~\eqref{scalaransatz} that 
\begin{align}
\label{Scalars_Down_U1}
& X_1 = \frac{f \, f^{-1}_{\text{min}}}{3} + h_1 \, f  \,, \\
& X_2 = \frac{f \, f^{-1}_{\text{min}}}{3} + h_2 \, f \,,\\
& X_3 = \frac{f \, f^{-1}_{\text{min}}}{3} - \left(h_1 + h_2 \right) \, f \,.
\end{align}
Furthermore, only $X_1$ and $X_2$ are independent since it holds the following constraint 
\begin{equation}
X_1 + X_2 + X_3 = f \, f^{-1}_{\text{min}}  \,.
\end{equation}
For the upper-index scalars $X^I$ we obtain instead that 
\begin{align}
\label{Scalars_Up_U1}
& X^1 = \left( f \, f^{-1}_{\text{min}} \right)^2 - 3 \, f^2 \, f^{-1}_\text{min} \, h_1 - 9 \, f^2 \, \left(h_1 + h_2 \right) \, h_2 \,, \\
& X^2 = \left( f \, f^{-1}_{\text{min}} \right)^2 - 3 \, f^2 \, f^{-1}_\text{min} \, h_2 - 9 \, f^2 \, \left(h_1 + h_2 \right) \, h_1 \,,\\
& X^3 =\left( f \, f^{-1}_{\text{min}} \right)^2 + 3 \, f^2 \, f^{-1}_\text{min} \, \left(h_1 + h_2 \right) + 9 \, f^2 \, h_1 \, h_2 \,,
\end{align}
and we have that only $X^1$ and $X^2$ are independent, since it results 
\begin{equation}
\label{Scalars_Up_constraint}
X^1 + X^2 + X^3 = 3 \left(f \, f^{-1}_{\text{min}} \right)^2 - 9 \, f^2 \, \left(h_1^2 + h_2^2 + h_1 \, h_2 \right) .
\end{equation} 

We now briefly discuss which conditions one should impose to reduce to previously known solutions of the $\mathcal{N} = 2$, $D=5$ Fayet-Iliopoulos gauged supergravity and to minimal gauged supergravity. To obtain the latter it is sufficient to take $H_1 = H_2 = 0$. Indeed, in this case the equations for $H_1$ and $H_2$~\eqref{eqforH},~\eqref{eqforK} are trivially satisfied, while the equation for $a$~\eqref{eqforaU1} becomes the same given in~\cite{Gutowski:2004ez}. All the physical relevant functions become the ones of minimal gauged supergravity. Indeed equation~\eqref{f_from_a_U1} gives $f = f_\text{min}$, therefore the scalars~\eqref{Scalars_Up_U1} become just constants, the $U^{1,2}$ functions are equal and provide the same gauge field of~\cite{Gutowski:2004ez} and equations~\eqref{w_from_a_U1},~\eqref{Scalars_Down_U1} reduce to the same form they take in minimal gauged supergravity.
We can also easily obtain the U$(1)^3$ version of the general family of solutions given in~\cite{Cassani:2018mlh}. Indeed the solutions of that paper are obtained by taking the simplifying ansatz $H_I \left(\rho \right) = q_I \, H \left( \rho \right)$. Further conditions discussed in~\cite{Cassani:2018mlh} fix the charges $q_I$ to assume in the U$(1)^3$ theory the values $q_1 = q_2 = \frac{1}{6} , \, q_3 = - \frac{1}{3}$ (or cyclic permutations). Therefore to reduce to this class of solutions we have to take the limit $H_1 = H_2 = \frac{1}{6} \, H , \, H_3 = - \frac{1}{3} \, H$ (or cyclic permutations). Doing so, equations~\eqref{eqforH} and~\eqref{eqforK} become equal and become the same as equation (2.58) of~\cite{Cassani:2018mlh}, while equation~\eqref{eqforaU1} reduces to (2.59) of the same paper. The Gutowski-Reall solution~\cite{Gutowski:2004yv} is also recovered, since it is just a particular limit of the more general solutions of~\cite{Cassani:2018mlh}. 

We conclude this section by noting that the supersymmetry equations~\eqref{eqforH},~\eqref{eqforK} and~\eqref{eqforaU1} possess a scaling symmetry~\cite{Cassani:2018mlh}: indeed rescaling the coordinates such that $\rho = \lambda^{-1} \tilde{\rho}$, $y = \lambda^2 \tilde{y}$, the functions $a(\rho)$, $H_1(\rho)$ and $H_2(\rho)$ become $\tilde{a}(\tilde{\rho}) = \lambda a(\lambda^{-1} \tilde{\rho})$, $\tilde{H_1}(\tilde{\rho}) = \lambda^2 H_1(\lambda^{-1} \tilde{\rho})$, $\tilde{H_2}(\tilde{\rho}) = \lambda^2 H_2(\lambda^{-1} \tilde{\rho})$ which still provide a solution for the supersymmetry equations. We shall use this scaling symmetry later to eliminate unphysical parameters and to help us interpolating the near-boundary and near-horizon perturbative solutions we will construct.

\subsection{First integrals and conserved charges} \label{sec:first_integrals}
The analysis performed in~\cite{Cassani:2018mlh} is able to find three first integrals for the black hole solutions constructed there (cf. equations~(3.45), (3.46), (3.47) in the above-mentioned paper). Two of them are straightforwardly obtained by considering the component of the Maxwell equation parallel to $\bar{X^I}$ and by the orthogonal ones; while the last first integral is derived by manipulating the component ${}^t{}_{ \psi}$ of the Einstein equations using the Maxwell equation and the supersymmetric conditions. In particular in the special ansatz considered in~\cite{Cassani:2018mlh} all the $H_I$ functions are equal (up to a constant) and therefore all the components of the Maxwell equation orthogonal to $\bar{X^I}$ are also equal (up to a constant), so they globally provide only one non-trivial first integral. Thus a total number of three first integrals is obtained: two coming from the Maxwell equation and one from the Einstein equations.

In the general case of $n_V + 1$ different $H_I$ functions, we find a total number of $n_V +  3$ first integrals: one coming from the component of the Maxwell equation parallel to $\bar{X}_I$, $n_V + 1$ coming from the orthogonal ones and finally one can be derived from the ${}^t{}_{\psi}$ component of the Einstein equations. These first integrals are given by
\begin{subequations}\label{KappaCharges}
	\begin{align}
	\CK_1 &=  a^3 a^\prime \left( f_{\rm min}^{-1} \right)' + \frac{1}{\ell} a^2 w  + \frac{\ell^2p^2}{18} + \frac{36}{\ell^2a^4}  C^{IJK}\bar{X}_{I} H_JH_K  \,, \\
	\CK^{(I)}_2 &=  H_I'' - \left(\frac{3a'}{a}+\frac{a''}{a'}\right) H_I' + \frac{2p}{3a^2}H_I + \frac{24}{\ell^2a^4} \left( \bar{Q}_{IJ} - \frac{3}{2}\bar{X}_I \bar{X}_J \right) (CHH)^J \,, \\
	\CK_3 &= \frac{a}{a' f^3} \left( f^3 w^2 - 4 a^2 (a')^2 \right)^2 \left[ \frac{f^3 w}{ f^3 w^2 - 4 a^2 (a')^2} \right]' \nonumber \\
	& \qquad - 12 A^I_\psi  \left( \CK_1 \bar X_I + \CK^{(I)}_2 q_I \right) + \frac{1}{3} \, C_{IJK} A^I_\psi A^J_\psi A^K_\psi \,,
	\end{align}
\end{subequations}
where the $q_I$ are defined such that $\bar X^I q_I = 0$ and they have to satisfy eq.~(2.55) of \cite{Cassani:2018mlh}. Note that in the general $n_V$ case, the $H_I$ functions satisfy the constraint $\bar{X}^I \, H_I = 0$, therefore one of the above first integrals is dependent from the others and thus there are $n_V + 2$ independent first integrals. 

As it is for the three first integrals of~\cite{Cassani:2018mlh}, also the generalized $n_V+3$ ones we found have an interpretation in terms of conserved charges. We introduce the conserved Page electric charges as\cite{Page:1984qv}
\begin{equation}
\label{Page_Charges_Definition}
P_I = \frac{1}{\kappa^2} \int_{\Sigma_\infty} \left( Q_{IJ} \star F^J + \frac{1}{4} \, C_{IJK} A^J \wedge F^K \right) ,\\
\end{equation}
where $\Sigma_\infty$ denotes the three-dimensional $\rho = \infty$ hypersurface obtained by the general family $\Sigma_\rho$ which foliates a generic Cauchy surface (a hypersurface of constant time).
By using the various definitions and relations given in the subsections above for the quantities appearing in \eqref{Page_Charges_Definition}, it is possible to show that the Page charges can be rewritten as
\begin{equation}
\label{Page_Charges_First_Integral}
P_I = - \frac{48 \pi^2 \ell^2}{\kappa^2} \left( \CK_1 \bar X_I + \CK_2^{(I)} q_I \right) \, ,
\end{equation}
with the overall factor introduced for convenience.
The Page charges are therefore described by all the first integrals which derive from the Maxwell equation. 
The last first integral can be interpreted by considering a conserved angular momentum $J$ referred to the symmetry generated by the killing vector $\frac{\partial}{\partial \, \psi}$. This is given by the following generalized Komar integral~\cite{Cassani:2018mlh}:
\begin{equation}
\label{Angular_Momentum_Komar_Definition}
J = \frac{1}{2\kappa^2} \int_{\Sigma_\infty}  \left[  \star\,  \diff K + 2 \iota_K A^I   \left( Q_{IJ} \star F^J + \frac{1}{4} \, C_{IJK} A^J \wedge F^K \right)  \right]  .
\end{equation}
Evaluating this angular momentum on the supergravity background we are considering, we find
\begin{equation}
\label{Angular_Momentum_First_Integral}
J = \frac{4 \pi^2 \ell^3}{\kappa^2} \, \CK_3 \,.
\end{equation}
so that $J$ is proportional to the last first integral. 

For the present paper, we are interested in the case $n_V = 2$, therefore we should have $5$ first integrals in total. By setting $n_V=2$ in \eqref{KappaCharges} and performing the needed contractions, we find the following relations for our U$(1)^3$ case: 
\begin{subequations}
	\label{KappaCharges_U1}
	\begin{align}
	\CK_1 &=  a^3 a^\prime \left( f_{\rm min}^{-1} \right)' + \frac{1}{\ell} a^2 w  + \frac{\ell^2p^2}{18} + \frac{24}{\ell^2a^4} \, \Sigma \,, \\
	\CK^{(1)}_2 &= H_1'' - \left(\frac{3a'}{a}+\frac{a''}{a'}\right) H_1' + \frac{2p}{3a^2}H_1 +  \frac{8}{\ell^2a^4} \left( H_1^2- 2 H_1 H_2 - 2H_2^2  \right) ,\\ 
	\CK^{(2)}_2 &= H_2'' - \left(\frac{3a'}{a}+\frac{a''}{a'}\right) H_2' + \frac{2p}{3a^2}H_2 +  \frac{8}{\ell^2a^4} \left( -2 H_1^2 - 2 H_1 H_2+ H_2^2  \right) ,\\ 
	\CK^{(3)}_2 &= - (H_1''+ H_2)'' + \left(\frac{3a'}{a}+\frac{a''}{a'}\right) (H_1'+H_2') - \frac{2p}{3a^2}(H_1+H_2)  \notag  \\ 
	& \qquad + \frac{8}{\ell^2a^4} \left( H_1^2+4 H_1 H_2 + H_2^2 \right) , \\
	\CK_3 &=  \frac{a}{a' f^3} \left( f^3 w^2 - 4 a^2 (a')^2 \right)^2 \left[ \frac{f^3 w}{ f^3 w^2 - 4 a^2 (a')^2} \right]' \nonumber \\
	& \qquad - 12  \left( \CK_1 A^I_\psi  \bar X_I + \frac{1}{6}  (\CK^{(1)}_2 A_\psi^1 + \CK^{(2)}_2 A_\psi^2) -\frac{1}{3} \, \CK_2^{(3)} A_\psi^3 \right) + 2  A^1_\psi A^2_\psi A^3_\psi \,.
	\end{align}
\end{subequations}
Note that the $q_I$ are fixed to be $q_1 = q_2 = \frac{1}{6}, q_3 = - \frac{1}{3}$, as it is for the U$(1)^3$ version of the solution of~\cite{Cassani:2018mlh}. Moreover, equations \eqref{Page_Charges_First_Integral} and \eqref{Angular_Momentum_First_Integral} are still valid with the $I$ index which runs from $1$ to $3$. In \eqref{KappaCharges_U1} we already used the constraint \eqref{H_constraint} to eliminate $H_3$ in favour of $H_1$ and $H_2$. As a consequence, we immediately see that $\CK^{(1)}_2$, $\CK^{(2)}_2$ and $\CK^{(3)}_2$ are not independent, but they satisfy the relation
\begin{equation}
\label{Kappa_Constraint}
\CK^{(1)}_2 + \CK^{(2)}_2 + \CK^{(3)}_2 = 0 \,,
\end{equation}
so that we have $4$ independent first integrals in total. We will always use \eqref{Kappa_Constraint} to trade $\CK^{(3)}_2$ with $\CK^{(1)}_2$ and $\CK^{(2)}_2$, so the set of independent first integrals we choose is $(\CK_1, \CK^{(1)}_2, \CK^{(2)}_2, \CK_3)$.
As we shall see later in the paper, these first integrals will also help us to connect the parameters of the perturbative near-boundary solution we will construct with the parameters of the near-horizon one.

The Page charges $P_I$ and the quantity $J$ defined by a Komar integral can be regarded as the electric charges and the angular momentum of the solution. However the procedure of holographic renormalization, which we will employ later in the paper, gives the possibility to define in a different manner analogous conserved quantities playing the same role; we will therefore compare them with the conserved Page charges and angular momentum we have defined in the present section. In particular, since the contribution provided by the Chern-Simons term to the holographic charges is different to the one for the Page charges, we should expect that those conserved quantities are not equal.

\subsection{Closing remark: uplift to type IIB supergravity} \label{sec:uplit}
Here we briefly review how we can embed ${\cal N}=2$, $D=5$ supergravity with U${(1)}^3$ gauge group in type IIB supergravity following~\cite{Cvetic:1999xp}. Starting from type IIB on AdS$_5 \times S^5$, we can have a consistent truncation turning on the $\tau = C_0 + i e^{-  \Phi}$, $C_4$ and $g_{MN}$ fields, where $x^M = (x^\mu, y^a)$ with $x^\mu$ being the AdS$_5$ coordinates and $y^a = (\tilde \theta, \tilde \psi , \tilde \phi_1, \tilde \phi_2, \tilde \phi_3)$ being the $S^5$ coordinates\footnote{In this set of coordinates the round $S^5$ is 
	\be
	\diff \Omega_5^2 = \sum_i \left( \diff \mu_i^2 + \mu_i \, \diff\tilde \phi_i^2 \right) .
	\ee
}. 

We can write %
\begin{subequations}\label{eq_IIB}
\begin{align}
	\diff s_{10}^2 &= \sqrt{\widetilde{\Delta}} \, \diff s^2_{5} + \frac{1}{\ell^2 \sqrt{\widetilde{\Delta}}} \, \diff \tilde  s^2_5\,, \qquad \widetilde{\Delta} = \sum_{i=1}^3 X_i \mu_i \,, \\
	\diff \tilde s^2_5 &= G_{ab} \diff y^a \diff y^b = \sum_i X_i^{-1} \left[ \diff \mu_i^2 + \mu_i \left(\diff \tilde \phi_i^2 + \ell A_i^{(1)}  \right)^2  \right] \,, \\
	\mu_1 &= \sin \tilde \theta\,, \quad \mu_2 = \cos \tilde\theta \sin \tilde\psi \,, \quad \mu_3 = \sin \tilde\theta \cos \tilde\psi \,, \\
	X_i &= e^{-\half \boldsymbol{a}_i \cdot \boldsymbol{\varphi} } \,, \quad X_1 X_2 X_3 = 1 \, , \quad F_{(2)}^i = \diff A_{(1)}^i \,, \\
	a_1 &=  \left( \frac{2}{\sqrt{6}} \,, \,+  \sqrt{2} \right) , \quad a_2 =  \left( \frac{2}{\sqrt{6}} \,, \, - \sqrt{2} \right) , \quad a_3 =  \left( -\frac{4}{\sqrt{6}} \,, \, 0 \right) , \\
	C_2 &= 0 = B_2 \,, \quad F_5 = \diff C_4 =  G_5 + \star_{10} G_5 \,,\\
	G_5 &= 2 \ell \sum_i \left( X_i \mu_i^2 - \widetilde{\Delta} \, X_i \right) {\rm vol}_5 - \frac{1}{2\ell} \, \sum_i \star_5 \diff \log X_i \wedge \diff \mu_i^2\notag \\
	& \qquad  + \frac{1}{2\ell^2} \sum_i \diff \mu_i^2 \wedge \left(\diff \tilde \phi_i^2 + \ell A_i^{(1)}  \right) \wedge \star_5 F_{(2)}^i \,,
\end{align}
\end{subequations}
here $\star_5 \equiv \star$ and ${\rm vol}_5$ are referred to the AdS$_5$ metric. 
Considering the only relevant part of the usual type IIB lagrangian
\be
{\cal L}_{\rm IIB} \supseteq  \sqrt{- g_E} \left[ R_E - \half \, \frac{\pd \tau \cdot \pd \overline{\tau}}{{\rm Im}\,  \tau} - \half \, F_5 \wedge \star_{10} F_5 \right] ,
\ee
and inserting the ansatz~\eqref{eq_IIB}, with an appropriate field-redefinition, we land to eq.~\eqref{Bulk_action}.


\section{Constructing the solution} \label{sec:Solution}

The three ODE's  \eqref{eqforH}, \eqref{eqforK}, and \eqref{eqforaU1} we obtained are difficult to solve analytically, therefore we resort to a numerical method to find new solutions. In order to do so, we series-expand the fields both in the near-boundary region $\rho \to \infty$ and in the near-horizon one $\rho \to 0$, and we fix the series coefficients by solving the ODE's order by order; after that we build our numerical solution by matching the two expansions in the bulk \emph{i.e.} in a region where they overlap. Our perturbative and numerical approaches are similar in spirit to the ones adopted in~\cite{Cassani:2018mlh}, however here we will not impose any ansatz on the scalar fields and we will therefore look for more general solutions. In the near-boundary region we find expansions that are compatible with AlAdS solutions, in the near-horizon one we note that there are solutions which possess the characteristics of black holes and, using a numerical procedure, we establish that there are well-behaved solutions interpolating between these two regimes. For ease of notations, we will set $\ell =1$ in the whole section and we also change the label of the functions $H_1$ and $H_2$ to:
\begin{equation}
H_1 \left(\rho \right) \to H \left(\rho \right) ,  \qquad H_2 \left(\rho \right) \to K \left(\rho \right) .
\end{equation}

\subsection{Near-boundary analysis}\label{sec:near_boundary}

We now perturbatively solve the equations~\eqref{eqforH},~\eqref{eqforK} and~\eqref{eqforaU1} around $\rho \to \infty$. This is the limit in which the solution approaches the conformal boundary. The unknown functions are $a$, $H$ and $K$. We assume for them the following asymptotic expansions 
\begin{align}
a(\rho) & = a_0 e^\rho \left[ 1 + \sum_{k \geq 1} \, \sum_{0 \leq n \leq k} \, a_{2k,n} \, \rho^n \, \left(a_0 \, e^\rho \right)^{-2k} \right] \notag \\
& = a_0 e^\rho \left[1 + \left(a_{2,0} + a_{2,1} \rho \right) \frac{e^{-2\rho}}{a^2_0} + \left(a_{4,0} + a_{4,1} \, \rho + a_{4,2} \, \rho^2 \right) \frac{e^{-4\rho}}{a^4_0}  + \dots \right] ,\\ \notag 
& \\
H(\rho) & = a_0^4 e^{4 \rho} \left[\sum_{k \geq 0} \, \sum_{0 \leq n \leq k} \, H_{2k,n} \, \rho^n \, \left(a_0 \, e^\rho \right)^{-2k} \right] \notag \\
& = a_0^4 e^{4\rho} \bigg[H_{0,0} + \left(H_{2,0} + H_{2,1} \rho \right) \frac{e^{-2\rho}}{a^2_0} + \left(H_{4,0} + H_{4,1} \, \rho + H_{4,2} \, \rho^2 \right) \frac{e^{-4\rho}}{a^4_0} + \dots \bigg] , \\ \notag
& \\ 
K(\rho) & = a_0^4 e^{4 \rho} \left[\sum_{k \geq 0} \, \sum_{0 \leq n \leq k} \, K_{2k,n} \, \rho^n \, \left(a_0 \, e^\rho \right)^{-2k} \right] \notag \\
& = a_0^4 e^{4\rho} \bigg[K_{0,0} + \left(K_{2,0} + K_{2,1} \rho \right) \frac{e^{-2\rho}}{a^2_0} + \left(K_{4,0} + K_{4,1} \, \rho + K_{4,2} \, \rho^2 \right) \frac{e^{-4\rho}}{a^4_0} + \dots \bigg] .
\end{align}
We furthermore assume $a_0 \neq 0$. We have included only odd powers of $e^\rho$ in the expansion for $a$: that is because any term weighted by an even power of $e^\rho$ would vanish because of the equations. For an analogous reason, the expansions for $H$ and $K$ involve only even powers of $e^\rho$.
We have obtained a perturbative solution for the three equations \eqref{eqforH}, \eqref{eqforK} and \eqref{eqforaU1} which is valid up to order $\CO \left(e^{-10 \, \rho} \right)$ and is controlled by the following eleven parameters\footnote{In principle, other solutions are possible. They have $H_{0,0} \neq 0$ or $K_{0,0} \neq 0$, so the $H$ and $K$ functions have a different leading behaviour. However these solutions present metrics which are not AlAdS, since their leading term is of order $\CO\left(e^{4 \, \rho} \right)$. We are interested only in AlAdS behaviours, therefore we will not discuss these solutions in the following.}:
\begin{align}
& \qquad \qquad a_0 \, , \, \qquad  a_2 = a_{2,0} \, , \qquad c = a_{2,1} \, , \qquad a_4 = a_{4,0} \, , \qquad a_6 = a_{6,0} \, , \notag \\ \vspace{0.6cm}
& \qquad \qquad \qquad  H_2 = H_{2,0} \, , \qquad H_4 = H_{4,0} \, , \qquad \tilde{H} = H_{2,1} \, , \notag \\
& \qquad \qquad \qquad K_2 = K_{2,0} \, , \qquad K_4 = K_{4,0} \, , \qquad \tilde{K} = K_{2,1}  \,. \notag
\end{align}
The explicit form of the first terms of the perturbative solutions for $a$, $H$ and $K$ are given by equations~\eqref{Sol_a_UV}, ~\eqref{Sol_H_UV} and~\eqref{Sol_K_UV} reported in app.~\ref{app_more_UV}, where we provide further details about the near-boundary solution.

Using the near-boundary solution we found, we can perturbatively evaluate all the other relevant functions. However before computing them, we introduce the following parameter
\be
\label{Squashing_from_c}
v^2 = 1-4c\ ,
\ee 
which will be related to the squashing of the three-sphere at the boundary. We will trade the parameter $c$ for $v^2$ when writing the main results of the present paper, since the latter has a clearer physical interpretation. We furthermore define the change of coordinates
\be
\label{Change_Coord}
y = t \ , \qquad\qquad
\hat{\psi} = \psi - \frac{2}{v^2} \, t \ ,
\ee
which trades $y, \hat \psi$ for $t, \psi$; in the latter set of coordinates it is easier to see that, at the conformal boundary, the metric is static. 
In the new set of coordinates, the metric and the gauge fields turn to:
\begin{subequations}
\begin{align} 
\diff s^2 &= g_{\rho \rho} \diff  \rho^2 + g_{\theta \theta} (\sigma_1^2 + \sigma_2^2) + g_{\psi \psi} \sigma_3^2 + g_{tt} \diff t^2 + 2 g_{t \psi} \, \sigma_3 \, \diff t\ , \label{comp_5dmetric} \\ 
A^I &= A^I_t \, \diff t + A^I_\psi \, \sigma_3\ ,\label{comp_5dgaugefield}
\end{align}
\end{subequations}
the $\sigma_i$ being defined in the same way as the $\hat \sigma^i$ with $\psi$ replacing $\hat \psi$.
 
The functions $f$ and $w$, which are independent on the change of coordinates \eqref{Change_Coord}, can be easily evaluated using \eqref{f_from_a_U1} and \eqref{w_from_a_U1}. From their explicit expansions~(\ref{eq_asymptf}, \ref{eq_asymptw}), we see that $f$ goes to $1$ in the near-boundary limit, while the $w$ function has a $e^{2 \, \rho}$ leading term. Both of these two near-boundary behaviours are fully consistent with an AlAdS solution.

We can proceed to compute the metric \eqref{comp_5dmetric} in order to verify that this is indeed static. We find that it can be rewritten as
\be
\diff s^2 = \diff \rho^2 + e^{2\rho}\,\diff s^2_{\rm bdry}+ \ldots\ ,
\ee
with $\diff s^2_{\rm bdry}$ being the metric at the boundary, which reads
\be\label{bdrymetric}
\diff s^2_{\rm bdry} \ = \  (2a_0)^2\left[ - \frac{1}{v^2} \, \diff t^2 + \frac{1}{4}\left(\sigma_1^{\,2} + \sigma_2^{\,2}  +   v^2\sigma_3^{\,2}\right) \right] ,
\ee
so it is indeed static, as wanted. Looking at~\eqref{bdrymetric} we can also see that the last term is the metric of a three-sphere whose squashing is controlled by the parameter $v$. 
The change of coordinates affects also the supersymmetric Killing vector $V$, which becomes:
\begin{equation}
\label{Killing_tpsi}
V = \frac{\partial}{\partial y} 
= \frac{\partial}{\partial t} + \frac{2}{v^2} \frac{\partial}{\partial \psi}\ .
\end{equation}

We present the asymptotic form of the scalars $X^I$ and the near-boundary expansions of all the components of both the metric and the gauge fields in app.~\ref{app_FG}, where we express the solution in Fefferman-Graham coordinates. The possibility to write our solution in Fefferman-Graham form will also confirm once more that it is indeed AlAdS$_5$ and the following analysis will show that four of the eleven free parameters, $a_0, c, \HL$ and $\KL$ determine the bulk fields at the boundary and therefore play the role of a source in the dual quantum field theory. In particular, looking at~\eqref{bdrymetric} and recalling that $v^2$ is related to $c$ as in~\eqref{Squashing_from_c}, it is possible to predict that $a_0$ and $c$ would control the metric at the conformal boundary, while $\HL$ and $\KL$ should determine the scalar fields. The analysis in Fefferman-Graham coordinates will reveal that this is indeed the case. We refer to app.~\ref{app_FG} for further details on the Fefferman-Graham form of the metric. 

We also recall that in sec.~\ref{sec:first_integrals} we have introduced the four independent first integrals $\CK_1, \, \CK_2^{(1)}, \, \CK_2^{(2)}$ and $\CK_3$. These are given by equations~\eqref{KappaCharges_U1}. Since the first integrals are obviously constants, we can evaluate them in the near-boundary region using the expansions~\eqref{Sol_a_UV},~\eqref{Sol_H_UV},~\eqref{Sol_K_UV} we have found. In this way, we find that the three first integrals coming from the Maxwell equation, $\CK_1, \, \CK_2^{(1)}$ and $\CK_2^{(2)}$, depend on the various near-boundary free parameters, among which the most subleading are $a_4, \, H_4$ and $K_4$. We can use these relations to express the latter free parameters with respect to the others and the first integrals. In the same fashion, we find that the first integral coming from the Einstein equations, $\CK_3$, depends on $a_6$; therefore we also get an expression for $a_6$ with respect the other near-boundary free parameters and the first integrals. These relations for $a_4, \, H_4, \, K_4$ and $a_6$ are rather involved and are thus reported in app.~\ref{app_First_Integrals}, in particular they are given by equations~\eqref{First_Integral_UV}.
The advantage we get from these equations is the following: we are able to evaluate all the first integrals both in the near-boundary region as well as in the near-horizon region, obtaining them as functions of, respectively, the near-boundary and the near-horizon parameters; combining the relations obtained in the near-horizon with the ones obtained in the near-boundary we will be able to express $a_4, \, H_4, \, K_4, \, a_6$ as functions of only the remaining near-boundary parameters and the near-horizon ones. This will allow us to replace the most subleading  parameters $a_4, \, H_4, \, K_4, \, a_6$ with the others.


\subsection{Near-horizon analysis} \label{sec:near_horizon}

Having shown that solutions compatible with an AlAdS$_5$ behaviour exist in the near-boundary, we proceed to solve the ODEs~\eqref{eqforH},~\eqref{eqforK} and~\eqref{eqforaU1} in the region near to $\rho \to 0$, which we identify with the interior region of our solution. It is reasonable to assume that the three unknown functions $K$, $H$ and $a$ can be written as a Taylor expansion near $\rho = 0$ 
\begin{align}
\label{expansion_IR_gen}
a(\rho) \,&=\, \air_0 + \air_1 \, \rho + \air_2 \, \rho^2 + \dots\ , \nn \notag \\[1mm]
H(\rho) \,&=\, \hir_0 + \hir_1 \, \rho + \hir_2 \, \rho^2 + \dots\ ,  \nn \notag  \\[1mm]
K(\rho) \, & = \, \kir_0 + \kir_1 \, \rho + \kir_2 \, \rho^2 + \dots  \, .
\end{align}
We want to search for either new black hole solutions or new soliton solutions; in the first case we should have an event horizon at $\rho = 0$, in the second one the solution should close off smoothly in the same point. Looking at the metric~\eqref{metric}, we see that both the types of solutions require $\air_0 = 0$, which we therefore assume. 
Moreover, due to the symmetries of the ODE's we can take $\air_1 > 0$ with no loss of generality\footnote{In principle, we could also assume $\air_1 = 0$ and search for solutions with $\air_2 \neq 0$. However we have verified that such solutions do not exist in minimal gauged supergravity theory, therefore we are not interested in this possibility for the purposes of the present paper, since we want to consider only solutions which have a minimal limit.}.

We proceed to solve~\eqref{eqforH},~\eqref{eqforK} and~\eqref{eqforaU1} perturbatively up to $\CO(\rho^{13})$. We find that equations~\eqref{eqforH} and~\eqref{eqforK} fix uniquely the form of $K$ and $H$ with the coefficients $\kir_0, \, \hir_0, \, \kir_1, \, \hir_1$ forced to vanish and all the others determined by the free parameters $\kir_2$, $\hir_2$ and by the coefficients of $a$. We find it convenient to define the new parameters
\be
\air\equiv\air_1\ ,\qquad \hir \equiv \frac{\hir_2}{\air_1^2}\ , \qquad \kir \equiv  \frac{\kir_2}{\air_1^2} \, ,
\ee
so that the first terms in the expansions of $K$ and $H$ are: 
{\small
\begin{align}
H(\rho) & \simeq \hir \, \air^2 \, \rho^2 - \frac{2 \, \air  \, \air_2}{\air^4 + 4 \, \air^2 - 6912 \left(\hir^2 + \hir \,  \kir + \kir^2 \right)+4} \times \notag \\
& \quad \times \left[192 \, \hir ^2 \left( \air ^2 + 36 \, \kir \right) - 384 \, \air^2 \, \kir^2 + \hir \, \left(3 \, \air^4 + \air^2 (4-384 \, \kir ) + 6912 \, \kir^2 -4 \right) + 6912 \, \hir^3\right] \rho^3 ,\\
K(\rho) & \simeq \kir \, \air^2 \, \rho^2 - \frac{2 \, \air  \, \air_2}{\air^4 + 4 \, \air^2 - 6912 \left(\kir^2 + \kir \,  \hir + \hir^2 \right)+4} \times \notag \\ 
& \quad \times  \left[192 \, \kir ^2 \left( \air ^2 + 36 \, \hir \right) - 384 \, \air^2 \, \hir^2 + \kir \, \left(3 \, \air^4 + \air^2 (4-384 \, \hir ) + 6912 \, \hir^2 -4 \right) + 6912 \, \kir^3\right] \rho^3 . 
\end{align}}%
Note that switching the parameters $\hir \leftrightarrow \kir$  we have that $H \leftrightarrow K$, as expected. 

The above expansions solve~\eqref{eqforH} and~\eqref{eqforK} without imposing any conditions on $a$. This function will then be constrained by the remaining equation~\eqref{eqforaU1} on which we now focus. The solution process of the latter equation brings us to distinguish between different cases. To solve the first non-trivial order of~\eqref{eqforaU1} we must satisfy the condition:%
\begin{equation}
\begin{split}\label{eq_choiceair2}
\air_2  \Big[ & 13 \, \air^6 + 60 \, \air^4 -12 \, \air^2 \left(6912 \left(\hir^2+\hir \, \kir +\kir^2 \right)-7\right) \\
& \qquad \quad - 32 (36 \, \hir  +1) (36 \, \kir +1) (36 \, \hir  +36 \, \kir - 1)\Big] = 0 \,,
\end{split}
\end{equation}%
which means that either $\air_2 = 0$ or the parenthesis vanishes. An analogous condition was found in \cite{Cassani:2018mlh}, while in the minimal gauged supergravity (which can be obtained by setting $\kir = \hir = 0$) the corresponding condition forces to choose $\air_2 = 0$ since the parenthesis cannot vanish in this theory \cite{Cassani:2014zwa}. Since we are interested in solutions which have a minimal gauged supergravity limit, we choose $\air_2 = 0$ and proceed further\footnote{We have also started exploring the opposite choice, in which $\iota = \iota (\alpha, \eta)$ is fixed by requiring the parentheses of eq.~\eqref{eq_choiceair2} to vanish. All the perturbative small-$\rho$ expansions we have constructed in this case have furnished unphysical solutions, so we do not discuss this possibility in the rest of the paper.}. 
The next order yields:
\begin{align}
\label{Soliton_Order}
\air_4 \bigg[ & 5819 \, \air^6 - 5244 \, \air^4 + 12 \, \air^2 \left(6912 \left(\hir^2 + \hir \,  \kir + \kir^2 \right) + 65 \right) \notag \\
& + 32 (36 \, \hir + 1) (36 \, \kir + 1) (36 \, \hir + 36 \, \kir -1) \bigg] = 0 \,,
\end{align}
this equation can be satisfied if $\air_4 = 0$ or the parenthesis vanishes. In \cite{Cassani:2018mlh} setting the corresponding parenthesis to zero led to the black hole solution studied there, while setting $\air_4$ to zero the only regular solution obtained was the one of \cite{Gutowski:2004yv}. In the minimal theory it is possible to obtain the black hole of \cite{Blazquez-Salcedo:2017ghg} by setting $\air = \sqrt{\frac{8}{11}}$ while the choice $\air_4 = 0$ leads either to the regular soliton of \cite{Cassani:2014zwa} or to the black hole of \cite{Gutowski:2004ez}. The soliton found in \cite{Cassani:2014zwa} is a solution of minimal gauged supergravity of special kind: indeed, considering the expression of $f_{\text{min}}$ in \eqref{f_from_a_U1}, we can write $f$ as
\begin{equation}
\label{f_soliton}
f(\rho) = \frac{12 \, a^3 \, a^\prime}{\left[\left(36 \, H^\prime + \mathcal{P} \right) \left(36 \, K^\prime + \mathcal{P} \right)  \left( - 36 (H + K)^\prime + \mathcal{P} \right) \right]^{1/3}} \, ,
\end{equation}
in order to describe a soliton, the $f$ function must start with a constant term in a small $\rho$-expansion, so that the solution closes smoothly. However from \eqref{f_soliton} we can argue that this is possible only if the numerator and the denominator have the same leading behaviour at small $\rho$. Plugging the expansions \eqref{expansion_IR_gen} in \eqref{f_soliton}, we can easily check whether this is possible or not; in particular we note that the numerator goes as $\rho^3$ at small $\rho$, so we have to impose the same behaviour to the denominator. In the minimal case $H = K = 0$ this is easily achieved by taking $\air = \pm \frac{1}{2}$, since
\begin{equation}
\mathcal{P} = a^{\prime \prime \prime} \, a^3 + a \, a^\prime \left(7 \, a \, a^{\prime \prime} +4 \left(a^\prime \right)^2-1\right) \, ,
\end{equation}
indeed starts with a $\rho^3$ term if and only if $\air = \frac{1}{2}$. This choice for $\air$ is the one taken in \cite{Cassani:2014zwa} and leads the author to find a soliton solution. In the general case we are considering in this paper, $H$ and $K$ are non vanishing, therefore recalling \eqref{expansion_IR_gen} it is evident that the denominator of \eqref{f_soliton} goes always as $\rho$, while the numerator begins with $\rho^3$. We then conclude that there is no possibility to find a soliton solution in the U$(1)^3$ theory with non trivial scalars. We have furthermore verified that, even in our general framework where we do not have imposed any ansatz on the scalar fields, the choice $\air_4 = 0$ leads only to the solution of \cite{Gutowski:2004yv} or to a singular solution\footnote{Indeed, setting $\air_4 = 0$ we find a near-horizon expansion which is compatible with a new black hole, but when integrated numerically towards $\rho \to \infty$ this solution presents divergences in the interior region for all the different initial integration conditions we tried.}.

We therefore proceed to analyse the only new solution we find, which is obtained by setting the parenthesis in~\eqref{Soliton_Order} to zero. This gives%
{\small 
\begin{equation}
\label{iota_BH}
\begin{split}
\kir &= -\frac{\hir}{2} \pm \frac{\sqrt{\left( 72 \, \hir -23 \, \air^2 +2 \right) \left(2 \, \air^2 + 36 \, \hir +1 \right) \left[253 \, \air^4 + \air^2 (792 \, \hir - 206) + 16 (1-18 \, \hir)^2 \right]}}{144 \sqrt{2} \left(2 \, \alpha ^2 + 36 \, \hir +1 \right)} \,.
\end{split}
\end{equation}}%
Here one has the possibility to choose either the plus or the minus sign; we leave this choice unspecified for now and proceed further.
Setting $\iota$ as in \eqref{iota_BH}, we continue to perturbatively solve the equation \eqref{eqforaU1} finding that the solution is uniquely determined in terms of the free parameters $\air, \hir, \air_3$ and $\air_4$. We report in app.~\ref{app_more_IR} the first terms of the expansions of $a$, $H$ and $K$. They are given by equation~\eqref{aHK_nearhorizon}. In the same appendix, we also give more information about the near-horizon solution. Note that looking at~\eqref{aHK_nearhorizon} it can appear that $K$ can be obtained from $H$ switching $\hir \leftrightarrow \kir$, however one must keep in mind that $\kir$ is not a free parameter anymore, being fixed as in \eqref{iota_BH}. 

We now briefly show how to reduce our general solution to the U$(1)^3$ version of the one constructed in \cite{Cassani:2018mlh}. As discussed at the end of sec. \ref{sec:susyU1}, we need to impose $H = K$, which means $\hir = \kir$. The condition \eqref{iota_BH} then becomes an equation for $\hir$ which gives:
\begin{equation}
\hir_{\text{ limit}} = \frac{1}{288} \bigg(-8 + 11 \air^2 \pm 9 \air\sqrt{8  - 11 \air^2}\, \bigg) = \frac{\hir_\text{ there}}{6} \ ,
\end{equation}
which is consistent with the fact that it must be $H_\text{limit} = \frac{1}{6} \, H_\text{there}$ as already stated above. As consequence, the expansions \eqref{aHK_nearhorizon} fully reproduce the ones of \cite{Cassani:2018mlh}. 

The perturbative solution near to $\rho = 0$ is characterized by four free parameters: $\air, \, \air_3, \, \air_4$ and $\hir$. However due to the scale symmetry of the supersymmetry equations discussed at the end of sec.~\ref{sec:susyU1}, we can rescale one of the parameters without changing the solution we found. In particular we note that $\air$ and $\hir$ are left invariant under the action of these symmetries, while $\air_3$ and $\air_4$ can be rescaled. We can therefore argue that only three of the free parameters we found are physical and we choose to consider $\air_3$ as an unphysical parameter. We will explicitly use the possibility to rescale $\air_3$ to numerically match our small-$\rho$ behaviour with the near-boundary one discussed in sec.~\ref{sec:near_boundary}, showing that an interpolating solution indeed exists for different values of the remaining physical free parameters. 

We also define a new parameter $\xi$ such that it is invariant under the scaling symmetry discussed at the end of sec.~\ref{sec:susyU1}:
\begin{equation}
    \label{defining_xi}
    \air_4 = \xi \, \air_3^{3/2} \, .
\end{equation}
We shall use this definition to trade $\air_4$ with $\xi$ wherever the former appears. From now on, our set of independent near-horizon free parameters will then be $(\air, \, \hir, \, \xi$). 

We now proceed to report the explicit near-horizon form of the metric, the gauge fields and the scalars, showing that they are compatible with a black hole which is a generalization of the U$(1)^3$ version of the solution presented in \cite{Cassani:2018mlh}. Here we present only the most relevant information about the near-horizon analysis and we refer to app.~\ref{app_more_IR} for further details. Although it is fixed as in~\eqref{iota_BH}, we will keep $\iota$ for compactness in the formulae below.
We begin by presenting the form the five-dimensional metric takes at leading order in the small $\rho$ expansion:
\begin{equation}
    \label{Metric_IR}
    \diff s^2 = - \frac{48 \, \air^6}{\Delta^2 \, \Theta} \, \rho^4 \, \diff t^2 +  \Delta \left[ \frac{\diff \rho^2}{12 \, \air^2 \, \rho^2} + \frac{1}{12} \left(\sigma_1^2 + \sigma_2^2 \right) + \Theta \left(\sigma_3 - \frac{2}{v^2} \, \diff t  \right)^2 \right] \, ,
\end{equation}
where we have defined as $\Delta$ and $\Theta$ the two following quantities
\begin{subequations}
\begin{align}
    \label{Delta_BH}
   \Delta &= \left[\left(4 \alpha ^2+72 \eta -1 \right) \left(4 \alpha ^2+72 \iota -1 \right) \left(4 \alpha ^2-72 (\eta +\iota )-1 \right) \right]^{1/3} \, , 
     \\
   \Theta &= \frac{1}{48 \, \Delta^3} \bigg\{  256 \alpha ^8-96 \alpha ^4 \left[1728 \left(\eta ^2+\eta \, \iota +\iota ^2\right)+1\right] -32 \alpha ^2 \left[186624 \, \eta \, \iota \, (\eta +\iota ) -1 \right] \notag \\ 
    & \qquad \qquad \quad   -3 \left[1 - 1728 \left(\eta^2+\eta \,  \iota + \iota ^2\right)\right]^2 \bigg\} \, .\label{Theta_BH} 
\end{align}
\end{subequations}
We proceed to report the near-horizon expansions of the scalar fields $X^1$ and $X^2$:
{\small
\begin{subequations}
\begin{align}
    X^1 & = \frac{\Delta}{4 \air^2 + 72 \, \eta -1} \notag \\
    & \quad +\frac{5184 \, \air \, \air_3 \left(4 \alpha ^2+72 \iota -1\right) \left(4 \alpha ^2-72 \left(\eta + \iota \right) -1\right) \left(\eta  \left(4 \alpha ^2-24 \eta -1\right)+48 \, \eta \, \iota +48 \iota ^2\right)}{\Delta ^5} \, \rho^2 \notag \\
    & \qquad + \CO(\rho^4) \, , \\
    X^2 & =\frac{\Delta }{4 \alpha ^2+72 \, \iota -1} \notag\\
    & \quad + \frac{5184 \, \air \,  \air_3 \left(4 \alpha ^2+72 \eta -1\right) \left(\iota  \left(4 \alpha ^2-24 \iota -1\right)+48 \, \eta ^2 + 48 \, \eta \, \iota \right) \left(4 \alpha ^2-72 \left(\eta + \iota \right) -1\right)}{\Delta ^5} \, \rho^2 \notag\\
    & \qquad + \CO(\rho^4) \, , 
\end{align}
\end{subequations}
}%
while the last scalar field $X^3$ is easily determined using the constraint between them, given by eq~\eqref{Scalars_Up_constraint}. We do not report the expansions for the scalars with lower indices, $X_I$, since they can be straightforwardly obtained from~\eqref{Def_XI_Down}.

Finally we show the near-horizon behaviour of the gauge fields, which is %
{\small 
\begin{subequations}
\begin{align}
\label{Gauge_Field_IR}
    A^1 & = - \frac{2}{v^2} \, A^1_\psi (\rho = 0) \, \diff t + \frac{16 \alpha ^4+8 \alpha ^2 (72 \eta -1)+5184 \left(\iota ^2+\iota \, \eta -\eta ^2\right)-144 \eta +1}{12 \left(4 \alpha ^2+72 \eta -1\right)} \, \sigma^3 + \CO(\rho^2) ,\\
    A^2 & = - \frac{2}{v^2} \, A^2_\psi (\rho = 0) \, \diff t + \frac{16 \alpha ^4+8 \alpha ^2 (72 \iota -1)+5184 \left(\eta ^2+\eta \, \iota -\iota ^2\right)-144 \iota +1}{12 \left(4 \alpha ^2+72 \iota -1\right)} \, \sigma^3 + \CO(\rho^2) \, ,
\end{align}
\end{subequations}}%
and, again, the third gauge field $A^3$ can be easily determined by the other two and will not be presented here.

Looking at the expansions of the various supergravity fields reported above, it is evident that the pertubative solution can be regarded as the near-horizon expansion of a black hole whose horizon is located at $\rho = 0$. Indeed, the metric has a divergent radial component which is $\CO(\rho^{-2})$ while its spatial part stays finite as the limit $\rho \to 0$ is approached. Furthermore, the supersymmetric Killing vector $V$, given by eq.~\eqref{Killing_tpsi}, is everywhere timelike but on the horizon, where its norm $-f^2$ vanishes\footnote{This can be easily seen by eq.~\eqref{f_IR} in app.~\ref{app_more_IR}, where the near-horizon expansion of $f$ is presented. From this expansion it is evident that $f \sim \rho^2$ at $\rho \sim 0$ and therefore it vanishes at the horizon.}. All the scalar fields stay regular as the limit $\rho \to 0$ towards the horizon is approached and the same do the gauge fields, which are furthermore transverse to $V$ in the gauge we have chosen\footnote{Indeed, it results $V^\mu \, A^I_\mu = 0$ as it is easy to verify using eq.~\eqref{Gauge_Field_IR}.}.

\begin{figure}
        \begin{subfigure}[b]{0.5\textwidth}
                \centering
                \includegraphics[width=1\linewidth]{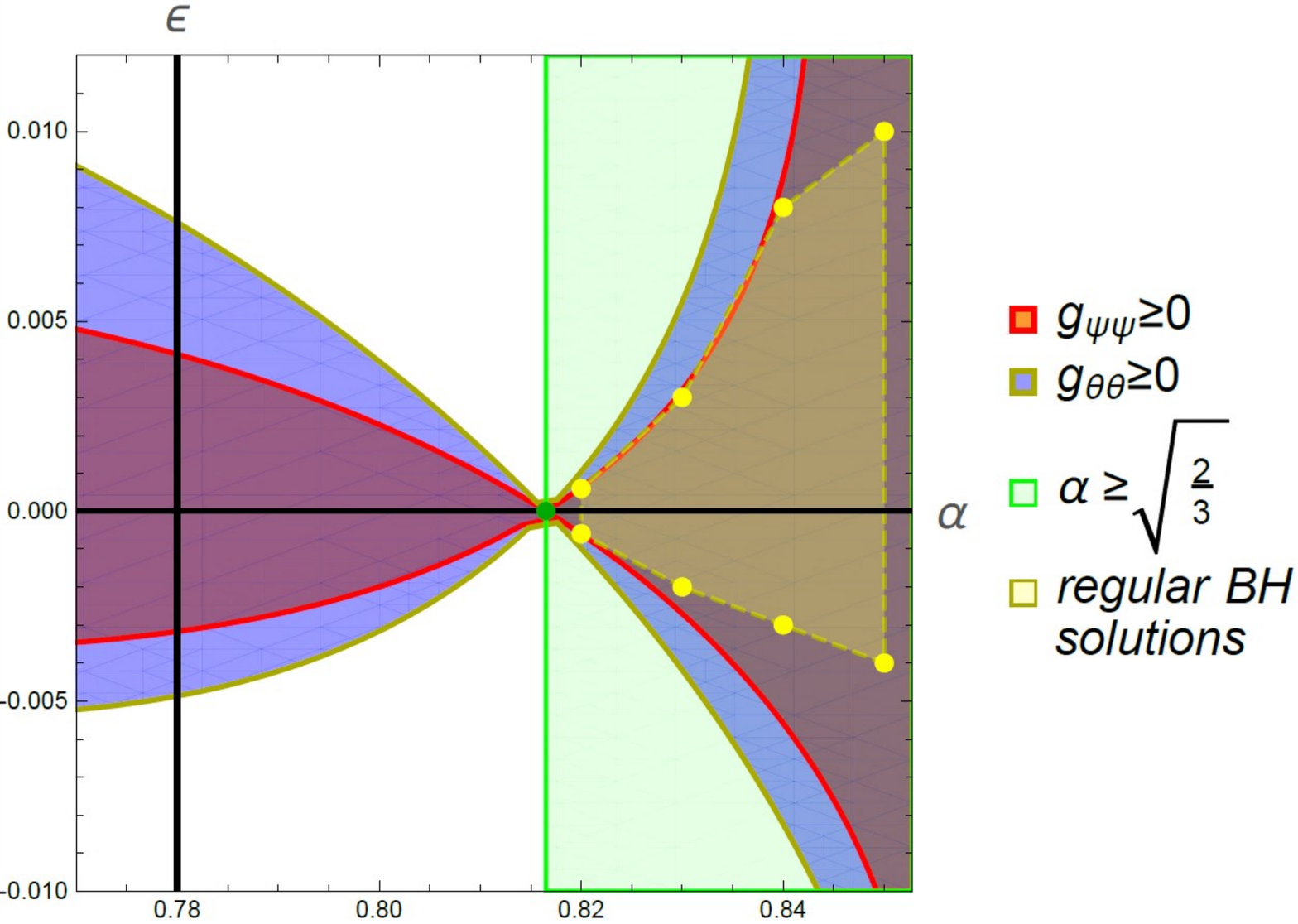}
               \caption{The parameter space for the minus sign.}
                \label{fig:TheTriangleMinus}
        \end{subfigure}%
         \begin{subfigure}[b]{0.5\textwidth}
                     \centering
                \includegraphics[width=1\linewidth]{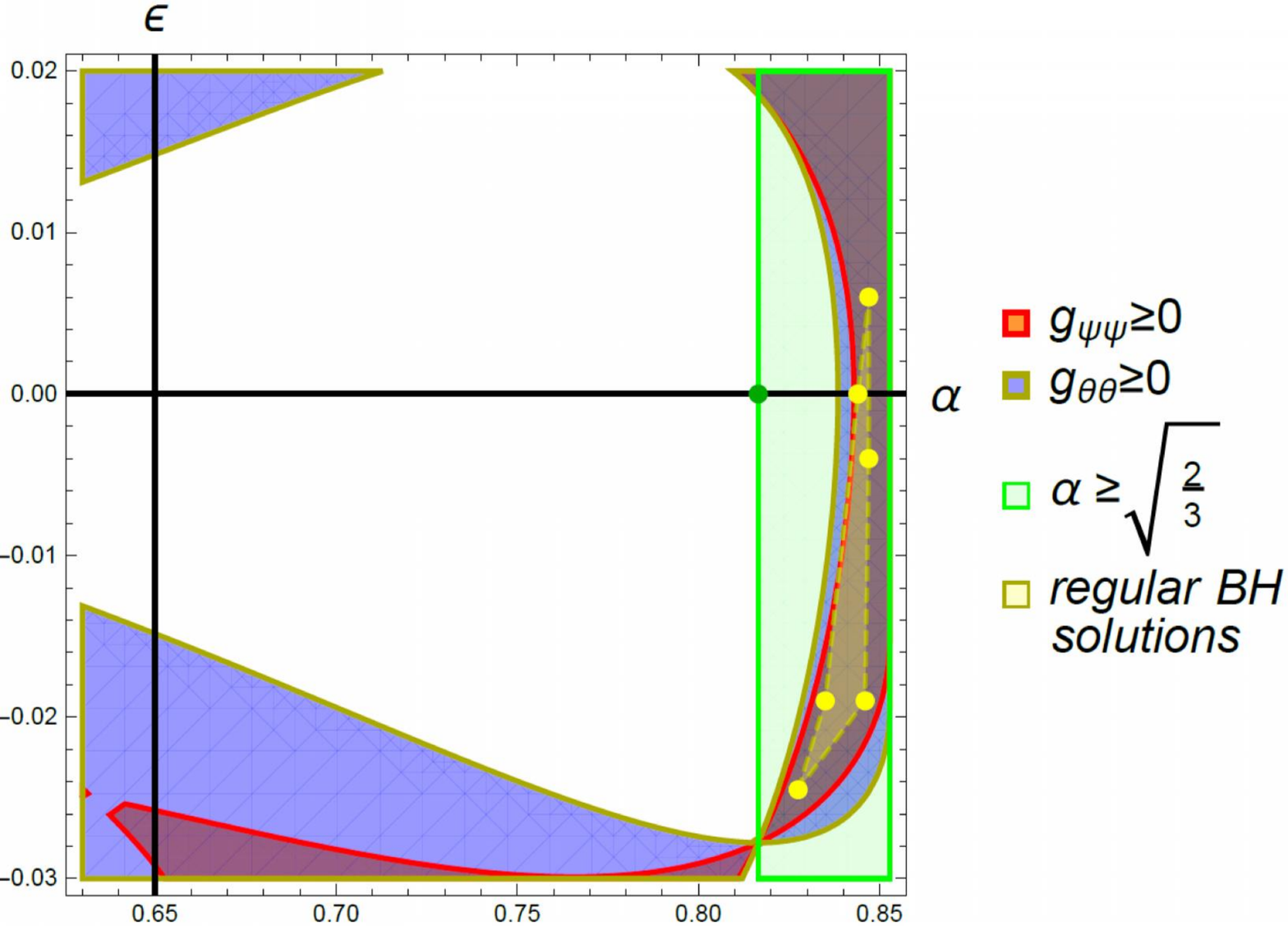}
               \caption{The parameter space for the plus sign. }
                \label{fig:TheTrianglePlus}        \end{subfigure}%
        \caption{ {\small On the left we show the parameter space with the minus sign choice in eq.~\eqref{iota_BH}, while on the right we show the parameter space for the plus sign choice. We shaded in yellow the region where regular black hole solutions are found.} }\label{fig:TheTrianglesPlusMinus}
\end{figure}

We now have to understand for which choice of parameters the solution has a well defined horizon at $\rho =0$. In order to ensure regularity of the horizon we need that all the spatial diagonal metric components in eq.~\eqref{Metric_IR} retain their sign for every value of the radial coordinate $\rho$. This means we must assure that $g_{ii} > 0$ for $i = \rho, \, \theta, \, \phi, \, \psi$\footnote{Actually to ensure regularity of the horizon we should also require that $g_{yy} \leq 0$, where the equal sign holds only at the horizon $\rho = 0$. However this is already guaranteed by the fact that $g_{yy} = -f^2$ with $f$ being, as already stated, a real function which vanishes at the horizon.}. From~\eqref{Metric_IR} it is easy to see that this translates into imposing the conditions $\Delta >0$ and $\Theta > 0$. Indeed, this ensures the positivity of $g_{ii}$ for every value of $\rho$. We are still left with the possible sign choice in eq.~\eqref{iota_BH}; both choices give a well defined black hole solution and we will analyze the parameter space for both of them, even if in the following we will report the numerical results only for the minus sign choice, which is the choice that leads to the largest space of regular solutions. In fig.~\ref{fig:TheTrianglesPlusMinus} we report the parameter space in terms of $\alpha$ and $\epsilon$, with the latter defined via
\begin{equation}
\label{Def_Epsilon}
    \hir = \frac{1}{288} \bigg(-8 + 11 \air^2 \pm 9 \air\sqrt{8  - 11 \air^2}\, \bigg) + \epsilon = \eta_{\text{ limit}} + \epsilon \,,
\end{equation}
so that the limit to $H = K$ case of \cite{Cassani:2018mlh} is simply reproduced by the choice $\epsilon = 0$. Note that we can trade $\hir$ with $\epsilon$ using~\eqref{Def_Epsilon} only if $\alpha \le \sqrt{\frac{8}{11}}$, that is the maximum value considered for $\alpha$ in fig.~\ref{fig:TheTrianglesPlusMinus}; we have analyzed the parameter space for $\alpha>\sqrt{\frac{8}{11}}$ and for generic values of $\eta$ finding that no regular black hole horizon with real coefficients  appears in this region\footnote{Note that the regularity conditions $\Delta > 0$ and $\Theta >0$ must be combined with the existence condition of the square root in~$\eqref{iota_BH}$. We found that these three requests are never simultaneosly verified when $\alpha > \sqrt{\frac{8}{11}}$.}.

In fig.~\ref{fig:TheTrianglesPlusMinus} we have reported the parameter space for both the sign choices. The region colored in red is the region where $g_{\psi\psi}>0$, while the region in blue is where $g_{\theta \theta}>0$. We have colored in yellow the regions where we managed to find numerically a regular black hole solution with real coefficients, by interpolating the near-horizon solution of the present section with the near-boundary one of the previous section. In particular, the yellow dots are the point characterized by the most extreme values for the parameters $\alpha\,, \epsilon$ for which we found a regular numerical solution. The points in the purple region that are not in the yellow one represent values of $\alpha\,, \epsilon$ for which the horizon is well behaved but a full solutions seems not to exist. This is because we find divergences in the bulk when we try to numerically interpolate the near-horizon region with the near-boundary one. Notice that we have reproduced the results of~\cite{Cassani:2018mlh} on the axis $\epsilon= 0$. We have a nice explanation for the peculiar behaviour appearing at the point $(\alpha \,, \epsilon)=(\sqrt{\frac{2}{3}}\,, 0)$, reported in fig.~\ref{fig:TheTriangleMinus} with a green dot, where in~\cite{Cassani:2018mlh} it emerges a non-analytic behaviour of $\Delta$: this is due to the peculiar structure of the $\Delta$ function in the $(\alpha \,, \epsilon)$ plane. 

Notice that, as it can be easily seen in fig.~\ref{fig:TheTrianglePlus}, the region of existence of regular black hole solutions with the plus choice in eq.~\eqref{iota_BH} is smaller than the one obtained with the minus choice; this is clearly visible from the form of the ``yellow triangle'' of solutions in the two cases. This is what is also found in the case $\epsilon = 0$ of~\cite{Cassani:2018mlh}. We also stress the fact that both $g_{\theta \theta}$ and $g_{\psi\psi}$ quickly drop to be negative outside the region of the parameter space we have shown in the figure, so no regular horizon can be found there. There is a possible exception only for $\alpha \in (0, \half)$, where instead we have found a region of regular positive $g_{\theta \theta}$ and $g_{\psi\psi}$, but there $H$ and $K$ becomes complex.

We conclude this section by observing that we can use the near-horizon solution to obtain the dependence of the first integrals~\eqref{KappaCharges_U1} on the near-horizon parameters $\air$ and $\hir$ \footnote{In principle also the parameter $\xi$ could appear in such relations, but it turns out that, since it is quite subleading, it is instead absent.}. In order to do this, we have just to plug our near-horizon expansions for the supergravity functions into~\eqref{KappaCharges_U1} and perform the computations. For conciseness we do not report the expressions such obtained here, but they are given by eq.~\eqref{First_Integral_IR} in app.~\ref{app_First_Integrals}, together with the relations between the near-boundary parameters and the $\CK_I$. As we mentioned at the end of sec.~\ref{sec:near_boundary}, confronting the near-boundary and the near-horizon expressions for the first integrals, we are able to write the most subleading near-boundary parameters $a_4, \, H_4, \, K_4$ and $a_6$ in terms of the remaining near-boundary ones and the near horizon ones. This allows us to eliminate these four parameters in all the expressions and we will proceed by doing it throughout the paper, as it simplifies many expressions. 

\subsection{The matching solution} \label{sec:numerics}

In this section we proceed to match the near-horizon perturbative solution constructed in sec.~\ref{sec:near_horizon} with the near-boundary one obtained in sec.~\ref{sec:near_boundary}, showing that a smooth interpolating solution indeed exist. This happens for all the points in the parameters space of fig.~\ref{fig:TheTrianglesPlusMinus} that are inside the yellow region. 

We begin by giving a brief explanation on how we construct the numerical solution. We use the near-horizon expansions of the previous section to set the initial conditions in the vicinity of the horizon, located at $\rho \simeq 0$, and then we numerically integrate\footnote{To numerically integrate we used the built-in \texttt{NDSolve} command in {\it Wolfram Mathematica}, with the option \texttt{ExplicitRungeKutta}. } the supersymmetry equations~\eqref{eqforH},~\eqref{eqforK} and~\eqref{eqforaU1} towards the near-boundary region, \emph{i.e.} towards large values of $\rho$. We recall that we found in the near-horizon an unphysical parameter, $\air_3$, which may be rescaled at will; we use this possibility to set the appropriate rescaling such that the AlAdS behaviour of $a$ holds in the near-boundary region. Obviously, in order to integrate the equations, we need to give numerical values to the near-horizon parameters $\air$, $\epsilon$ and $\xi$. We tried many different values for the parameters $\air$ and $\epsilon$ in the whole possible region of regularity of the solution (which coincides with the region colored in purple in fig.~\ref{fig:TheTrianglesPlusMinus}) finding regularity in the interior only in the points $(\air, \, \epsilon)$ inside the yellow region. This means that for every point in the yellow region there is an interval of allowed values of $\xi$ for which all the components of the metric, the scalars and the gauge fields are regular. The allowed interval of $\xi$ depends on $\air$ and $\epsilon$ and is determined by regularity of the boundary geometry. All the points outside the yellow region lead to solutions which present fields that are not regular in the bulk; in particular for such solutions the function $f$ turns out to have always a divergence at finite $\rho$. We shall therefore discard such solutions.

We found that the region of regularity of the solution corresponding to the minus sign choice in~\eqref{iota_BH} is inside 
\begin{equation}
    \label{Num_Para_Space}
    \sqrt{\frac{2}{3}} \leq \air \leq \sqrt{\frac{8}{11}} \, \qquad \text{and}  \qquad \, 
    -0.005 \leq \epsilon \leq  0.008 \, ,
\end{equation}
while a similar, but smaller, range is found for the plus sign solution. From now on we will specialize on the minus sign choice, but all the characteristics of the solutions we will discuss are present also in the ones obtained choosing the plus sign.

We constructed the full interpolating solution for many values of the near-horizon parameters inside the bounds reported in~\eqref{Num_Para_Space}. As illustrative examples, we discuss in the following two different analyses performed on the solution: the first is made by fixing $\alpha$ and $\xi$ and studying solutions with different $\epsilon$, the second one is made by fixing $\alpha$ and $\epsilon$ and studying solutions for various values of $\xi$. The first analysis gives us the possibility to compare the characteristics of the new solutions we have found with the ones of~\cite{Cassani:2018mlh}, which are obtained by setting $\epsilon = 0$. Since, as we know from~\cite{Cassani:2014zwa,Blazquez-Salcedo:2017ghg,Cassani:2018mlh}, the parameter $\xi$ is related to the squashing at the boundary, the second analysis allows us to study the new solutions (which present $\epsilon \neq 0$) with different squashed boundary geometries.

We begin by presenting the solutions with different $\epsilon$. We choose $\air = 0.84$ and $\xi = - \frac{1}{4}$. 

\begin{figure}[!ht]
	\label{Fig:a_H_solution}
	\begin{minipage}{.49\textwidth}
		\includegraphics[width=7 cm]{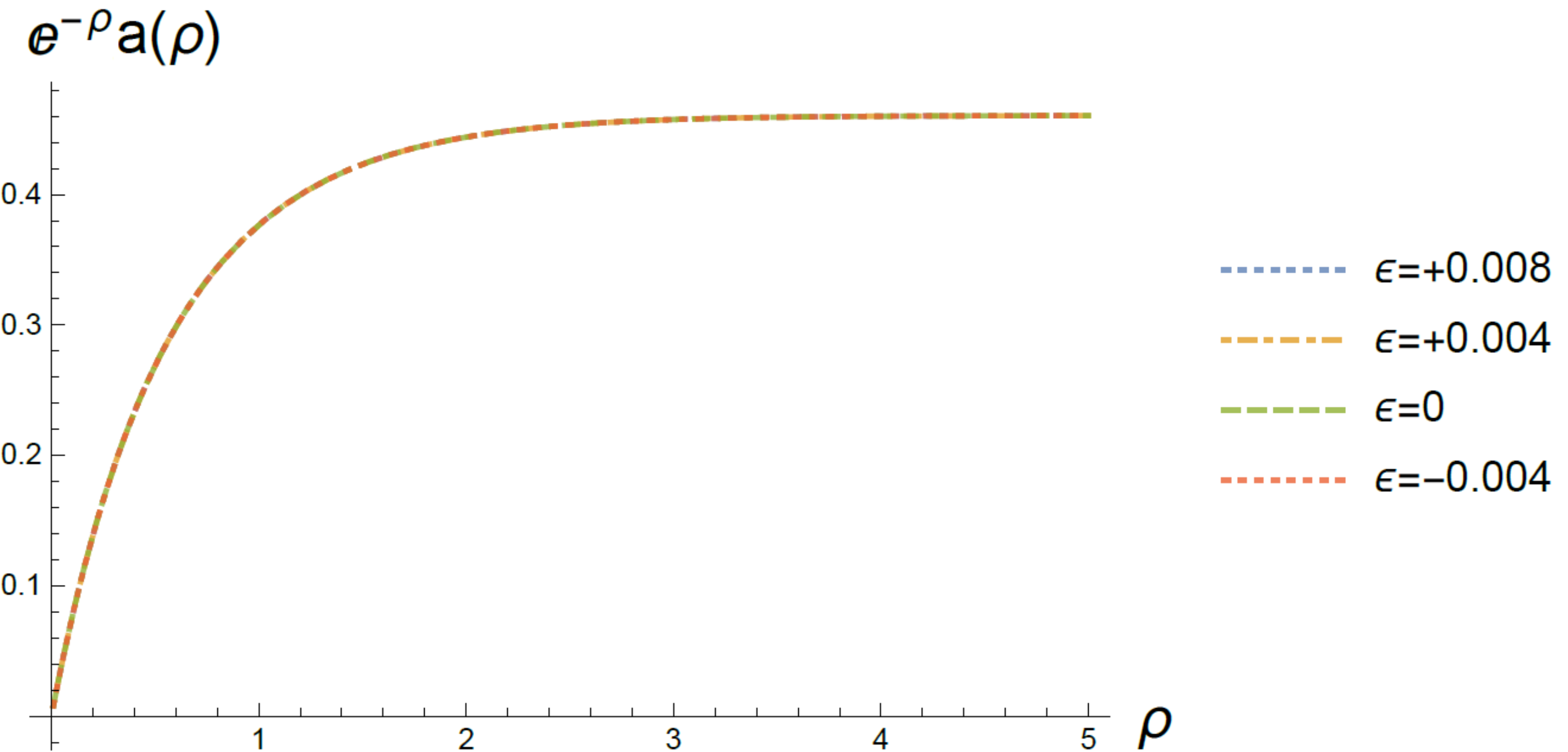}
		\text{(a) The solution $a$.}
	\end{minipage}
	\quad 
	\begin{minipage}{.49\textwidth}
		\includegraphics[width=7 cm]{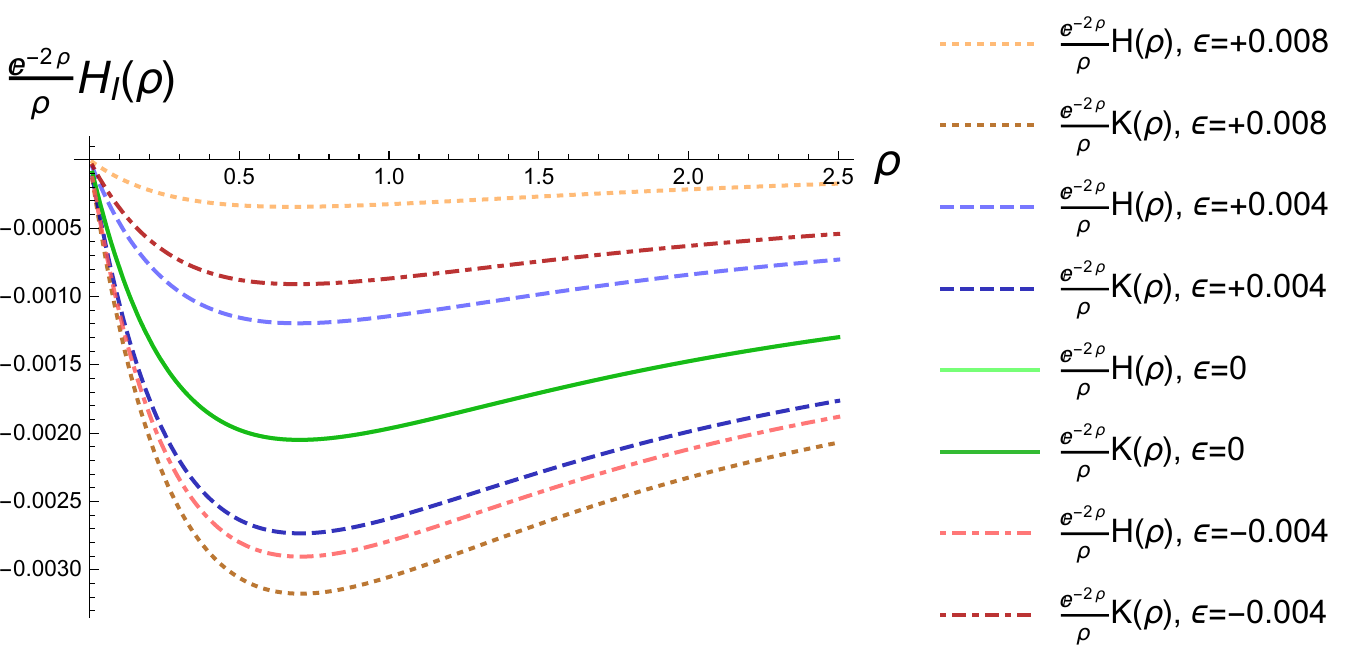}
		\text{(b) The solution $H$, $K$.}
	\end{minipage}
	\bigskip
	
	\begin{minipage}{.49\textwidth}
		\includegraphics[width=7cm]{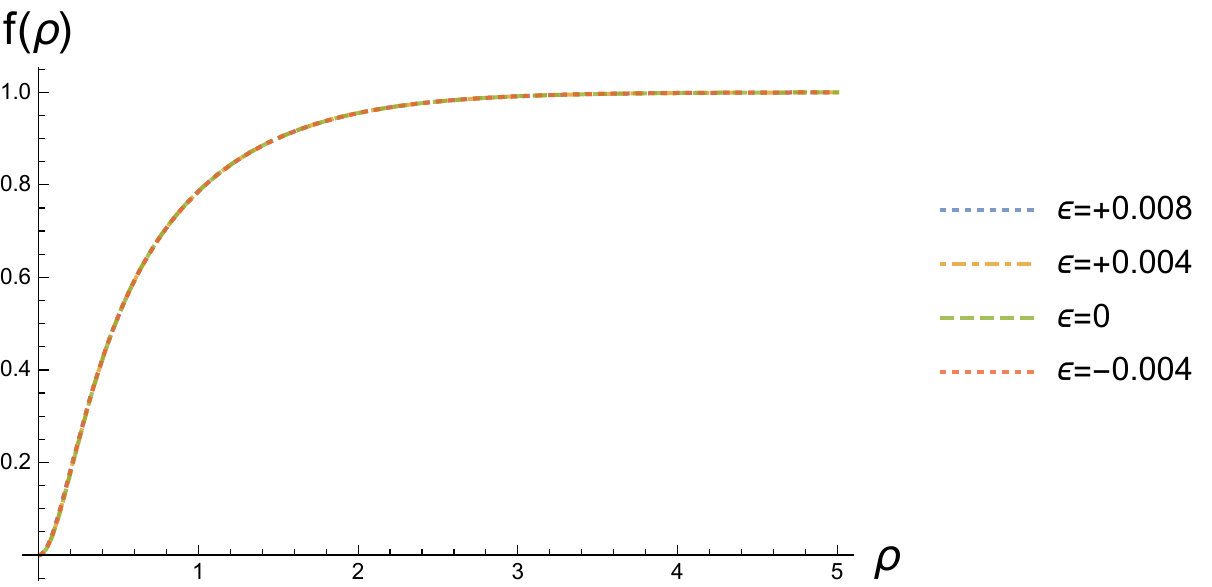}
		\text{(c) The function $f=g_{\rho \rho}^{-1}$.}
	\end{minipage}
	\bigskip
	\begin{minipage}{.49\textwidth}
		\includegraphics[width=7cm]{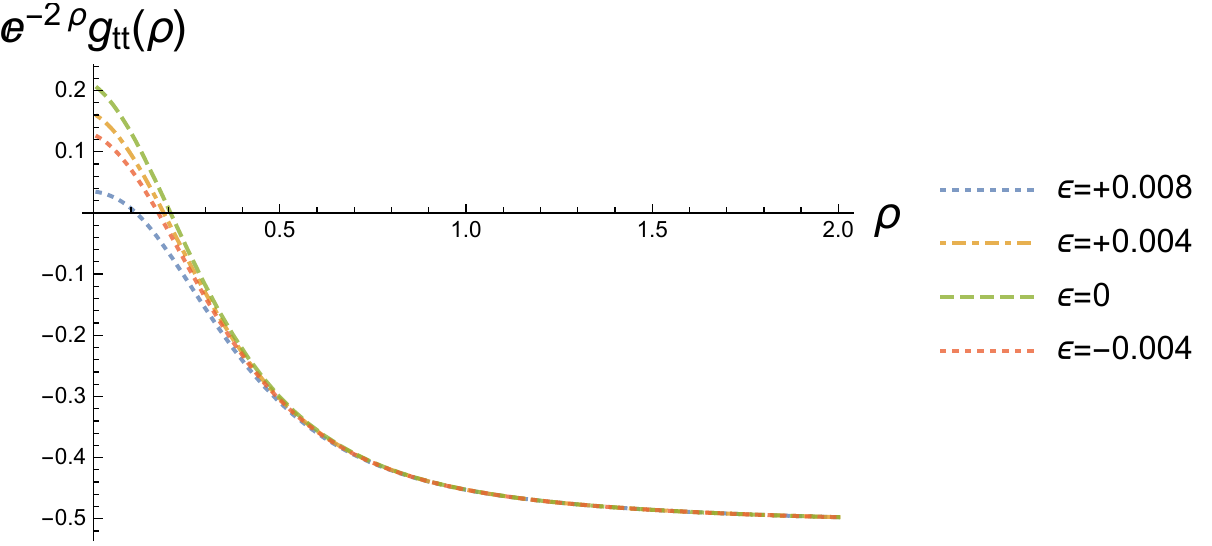}
		\text{(d) The component $g_{tt}$.}
	\end{minipage}
	\begin{minipage}{.49\textwidth}
		\includegraphics[width=7cm]{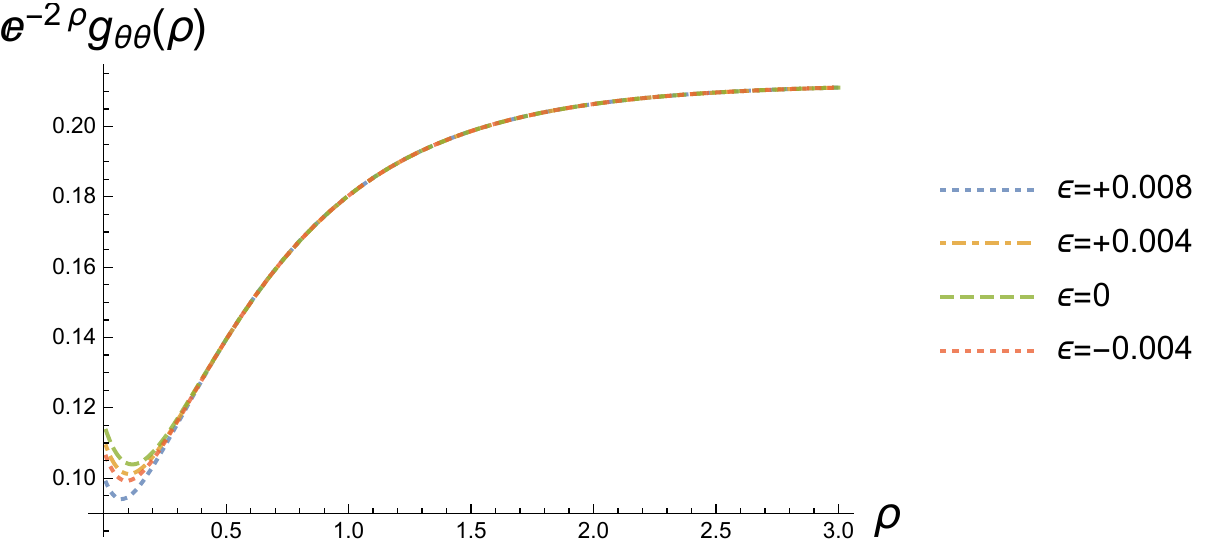}
		\text{(e) The component $g_{\psi \psi}$.}
	\end{minipage}
	\begin{minipage}{.49\textwidth}
		\includegraphics[width=7cm]{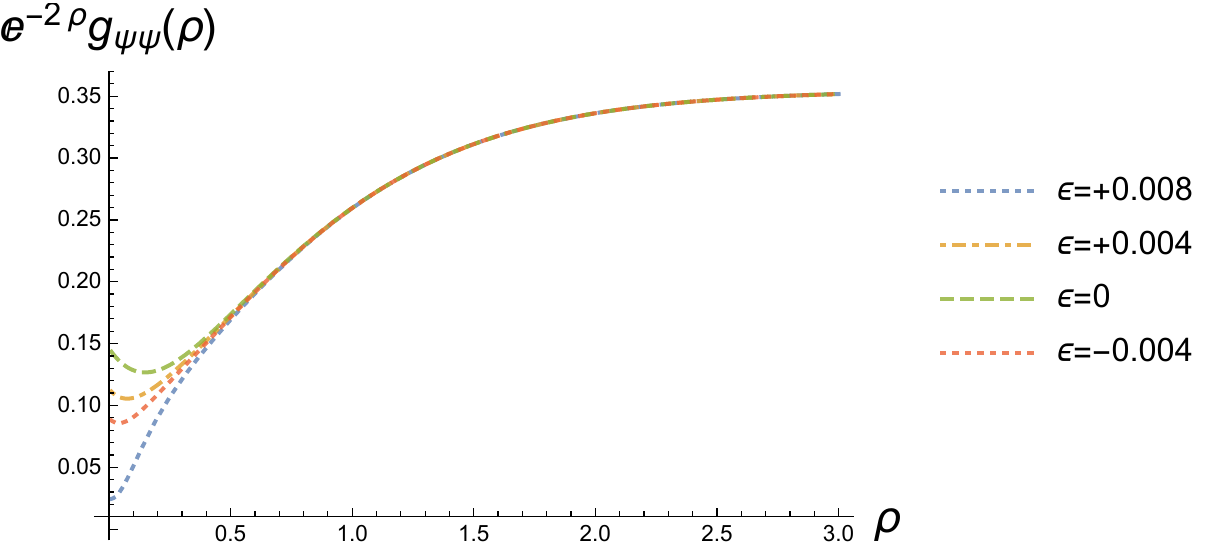}
		\text{(f) The component $g_{\theta \theta}$.}
	\end{minipage}
	\caption{ {\small	Relevant functions and metric components of our solution for $\alpha = 0.84$, $\xi = - \frac{1}{4}$ and different values of $\epsilon$, reported in the label. Each function is rescaled by its asymptotic behaviour at large $\rho$. We emphasize that both $g_{\theta \theta}$ and $g_{\psi\psi}$ are positive in all the $\rho \ge 0$ region, so our solution does not have any CTCs. Since instead $g_{tt}$ assumes positive values near the horizon, our solution does have an ergoregion. } } \label{fig:metricDelta}
\end{figure}

In fig.~\ref{fig:metricDelta} we show the numerical behaviour of the metric components and of the basic functions $a, H, K$. It is easy to see that in the near-horizon region their behaviour is in general different for the various choices of $\epsilon$; an exception are the $f$ and $a$ functions for which the differences are very small. We should notice that $g_{\theta \theta}$ and $g_{\psi\psi}$ tend to a positive non-zero value for $\rho=0$ and are always positive; this means that our solution has no {\it Closed Timelike Curves} (CTCs) in the whole region $\rho \ge 0$. We have verified that the same happens for many different values of the parameters in the yellow region of regularity of fig.~\ref{fig:TheTriangleMinus}. Also, since our solutions are rotating solutions, it is clear from the plot of $g_{tt}$ that an ergoregion emerges\footnote{More precisely, the fact that $g_{tt}$ becomes positive implies that the vector $\frac{\partial}{\partial t}$ becomes spacelike. If we regard this vector as the generator of time translations, then an ergoregion does exist in our solution. However we may also choose the vector~\eqref{Killing_tpsi} to be the generator and in this case there is no ergoregion. Similar features appear often in supersymmetric AdS black holes and were first discussed in~\cite{Gutowski:2004ez}.}. In fig.~\ref{fig:metricDelta}b we display both $H$ and $K$, opportunely rescaled with a $\rho^{-1} e^{-2\rho}$ prefactor, to show that they are indeed different for all the choices of $\epsilon$ but the case $\epsilon=0$, where we have $H=K$. 

\begin{figure}[!htb]
	\label{Fig:Gauge_Field_Components}
	\begin{minipage}{.49\textwidth}
		\includegraphics[width=7cm]{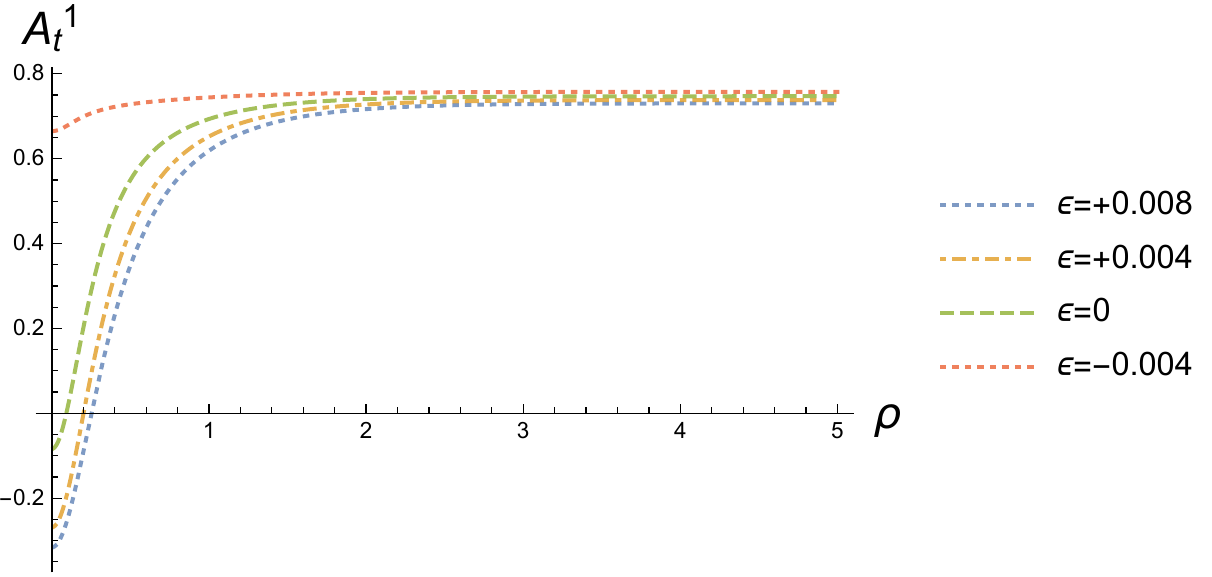}
		\begin{center}
			\text{(a) The component of $A^1_t$.}	
		\end{center}
	\end{minipage}
	\bigskip
	\begin{minipage}{.49\textwidth}
		\includegraphics[width=7cm]{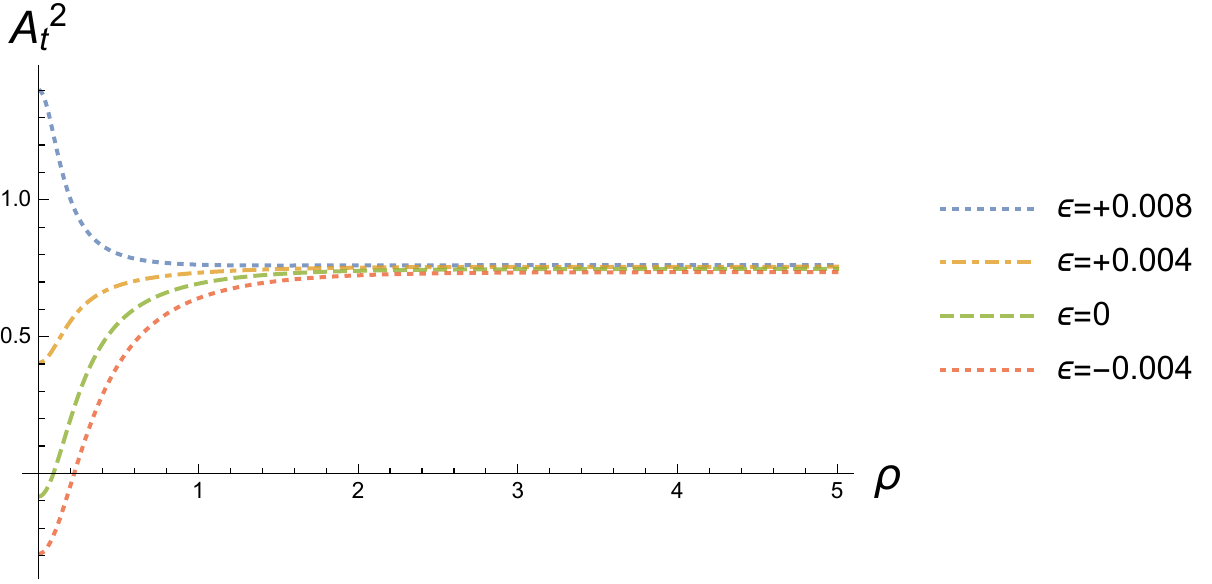}
		\begin{center}
			\text{(b) The component of $A^2_t$.}	
		\end{center}
	\end{minipage}
	
	\begin{minipage}{.49\textwidth}
		\includegraphics[width=7cm]{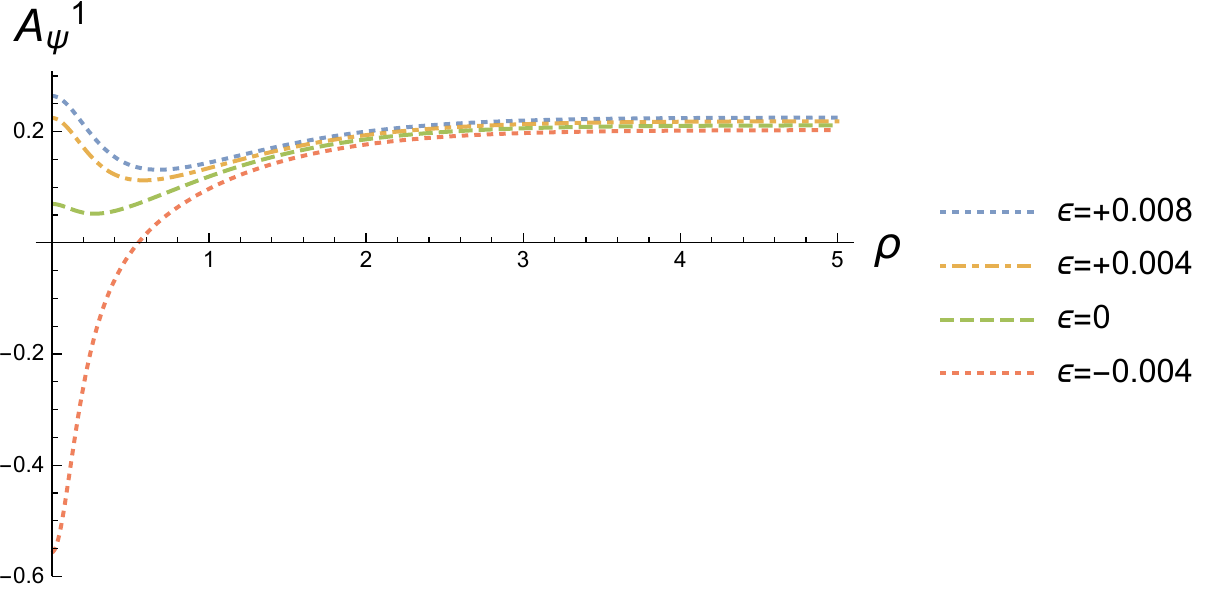}
		\begin{center}
			\text{(c) The component of $A^1_\psi$.}	
		\end{center}
	\end{minipage}
	\begin{minipage}{.49\textwidth}
		\includegraphics[width=7cm]{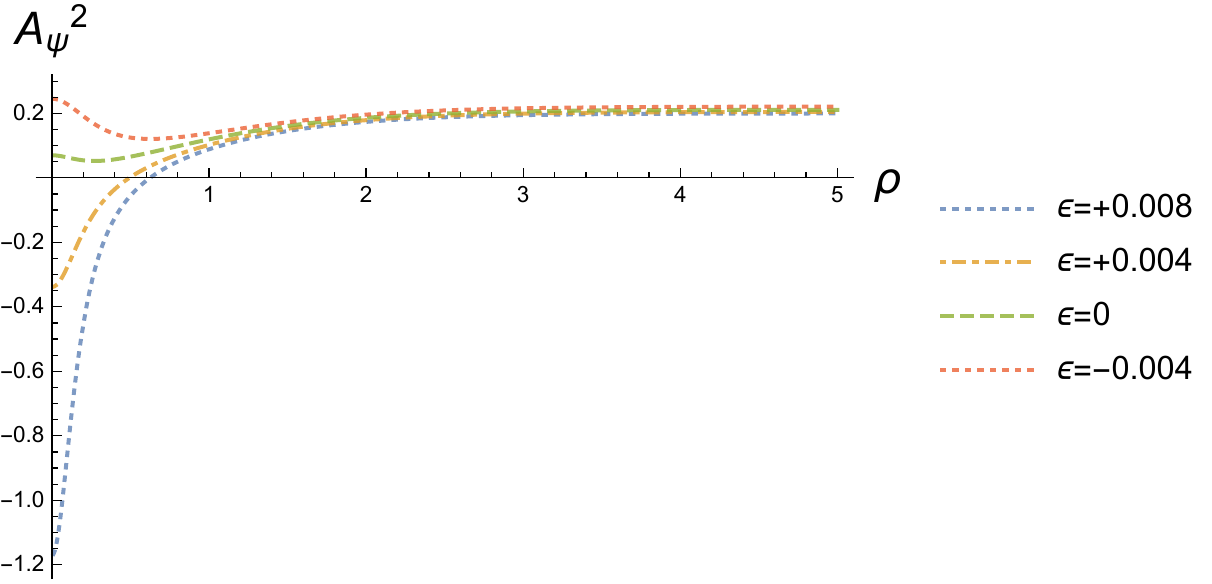}
		\begin{center}
			\text{(d) The component $A^2_\psi$.}	
		\end{center}
	\end{minipage}
	\bigskip
	\label{Fig:Scalar_Fields_Components}
	\begin{minipage}{.49\textwidth}
		\includegraphics[width=7cm]{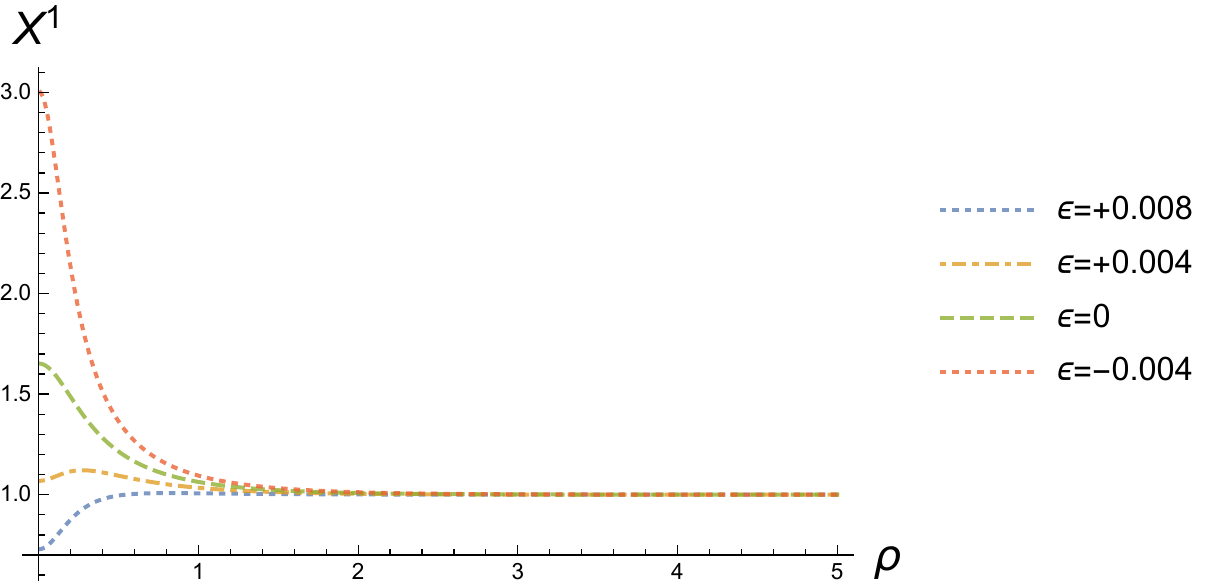}
		\begin{center}
			\text{(e) Scalar fields $X^1$. }	
		\end{center}
	\end{minipage}
	\bigskip
	\begin{minipage}{.49\textwidth}
		\includegraphics[width=7cm]{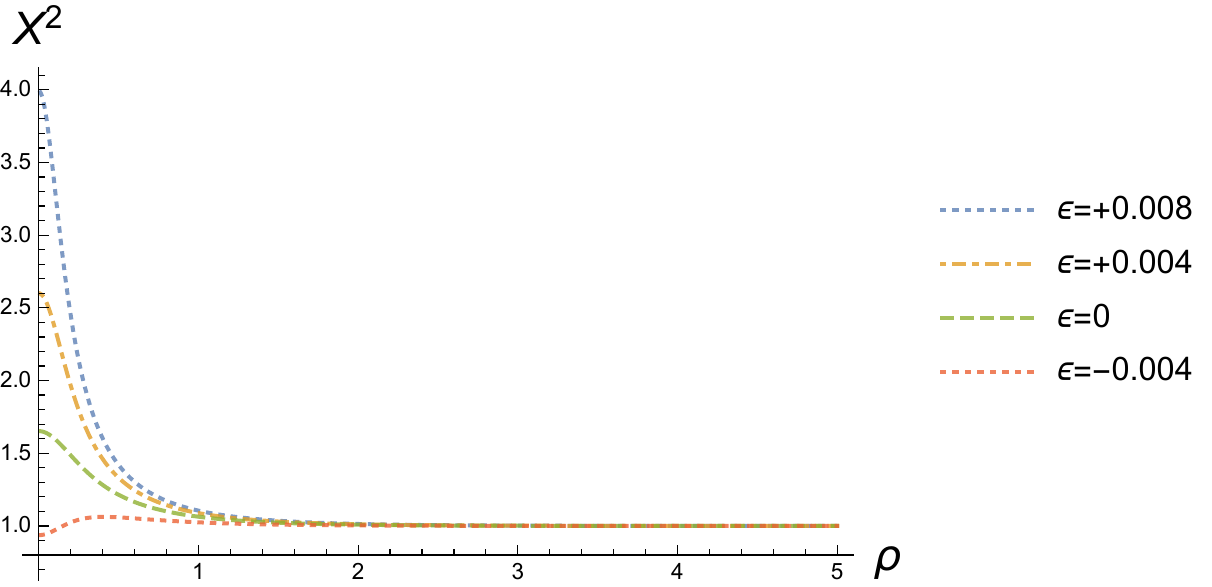}
		\begin{center}
			\text{\quad (f) Scalar fields $X^2$.}	
		\end{center}
	\end{minipage}
	\caption{ {\small{Components of the gauge fields $A^I$ and scalar fields $X^I$ at $\alpha = 0.84$, $\xi=-\frac{1}{4}$ for various $\ep$. It is evident that in the near-horizon region the differences among the fields for the various $\epsilon$ are quite large.} }}
		\label{fig:scalar_plot}
\end{figure} 

In fig.~\ref{fig:scalar_plot} we show the numerical solutions for the scalar and gauge fields. From this picture is quite evident how the change in $\epsilon$ affects the global structure of the solution, since in the near-horizon region the fields get attracted to different asymptotic values, while, as for the metric components, they are all attracted to the same large $\rho$ value.

We now proceed to study a particular solution for some different values of $\xi$. We choose for the other parameters the values $\air = 0.84$ and $\epsilon = 0.008$.

\begin{figure}
	{\centering
	\begin{minipage}[c]{.49\textwidth}
		\includegraphics[width=7 cm]{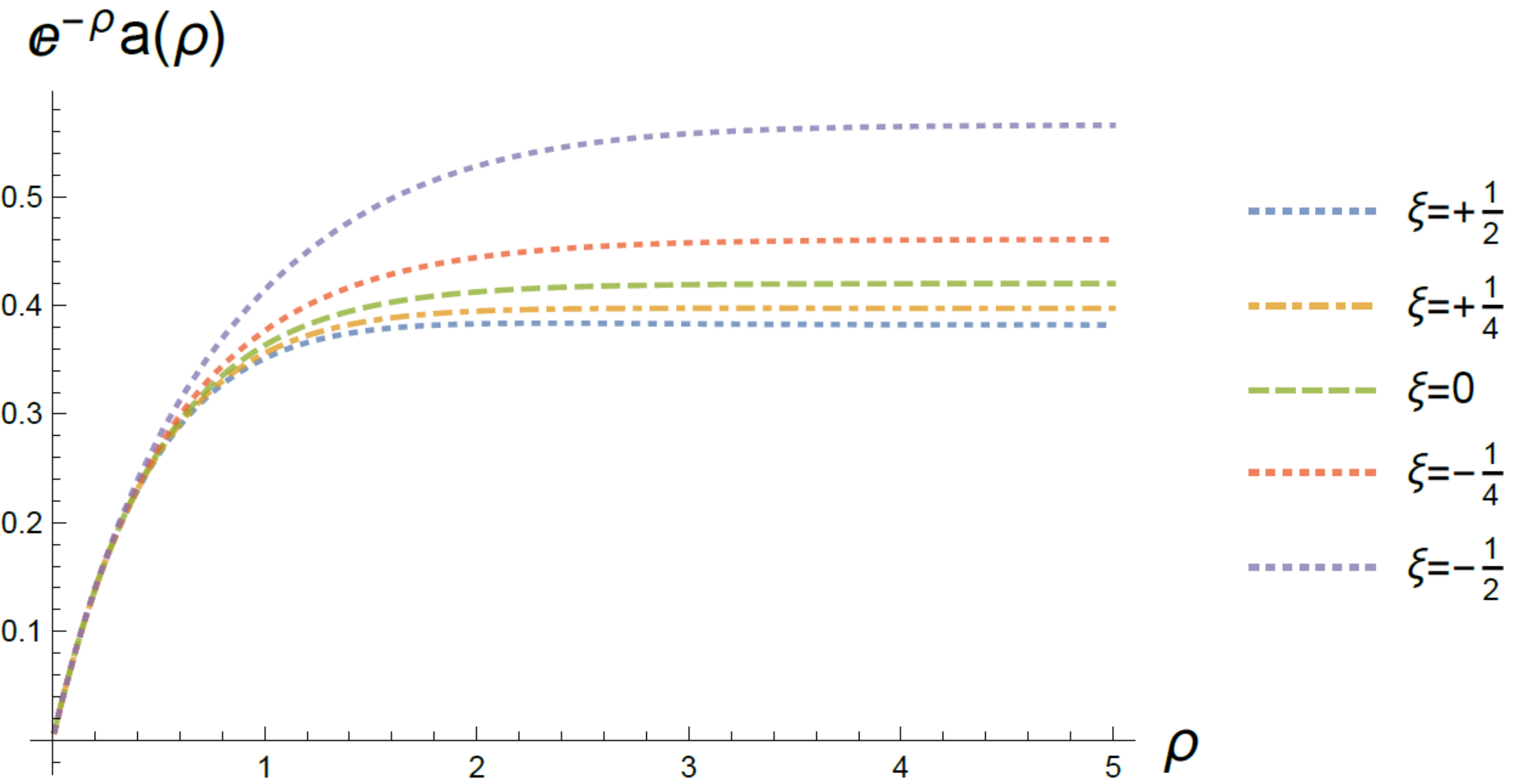}
		\text{(a) The solution $a$.}
	\end{minipage}
    }
    
	\begin{minipage}{.49\textwidth}
		\includegraphics[width=7 cm]{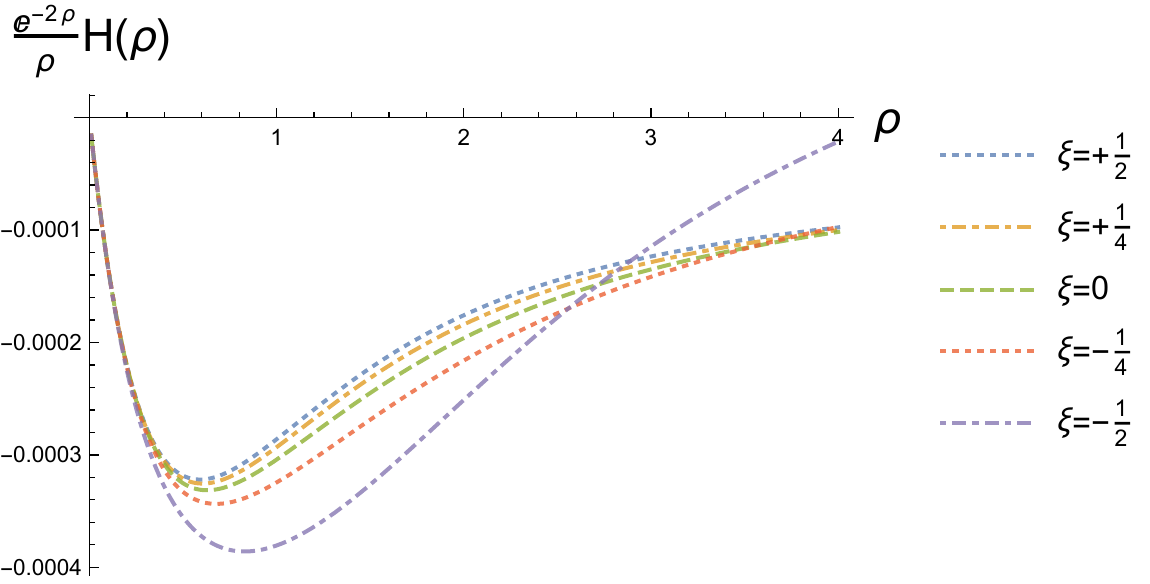}
		\text{(b) The solution $H$.}
	\end{minipage}
	\quad 
	\begin{minipage}{.49\textwidth}
		\includegraphics[width=7 cm]{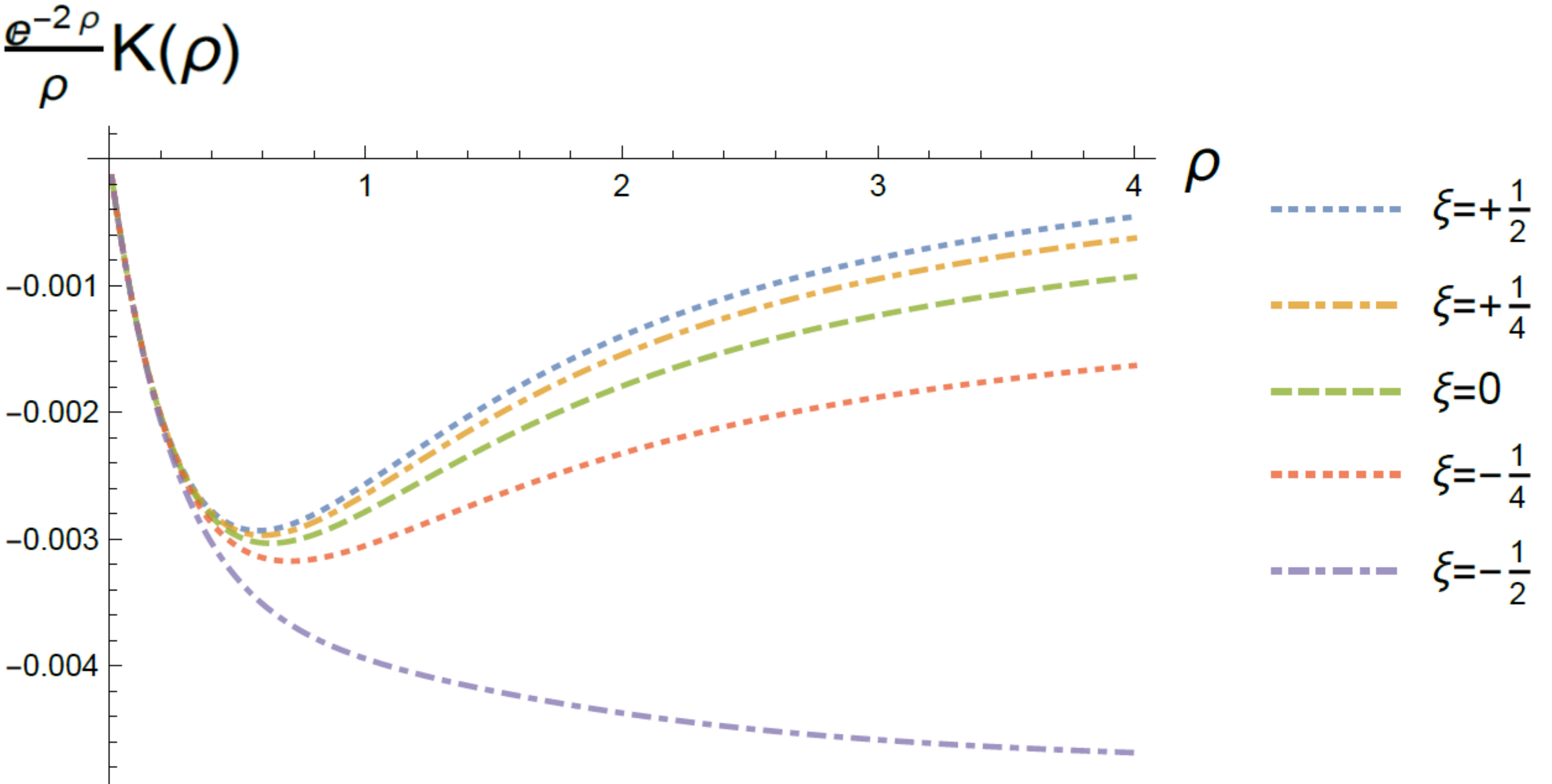}
		\text{(c) The solution $K$.}
	\end{minipage}
	\bigskip
	\begin{minipage}{.49\textwidth}
		\includegraphics[width=7cm]{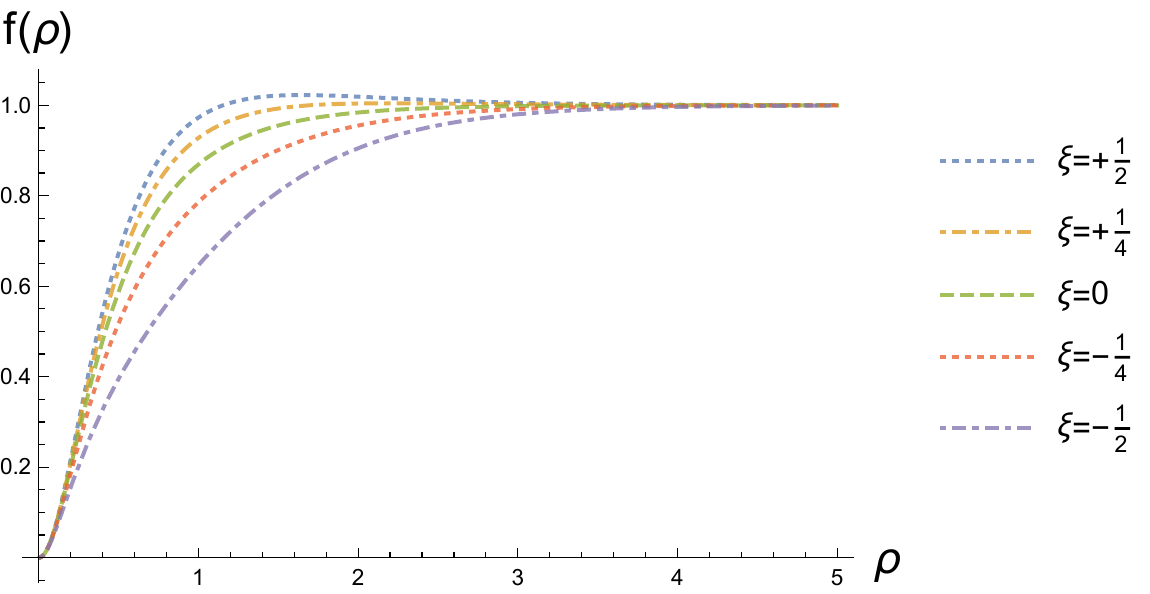}
		\text{(d) The function $f=g_{\rho \rho}^{-1}$.}
	\end{minipage}
	\bigskip
	\begin{minipage}{.49\textwidth}
		\includegraphics[width=7cm]{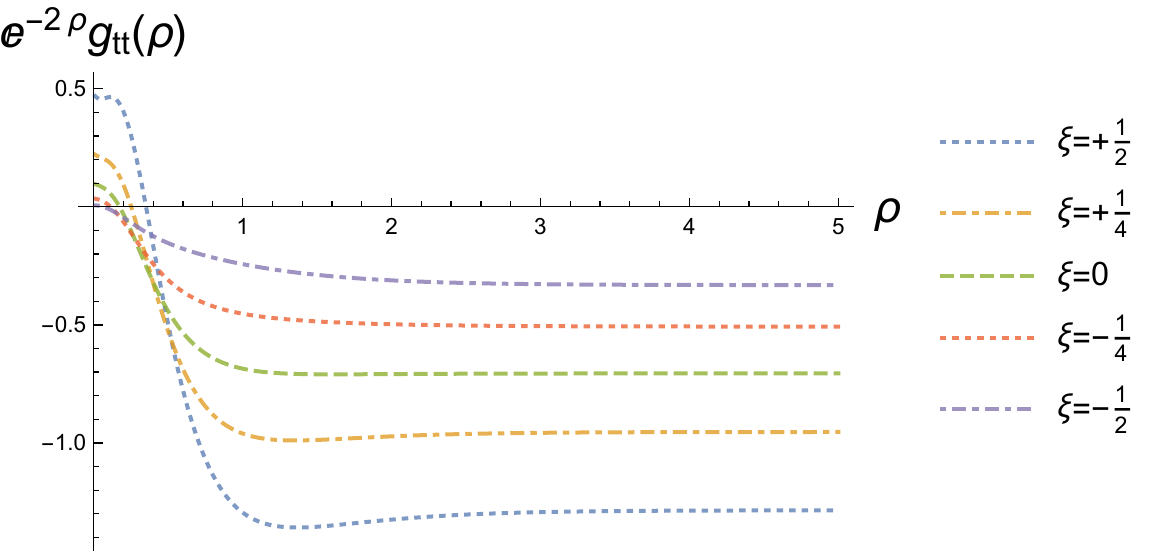}
		\text{(e) The component $g_{tt}$.}
	\end{minipage}
	\begin{minipage}{.49\textwidth}
		\includegraphics[width=7cm]{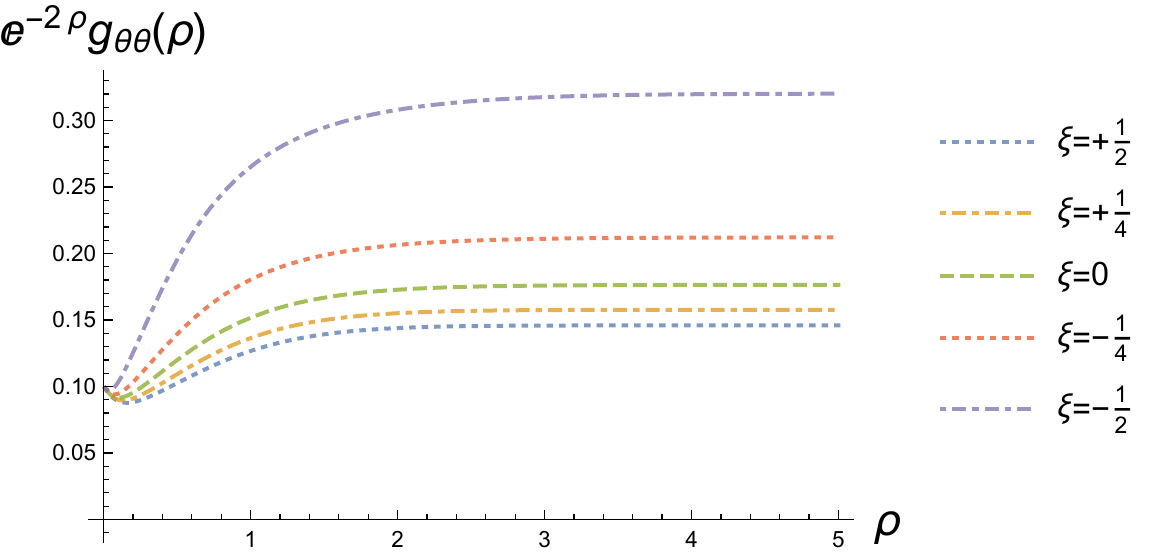}
		\text{(f) The component $g_{\psi \psi}$.}
	\end{minipage}
	\begin{minipage}{.49\textwidth}
		\includegraphics[width=7cm]{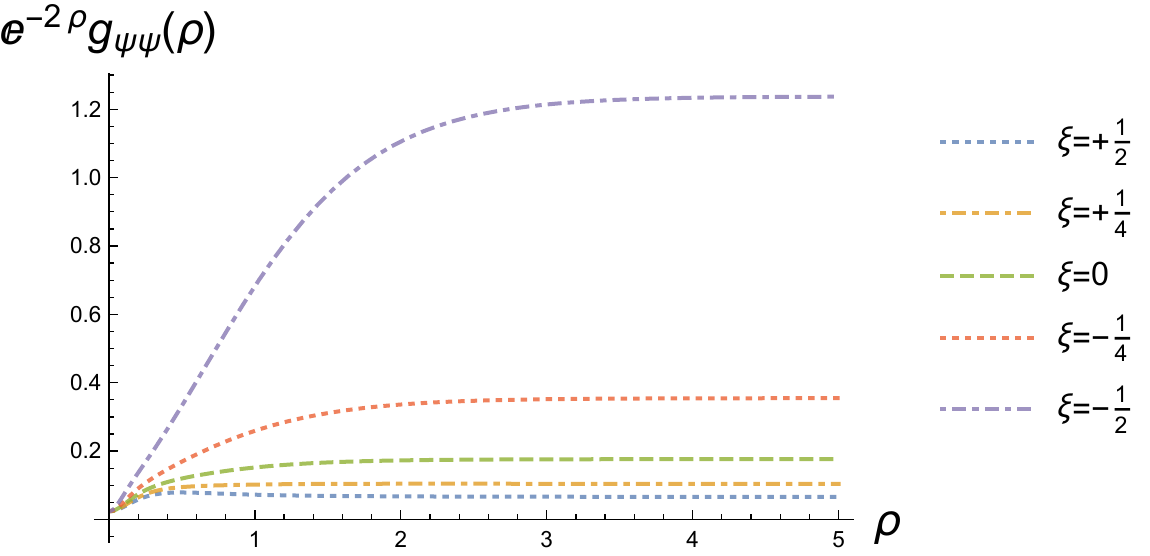}
		\text{(g) The component $g_{\theta \theta}$.}
	\end{minipage}
	\bigskip
	\caption{ {\small	Relevant functions and metric components of our solution for $\alpha = 0.84$, $\ep = + 0.008$ and different values of $\xi$, reported in the label. Each function is rescaled by its asymptotic behaviour at large $\rho$. We emphasize that both $g_{\theta \theta}$ and $g_{\psi\psi}$ are positive in all the $\rho \ge 0$ region, so our solution does not have any CTCs. Since instead $g_{tt}$ assumes positive values near the horizon, our solution does have an ergoregion. } } \label{Fig:a_H_solution_xi}
\end{figure}

We reported in fig.~\ref{Fig:a_H_solution_xi} the relevant functions $a$, $H$, $K$ and the metric components. As opposed to the fixed $\xi$ case, here the components go for $\rho \to \infty$ to different values. Furthermore, their behaviour is very similar in the near-horizon region. This is because the effect of having a different $\xi$ is almost negligible in the near-horizon, since the horizon geometry is controlled by $\air$ and $\epsilon$, while in the near-boundary region the same effect is relevant, being the $\xi$ parameter related to the squashing at the boundary. Again, we have an ergoregion, where $g_{tt}$ becomes positive, and no CTCs, since both $g_{\theta \theta}$ and $g_{\psi\psi}$ are positive everywhere.

\begin{figure}[!htb]
	\begin{minipage}{.49\textwidth}
		\includegraphics[width=7cm]{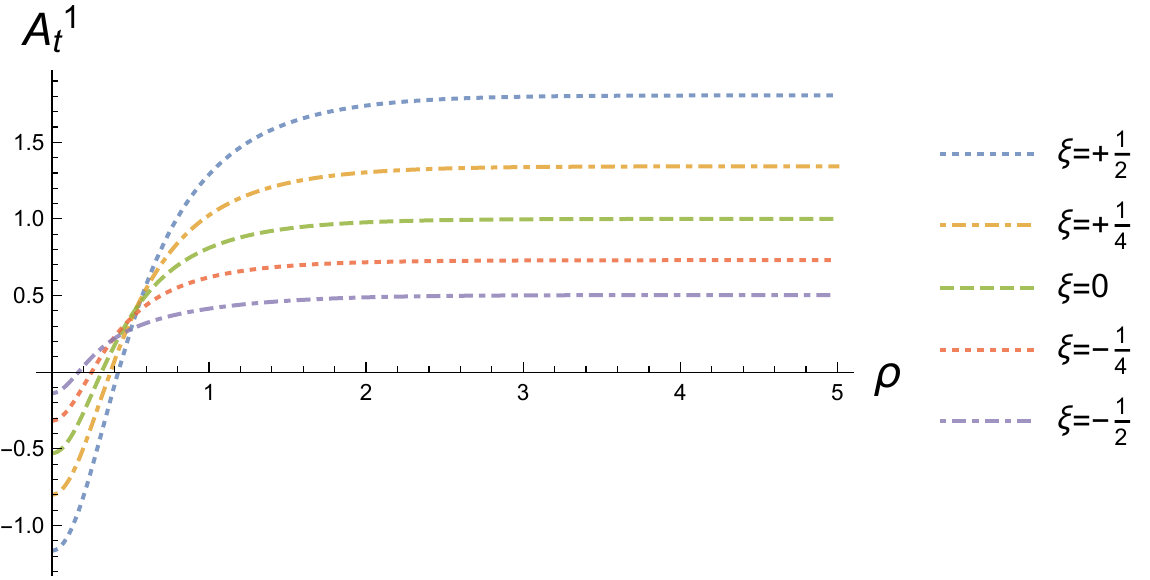}
		\begin{center}
			\text{(a) The component of $A^1_t$.}	
		\end{center}
	\end{minipage}
	\bigskip
	\begin{minipage}{.49\textwidth}
		\includegraphics[width=7cm]{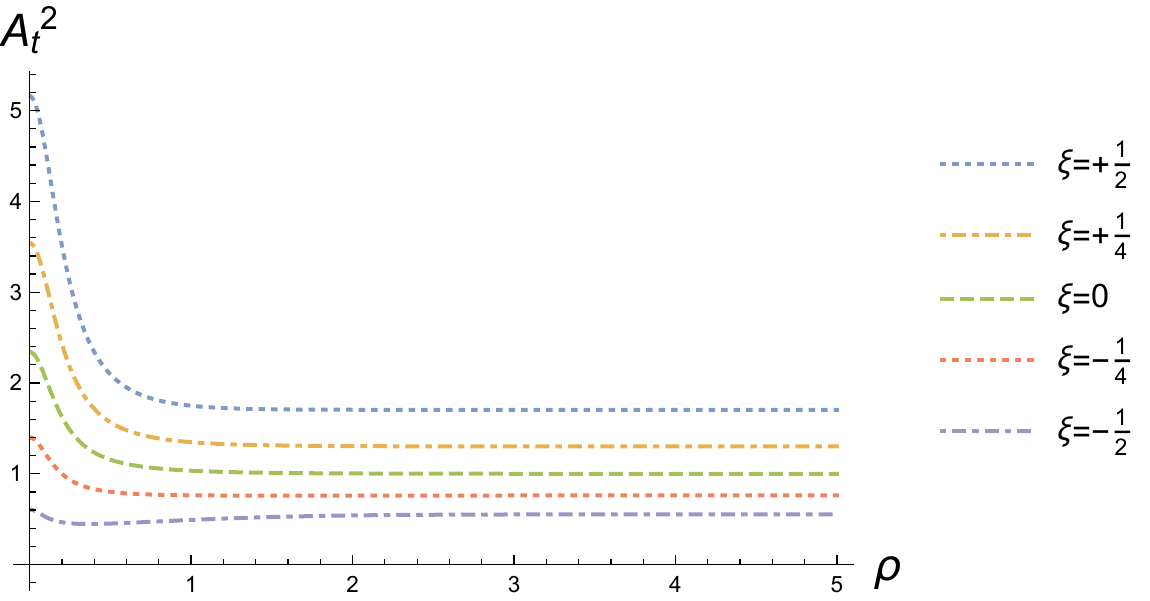}
		\begin{center}
			\text{(b) The component of $A^2_t$.}	
		\end{center}
	\end{minipage}
	
	\begin{minipage}{.49\textwidth}
		\includegraphics[width=7cm]{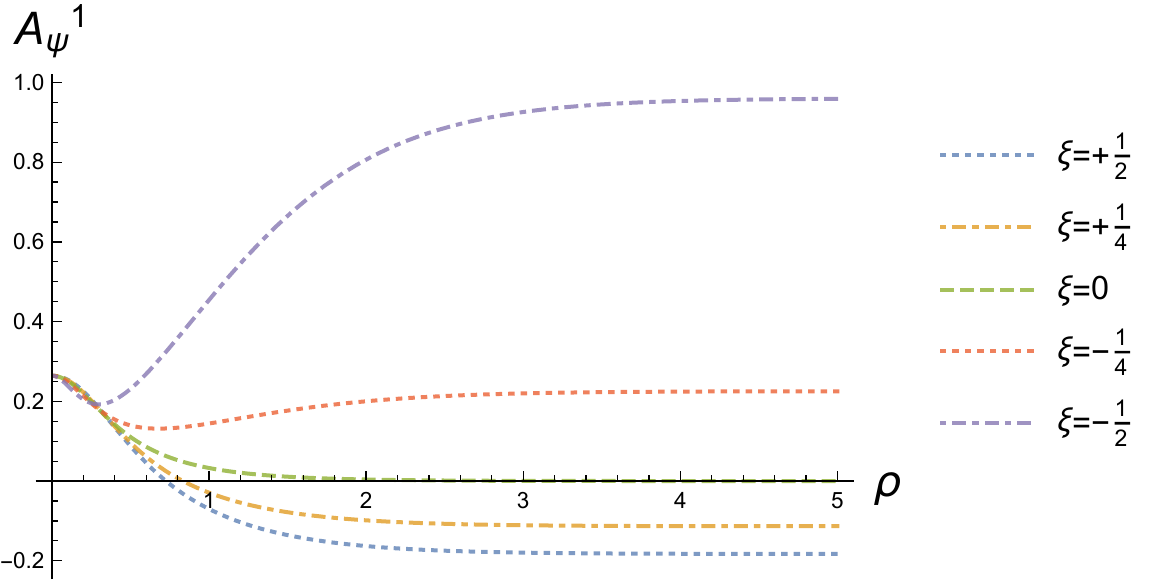}
		\begin{center}
			\text{(c) The component of $A^1_\psi$.}	
		\end{center}
	\end{minipage}
	\begin{minipage}{.49\textwidth}
		\includegraphics[width=7cm]{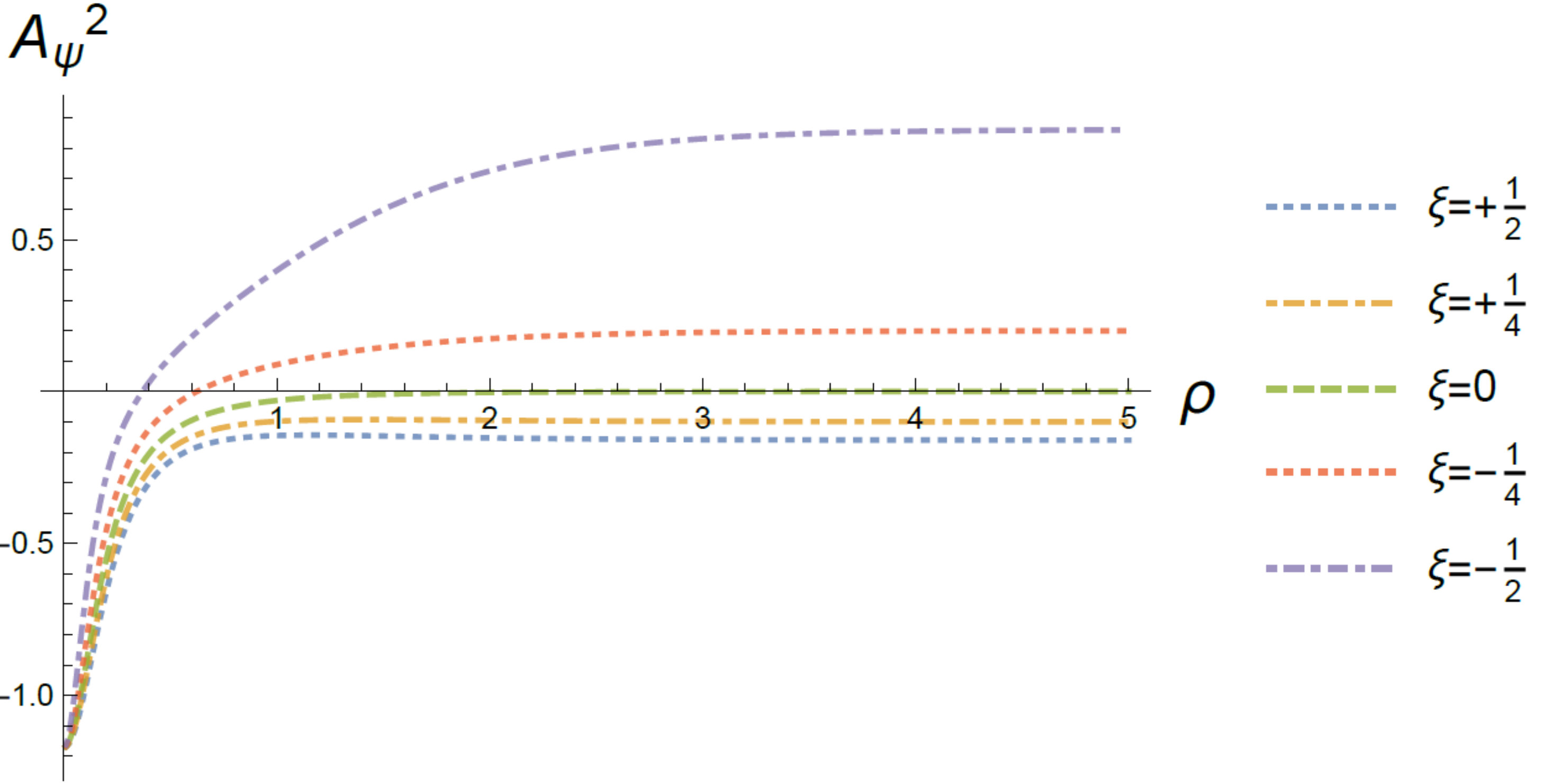}
		\begin{center}
			\text{(d) The component $A^2_\psi$.}	
		\end{center}
	\end{minipage}
	\bigskip
	\begin{minipage}{.49\textwidth}
		\includegraphics[width=7cm]{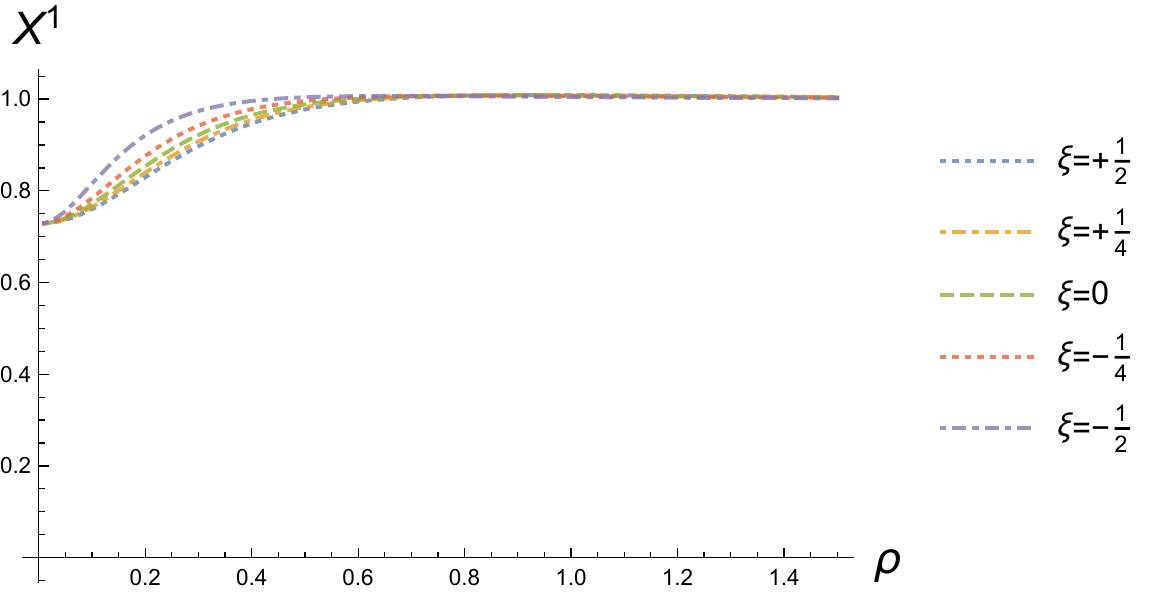}
		\begin{center}
			\text{(e) Scalar fields $X^1$. }	
		\end{center}
	\end{minipage}
	\bigskip
	\begin{minipage}{.49\textwidth}
		\includegraphics[width=7cm]{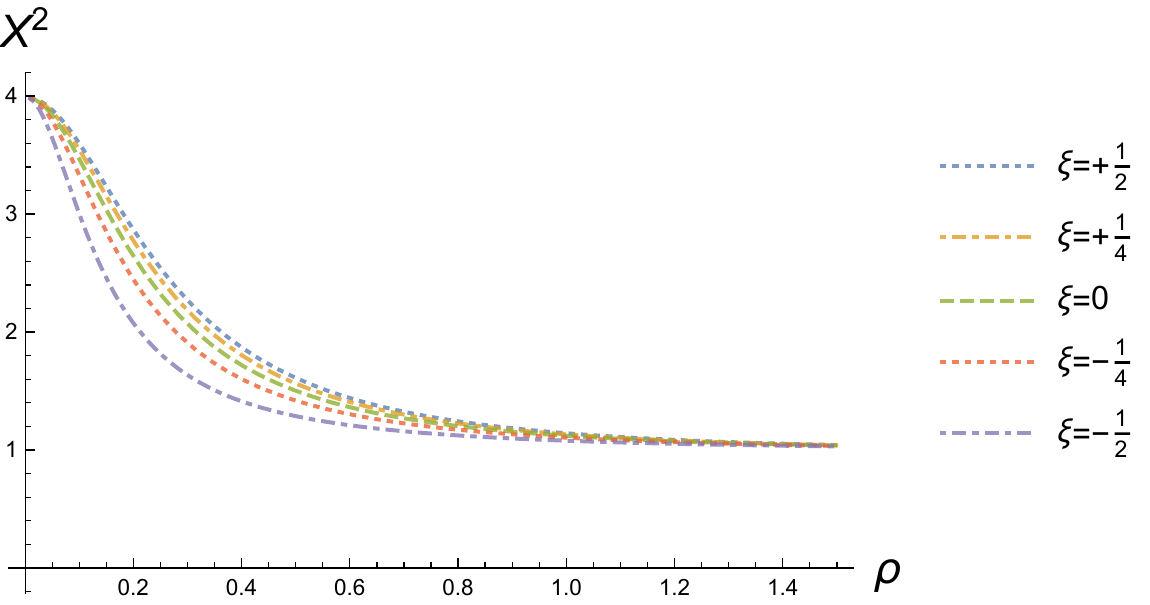}
		\begin{center}
			\text{\quad (f) Scalar fields $X^2$.}	
		\end{center}
	\end{minipage}
	\caption{ {\small Components of the gauge fields $A^I$ and scalar fields $X^I$ at $\alpha = 0.84$, $\ep= +0.008$ for various $\xi$.
	}}
		\label{fig:scalar_plotxi}
\end{figure} 

We then show fig.~\ref{fig:scalar_plotxi} where we reported the gauge and scalar fields for various $\xi$; again, we see that, in contrast with the fixed $\xi$ case, their behaviour in the near-horizon region is similar for all the $\xi$, while they go to different values in the near-boundary region. The only exception is the value of $A_t^I$, that also differs in the near-horizon region. This is due to the fact that in the coordinates $(t, \, \psi)$ we are using the time component of the gauge fields explicitly depends on the squashing $v$, as it is clearly visible by~\eqref{Gauge_Field_IR}. If we had used instead the coordinates $(y, \, \hat\psi)$, the time component $A_t^I$ would vanish at the horizon and would not be influenced by $\xi$. 

We end this section by summarizing the main characteristics of the family of solutions we have constructed. Both the near-horizon analysis and the numerical one prove that our solutions are black hole solutions whose horizon geometry is controlled by two of the three near-horizon parameters, $\air$ and $\hir$. The last near-horizon parameter, $\xi$, is related to the squashing at the boundary, and is therefore related with the parameter $v^2$ controlling the squashing of the boundary three-sphere. Both the near-boundary analysis and the numerical one show that our solutions are AlAdS, a conformally flat boundary being obtained only when the $S^3$ is round ($v^2 = 1$). In the near-boundary region, the solution is controlled by eleven free parameters ($a_0, \, a_2, \, a_4, \, a_6, \, v^2, \, H_2, \, H_4, \, \HL, \, K_2, \, K_4, \, \KL$), these should be connected with the near-horizon ones and should be dependent on them. We were able to trade four of them ($a_4, \, a_6, \, H_4, \, K_4$) with the four linearly-independent first integrals ($\CK_1, \, \CK_2^{(1)}, \, \CK_2^{(2)}, \, \CK_3$) we have found in sec.~\ref{sec:first_integrals}; these first integrals are useful since they are immediately connected with the interior of the solution, being possible to express them with respect to the near-horizon parameters only. All the other near-boundary free parameters can be related to the near-horizon ones using a numerical procedure, as it was done in~\cite{Blazquez-Salcedo:2017ghg,Cassani:2014zwa,Cassani:2018mlh}. We have numerically shown that the near-boundary and near-horizon behaviours we have found interpolate smoothly in the bulk, giving rise to regular solutions which are free from CTCs.


\section{Physical properties of the solution} \label{sec:Properties}

In this section we compute the relevant physical quantities that characterize the family of solutions we have built. These are the energy, the angular momentum, the holographic and Page charges, which can be computed using the near-boundary perturbative solution, the chemical potentials and the entropy, which instead can be derived by means of the near-horizon expansions. Once these quantities are known, we can perform some consistency checks, for example by verifying the quantum statistical relation. 

In order to compute some of the above physical properties, we will use the technology of holographic renormalization \cite{Witten:1998qj,Henningson:1998gx,Balasubramanian:1999re, deHaro:2000vlm, Bianchi:2001kw, Bianchi:2001de}. We perform such computations using the Fefferman-Graham radial coordiante $r$, introduced in app.~\ref{app_FG}, instead of the usual one $\rho$. This is because the use of the Fefferman-Graham coordinate is standard in holography and may help to compare our results with other references. Moreover, we write the general Fefferman-Graham metric as
\begin{equation}
    \label{Fefferman_Graham_Induced}
    \diff s^2 = \ell^2 \, \frac{\diff r^2}{r^2} + h_{ij} \, \diff x^i \, \diff x^j \, ,
\end{equation}
where the five-dimensional coordinates split as $x^\mu = (r, x^i)$ with $x^i=\{ t, \theta, \phi, \psi\}$ and where $h_{ij}$ is the induced metric at the boundary of the spacetime. Similarly we define the boundary gauge fields $A^I_i$ and the boundary field strengths $F^I_{ij}$.
In this section we only report the results we got using holographic renormalization while we refer to app.~\ref{app_HR} for a more detailed discussions about how these results are obtained. We remark that we performed holographic renormalization using a minimal subtraction scheme; all the physical quantities evaluated by means of this formalism are therefore refereed to this renormalization scheme. We also underline that the application of holographic renormalization to five-dimensional Fayet-Iliopoulos gauged supergravity has been discussed in detail in~\cite{Cassani:2018mlh}.

We can compute via holographic renormalization the stress-energy tensor of our family of solutions\footnote{In the stress-energy tensor formula, as well as in all the formulae below, the quantity $r_0$ is the cutoff we used to regulate the large-distance divergences which appear. At the end of the computation it is removed by sending it to infinity.}. Its expression is quite standard
\be
\label{Stress_Energy_Tensor}
\begin{split}
\langle T_{ij} \rangle = -\frac{1}{\kappa^2} \lim_{r_0\to \infty} \frac{r_0^2}{\ell^2} & \Bigg[ K_{ij} - (K - {\cal W}) h_{ij} -  \frac{{\cal W} - 3\ell^{-1}}{\log \frac{r_0^2}{\ell^2}}  \, h_{ij}  - \frac{\ell}{2} \left( R_{ij} - \half \, R\, h_{ij} \right) \\
& \quad - \frac{\ell^3}{4} \log \frac{r_0^2}{\ell^2} \left( - \half \, B_{ij} - \frac{2}{\ell^2} \, Q_{IJ} F^I_{i k} F^J{}_j{}^k + \frac{1}{2\ell^2} \, h_{ij} Q_{IJ} F^I_{kl} F^{J \, kl}   \right)\!  \Bigg],
\end{split}
\ee
where $R_{ij} $, $R$, $B_{ij}$ are the Ricci tensor, the Ricci scalar and the Bach tensor of the induced metric $h_{ij}$. The other ingredients appearing in the expression~\eqref{Stress_Energy_Tensor} for the stress-energy tensor are the extrinsic curvature $K_{ij}$ of the induced metric $h_{ij}$, its trace $K$ and the superpotential $\CW = 3 \ell^{-1} \bar{X}_I \, X^I$ which derives from the scalar potential $\CV$ given in~\eqref{scalarpot}.
The conserved electric current also arises from holographic renormalization and it is given by
\be
\langle j_I^i \rangle = - \frac{1}{\kappa^2}  \lim_{r_0\to \infty} \frac{r_0^2}{\ell^2} \Bigg[  \ep^{ijkl} \left( Q_{IJ} \star F^J +\frac{1}{6} C_{IJK}A^J \wedge F^K \right)_{jkl} + \ell \, \nabla_j \left( Q_{IJ} F^{J \, ji}\right) \log \frac{r_0}{\ell}  \Bigg] .
\ee
Once both the stress-energy tensor and the electric current are evaluated, we are in the position to compute the energy and the angular momentum, which are visible as the charges associated to the two Killing vectors of the metric $\frac{\partial}{\partial t}$ and $- \frac{\partial}{\partial \psi}$
\be
\label{Energy_Holo_Ren}
\begin{split}
E &= Q_{\frac{\pd}{\pd t}} = + \int_{\Sigma_\infty} {\rm vol}_\Sigma\,  u_i \left( \langle T^i{}_{t} \rangle + A^I_t \langle j_I^i \rangle \right)  \,,  \\
J  &= Q_{- \frac{\pd}{\pd \psi}} = - \int_{\Sigma_\infty}{\rm vol}_\Sigma\,  u_i \left( \langle T^i{}_{\psi} \rangle + A^I_\psi \langle j_I^i \rangle \right)  \,,  \\
\end{split}
\ee
where $u^i \, \partial_{i} = \frac{v}{2 \, a_0} \, \partial_t$ is a unit timelike vector for the metric on the conformal boundary. The conserved charges~\eqref{Energy_Holo_Ren} explicitly are
\be
\label{Angular_Momentum_Holo_Ren}
\begin{split}
E &= \frac{\pi^2 \ell^2}{\kappa^2}  \left[ \frac{16}{9} - \frac{14}{9} \, v^2 + \frac{19}{36} \, v^4 + \frac{8}{v^2} \, {\cal K}_3 - 192 (\tilde H^2 +\tilde H \tilde K + \tilde K^2) \right]\, , \\
J  &= \frac{4 \pi^2 \ell^3 }{\kappa^2} \, {\cal K}_3  \,. \\
\end{split}
\ee
Note that the angular momentum precisely coincides with the generalized Komar integral~\eqref{Angular_Momentum_First_Integral}.
We can also compute the conserved electric charges as
\be\label{eq:Qcharges}
\begin{split}
Q_I &= \int_{\Sigma_{\infty}} {\rm vol}_\Sigma \, u_i \langle j^i_I \rangle \\
&=  -\frac{1}{\kappa^2} \int_{\Sigma_{\infty}} \left(  Q_{IJ} \star F^J + \frac{1}{6} \, C_{IJK} A^J \wedge F^K \right) ,
\end{split}
\ee
and it is fundamental for our discussion to remark that these differ on our solutions from the Page charges 
\be\label{eq:PCharges}
P_I = \frac{1}{\kappa^2} \int_{\Sigma_{\infty}} \left(  Q_{IJ} \star F^J + \frac{1}{4} \, C_{IJK} A^J \wedge F^K \right) .
\ee
In fact, on our solutions, we have
\be
\label{Holo_Charges_Computed}
\begin{split}
Q_1 &= - P_1 -  \frac{16 \pi^2 \ell^2}{\kappa^2} \frac{1}{54} \left[ \left(1- v^2-18 \tilde K \right) \left(1-v^2 + 18 (\tilde H+\tilde K)\right)  \right] \,, \\
Q_2 &= - P_2 -  \frac{16 \pi^2 \ell^2}{\kappa^2} \frac{1}{54} \left[ \left(1 - v^2 -18 \tilde H\right) \left(1 - v^2 + 18 (\tilde H+\tilde K)\right)  \right]    \,, \\
Q_3 &= - P_3 -  \frac{16 \pi^2 \ell^2}{\kappa^2} \frac{1}{54} \left[ \left(1-v^2-18 \tilde H\right) \left(1-v^2 -18 \tilde K\right)  \right]  \, ,
\end{split}
\ee
while
\be
\begin{split}
P_1 &=  - \frac{16 \pi^2 \ell^2}{\kappa^2} ({\cal K}_1 +3 {\cal K}_2^{(1)})   \,, \\
P_2 &= - \frac{16 \pi^2 \ell^2}{\kappa^2} ({\cal K}_1 +3 {\cal K}_2^{(2)})   \,, \\
P_3 &= -P_1 - P_2  - \frac{48 \pi^2 \ell^2}{\kappa^2}   {\cal K}_1 \, .
\end{split}
\ee
We want to stress that this fact is a consequence of the squashing of the boundary: it is trivial to see from eqs.~(\ref{eq:Qcharges}, \ref{eq:PCharges}) that the difference between holographic and Page charges is related to the Chern-Simons term which gives a different contribution to the two quantities. In usual non-squashed solutions like \cite{Gutowski:2004yv, Gutowski:2004ez} the same term gives no contribution, since the field strength $F^I$ vanishes asymptotically, and the two different types of charges are therefore equal. This implies that we may have some relevant departure from the equation of~\cite{Kim:2006he} that relates the entropy and the charges, since we have no unique way to choose which charge is the correct one for reproducing the entropy; in fact, it will turn out later that the entropy is indeed reproduced in terms of the Page charges, instead of the holographic charges obtained in~\eqref{Holo_Charges_Computed}.

We remark that our results for the holographic and Pages charges, the energy $E$ and the angular momentum $J$ are valid for every AlAdS solution to five-dimensional Fayet-Iliopoulos gauged supergravity with $n_V = 2$ which satisfies the supersymmetry equations~(\ref{eqforH}, \ref{eqforK}, \ref{eqforaU1}). In particular all the conserved charges depend only on the first integrals $\CK_1, \, \CK_2^{(1)}, \, \CK_2^{(2)}, \, \CK_3$, on the squashing at the boundary $v$ and on the two scalar sources $\HL, \, \KL$. As we have already discussed in sec.~\ref{sec:near_horizon} and as we will show explicitly in app.~\ref{app_First_Integrals}, the first integrals are completely determined by the interior of the solution, since they can be expressed with respect the near-horizon parameters $\air$ and $\hir$ (see~\eqref{First_Integral_IR} for their explicit expressions).

We now turn to examine the near-horizon properties of our solution. We can easily compute the entropy by looking at the horizon metric~\eqref{Metric_IR}:
\be
\begin{split}
{\cal S} &= \frac{2\pi}{\kappa^2} \, {\rm Area} = \frac{8\pi^3 \ell^3}{\kappa^2} \, \sqrt{ 48 {\cal K}_1 - 144\left[ \left({\cal K}_2^{(1)} \right)^2 + {\cal K}_2^{(1)} {\cal K}_2^{(2)} + \left( {\cal K}_2^{(2)} \right)^2   \right] - {\cal K}_3 } \\
&= 2 \pi \ell \sqrt{ \frac{3}{2} \, C^{IJK} \bar X_I P_J P_K - \frac{4\pi^2 \ell}{\kappa^2} \, J }  \,.
\end{split}
\ee
This is the anticipated result: the entropy of the black hole solutions with squashed boundary can be reproduced by a simple combination of the conserved charges; however the formula does \emph{not} involve the usual holographic electric charges but rather the Page charges, signaling a relevance for the entropy counting of the Chern-Simons term that was previously unnoticed by the non-squashed solutions of~\cite{Gutowski:2004ez,Gutowski:2004yv}. This means that the entropy formula found for example in~\cite{Kim:2006he} for asymptotically AdS$_5$ black holes retains its validity for these AlAdS$_5$ solutions if we identify the charges appearing therein with the Page ones instead of the holographic ones. The result we have found was already obtained in~\cite{Cassani:2018mlh} and is somehow anticipated by the explicit form the different types of charges take: indeed the $Q_I$ depend on both the squashing $v$ and the scalar sources $\HL, \KL$ on which the horizon geometry is independent and therefore appear to be inadequate to describe an horizon quantity like the entropy. The Page charges $P_I$, instead, depend only on the first integrals which are immediately related to the horizon geometry only and seem thus the right charges to describe the entropy.

The angular velocity $\Omega$ of our family of solutions is easily read from the supersymmetric Killing vector~\eqref{Killing_tpsi} and it results:
\begin{equation}
    \Omega = \frac{2}{\ell \, v^2} \, ,
\end{equation}
the electric potential is instead found to be\footnote{The definition of the electric potential we used has been provided in~\cite{Papadimitriou:2005ii}. It measures the electric potential just at the horizon. This definition agrees with the fact that our conserved charges $E$ and $J$, given in eqs.~\eqref{Energy_Holo_Ren},~\eqref{Angular_Momentum_Holo_Ren}, are evaluated by considering also the contributions given by the gauge fields.}:
\begin{equation}
    \Phi^I = V^\mu \, A^I_\mu \rvert_{\text{hor}} = 0 \, .
\end{equation}
Our next aim is to use all the quantities we have evaluated in this section to verify the extremal limit of the quantum statistical relation valid for general AlAdS spacetimes. The quantum statistical relation is given by
\begin{equation}
\label{Quantum_Statistical_Relation}
\frac{I}{\beta} = E - T \, \CS - \Omega \, J - \Phi^I \, Q_I \, ,
\end{equation}
where $I$ is the Euclidean on-shell action, $T$ is the temperature of the black hole and $\beta = \frac{1}{T}$. We can take the extremal limit of this relation by recalling that for extremal black holes we have $T = 0$. Sending the temperature to zero, we obtain at the leading order the relation:
\begin{equation}
\label{Quantum_Statistical_Relation_Extremal}
\frac{I}{\beta} = E - \Omega \, J - \Phi^I \, Q_I \, ,
\end{equation}
which is valid for extremal black holes and should therefore be satisfied by our family of solutions. 

To verify eq.~\eqref{Quantum_Statistical_Relation_Extremal} we are missing the Euclidean on-shell action. This can be computed using again holographic renormalization. In particular it is possible to define a renormalized Lorentzian on-shell action as
\be
S_{\rm ren} = \lim_{r_0 \to \infty} S_{\rm reg} \,,
\ee
where the regularized action is
\be
 S_{\rm reg}  = S_{\rm bulk} + S_{\rm GH} + S_{\rm ct} \,.
\ee
In the equation above, $S_{\rm bulk}$ is the bulk action~\eqref{Bulk_action}, while $S_{\rm GH}$ and $S_{\rm ct}$ are the Gibbons-Hawking term and the counterterms piece respectively. We show in app.~\ref{app_HR} the explicit form of those pieces as well as how the whole computation of the renormalized action is performed; here we report instead only the final result, which is
\be
\label{Renormalized_Action_Computed}
S_{\rm ren} = - \frac{\pi^2 \ell^2}{\kappa^2} \Big[ \frac{16}{9} - \frac{14}{9} \, v^2 + \frac{19}{36}\, v^4 - 192 (\tilde H^2 + \tilde H \tilde K + \tilde K^2 )  \Big] \int \diff t\, .
\ee
This action is evaluated in a minimal subtraction scheme, as all the other quantities computed via holographic renormalization reported in this section.
We immediately notice that this depends only on the squashing at the boundary $v$ and on the scalar sources $\HL, \, \KL$. Moreover, we should remark that the regularized action is gauge-dependent due to the Chern-Simons term in the bulk action~\eqref{Bulk_action}; the result we reported above is valid when the gauge condition $V^\mu \, A_\mu^I = 0$ is imposed at the horizon. This particular gauge is justified since it ensures regularity of the solution by avoiding divergences in the square norm of the gauge fields.

The Euclidean continuation of the Lorentzian regularized action can now be obtained by performing a Wick rotation on the time and by making the latter periodic of period $\beta$, so that we have
\begin{equation}
    \label{Euclidean_Action}
    \frac{I}{\beta} = - \frac{S_\text{ren}}{\int \diff t} \, .
\end{equation}
Now we finally computed the last quantity involved in the quantum statistical relation. Plugging all the ingredients into eq.~\eqref{Quantum_Statistical_Relation_Extremal}, we see that it is indeed verified. Note also that by recalling that $\Phi^I = 0$ for our family of solutions and by defining the holographic charge associated to the Killing vector~\eqref{Killing_tpsi} as:
\begin{equation}
    Q_V = E - \frac{2}{\ell \, v^2} \, J \, ,
\end{equation}
the quantum statistical relation assumes the form:
\be
\frac{I}{\beta} = Q_V \,,
\ee
which can also be seen as the BPS relation between the holographic charges, the anomalous contribution of~\cite{An:2017ihs,Papadimitriou:2017kzw} being already included.


\section{Conclusions} \label{sec:Conclusion}

In the present paper we have constructed a new family of supersymmetric AlAdS$_5$ black holes with a boundary geometry containing a squashed $S^3$. These black holes generalize the solutions previously found in minimal gauged supergravity~\cite{Blazquez-Salcedo:2017kig,Blazquez-Salcedo:2017ghg} and also the one of~\cite{Cassani:2018mlh} for $n_V = 2$ (since we have not imposed any ansatz on the scalar fields), and can be uplifted to be solutions of ten-dimensional type IIB supergravity. Our family of solutions depends on three-parameters of which two regulate the horizon geometry, the angular momentum and the Page charges while the remaining one determines the squashing at the boundary. The horizon properties are totally independent on the squashing; therefore if we set a particular horizon geometry by fixing the two former parameters, whatever the squashing the $S^3$ metric will flow to a fixed one at the horizon. This has somehow the flavour of the attractor mechanism for scalar fields in four dimensions~\cite{Cabo-Bizet:2017xdr}.

Let us compare the number of parameters describing the horizon geometry we obtained with the number one should expect by a theoretical counting. For the $n_V = 2$ model we are examining, we have five conserved charges, which are the energy, one angular momentum and three electric charges; however only four of these five total charges are independent since supersymmetry imposes one linear constraint among them. One can therefore expect to find black hole solutions with four parameters regulating the horizon geometry, but already in the solution of~\cite{Gutowski:2004yv} one of these is constrained by the requirements to be imposed to avoid causal pathologies~\cite{Cvetic:2005zi}, so the independent parameters are three. We should therefore expect to be possible to find squashing solutions with three independent parameters regulating the horizon geometry, in addition to the one determining the squashing at the boundary. Our family of solutions presents an horizon geometry described by two parameters, generalizing the solutions constructed in~\cite{Cassani:2018mlh} characterized by only one parameter regulating the horizon geometry due to the ansatz for the scalar fields adopted there. The black holes presented here are thus the most general squashed solutions found in the U$(1)^3$ theory we are studying; however we are still missing squashed solutions with an horizon geometry regulated by three parameters, which should be the most possible general ones according to the theoretical counting arguments reported above. It could be that the general three parameter solution breaks the SU(2) $\times$ U(1)$^4$ symmetry in the bulk and should thus be searched in a more general setup than Fayet-Iliopoulos gauged supergravity.

We have seen in sec.~\ref{sec:Properties} that the entropy of our family of black holes is reproduced using the Page charges instead of the holographic charges. The two types of charges are different for AlAdS$_5$ solutions due to the presence of the Chern-Simons term, which does not vanish asymptotically like in the case of non-squashed solutions. The formula thus obtained for the entropy in terms of the conserved charges is in agreement with the typical one for asymptotically AdS$_5$ black holes reported for example in~\cite{Kim:2006he}, provided the fact that for AlAdS$_5$ solutions the charges appearing there must be identified with the Page charges and not with the holographic ones. According to the extremization principle proposed in~\cite{Hosseini:2017mds}, the entropy of supersymmetric asymptotically AdS$_5$ black holes can be obtained by Legendre-transforming a certain function of chemical potentials conjugated to the black hole conserved charges. This has been further discussed in~\cite{Cabo-Bizet:2018ehj}, where the authors identified the function to be Legendre-transformed as the on-shell action of the black hole and managed to reproduce the entropy of~\cite{Chong:2005hr} using the extremization principle. The result for the entropy we have obtained in this paper suggests that the same extremization principle would work for our family of black holes if one takes into account the Page charges and their conjugate chemical potentials instead of the electric holographic charges. This distinction cannot be established looking at asymptotically globally AdS$_5$ solutions since the holographic charges and the Page charges coincide for them, due to the fact that the Chern-Simons term vanishes at the boundary.

\section*{Acknowledgments} 

The authors would like to thank Davide Cassani for the useful comments and discussions. The authors acknowledge hospitality to the Galileo Galilei Instutue for Theoretical Physics during the completion of the present work.
AB is supported by the ANR grant Black-dS-String ANR16-CE31-0004-01.
AB wishes to thank the Institut de Physique Th\'eorique, Universit\'e Paris Saclay, CEA, CNRS for the hospitality while the final part of this project was realized. 

\appendix

\section{More on the perturbative solution} \label{app_more_perturbative}

\subsection{More on the near-boundary solution}\label{app_more_UV}

In this appendix we provide further details on the near-boundary analysis of sec.~\ref{sec:near_boundary}. We give some information and results about the main functions characterizing the solution in this region and we write it in Fefferman-Graham coordinates. These are the well-suited coordinates to be used to describe the near boundary behaviour of our solution, since they both allow to check that it is indeed AlAdS and to provide an holographic interpretation, helping us to understand the role of the different parameters controlling the expansions of the supegravity fields from a field theoretic point of view. All the functions presented in this appendix are evaluated in the coordinates $(t , \psi)$ defined in~\eqref{Change_Coord} as
\be
\label{change_psi_appendix}
y = t \ ,
\qquad \hat{\psi} = \psi + \chi \, t \ , \quad \text{where}\quad \chi \equiv \frac{2}{4c-1} \, .
\ee
We begin by reporting the first terms of the perturbative solutions for $a$, $H$ and $K$:%
{\footnotesize
	\begin{subequations}
	\begin{align}
	a(\rho) & = a_0 \, e^\rho + \left( a_2 + c \, \rho \right) \frac{e^{-\rho}}{a_0} + \bigg[ a_4 +  \left(2 - 16 \, a_2 - 5 \, c \right) \, \frac{c}{12} \, + 54 \, \bigg( (2 \, H_2 + 3 \, \HL + K_2) \, \HL + \notag \\
	&  + (H_2 + 3 \, \HL + 2 \, K_2) \, \KL + 
	3 \, \KL^2 \bigg) \, \rho  + \left(-\frac{2 \, c^2}{3} + 
	\frac{9}{2} (\HL^2 + \HL \, \KL + \KL^2) \right) \rho^2 \bigg ] \frac{e^{-3 \, \rho}}{a^3_0} + \CO(e^{-4 \, \rho}) \, , \label{Sol_a_UV} \\
	H(\rho) & = \left(H_2 + \HL \, \rho \right) a_0^2 \, e^{2 \rho} + H_4 + \frac{1}{6} \, \bigg[ \!\! \left(4 \, H_2 - 2 \, \HL \right) c + \HL + 4 \, a_2 \, \HL + 24 \bigg( \! \left(- H_2 - \HL + K_2 \right) \, \HL  \notag \\
	& + \left(H_2+ 2 \, \left(\HL + K_2 \right) \right) \KL + 2 \,\KL^2 \bigg)  \bigg] \, \rho + \left( \frac{2}{3} (c - 3 \, \HL) \, \HL + 4 \, \HL \, \KL + 4 \, \KL^2 \right) \rho^2  + \CO(e^{-2 \, \rho}) \, ,\label{Sol_H_UV} \\
	K(\rho) & = \left(K_2 + \KL \, \rho \right) a_0^2 \, e^{2 \rho} + K_4 + \frac{1}{6} \, \bigg[ \!\! \left(4 \, K_2 - 2 \, \KL \right) c + \KL + 4 \, a_2 \, \KL + 24 \bigg( \! \left(- K_2 - \KL + H_2 \right) \, \KL  \notag \\
	& + \left(K_2+ 2 \, \left(\KL + H_2 \right) \right) \HL + 2 \,\HL^2 \bigg)  \bigg] \, \rho + \left( \frac{2}{3} (c - 3 \, \KL) \, \KL + 4 \, \KL \, \HL + 4 \, \HL^2 \right) \rho^2  + \CO(e^{-2 \, \rho}) \, . \label{Sol_K_UV}
	\end{align}
	\end{subequations}
}%
{\normalsize From here we can see that, already at the displayed order, the coefficients of both $H$ and $K$ enter in the solution for $a$; this case is thus different from both the minimal gauged supergravity one~\cite{Cassani:2014zwa} and the solution of~\cite{Cassani:2018mlh}, where $K = H$. Furthermore we have verified that the expansions reduce to the ones of~\cite{Cassani:2018mlh} when the appropriated limit, discussed in sec.~\ref{sec:susyU1}, is taken. 

Using the above expansions and equations~\eqref{f_from_a_U1},~\eqref{w_from_a_U1}, we find the following asymptotic behaviour for the $f$ and $w$ functions} %
{\footnotesize
\begin{subequations}
	\begin{align}\label{eq_asymptf}
	f(\rho) & = 1 + \frac{1+ 4 c + 16 a_2 + 16 c \rho }{12 a_0^2} \, e^{-2\rho} \nonumber\\
	& \quad + \frac{1}{144 a_0^4} \Big\{ \! \Big[ 1 - 128 \,  a_2^2 + 8 \, (3 - 10 \, c) + a_2 (8 + 96 \, c)   \nonumber\\
	& \quad +  216 \, \big\{ 8 \, (H_2^2 + H_2 \,  K_2 + K_2^2) + 9 ( \tilde H^2 +\tilde H \, \tilde K+ \tilde K^2)  + 12 (H_2 \, \tilde H + K_2 \, \tilde K) + 6 ( \tilde H \, K_2 +  \tilde K \, H_2 ) \Big] \nonumber\\
	&\quad  + 8\Big[ c \, ( 1-32 \, a_2 + 12 \, c) + 108 \, \Big( 3 \, (\tilde H^2 + \tilde H \, \tilde K + \tilde K^2) + 4 \, ( H_2 \, \tilde H + K_2 \, \tilde K) + 2 \, (\tilde H \, K_2 + H_2 \, \tilde K ) \Big) \Big] \rho  \nonumber\\
	& \quad - 64 \, \Big[  2 \, c^2 - 27 ( \tilde H^2 + \tilde H \, \tilde K + \tilde K^2 )\Big] \rho^2  \big\} \Big\} e^{-4 \rho}  + \cdots \ ,
	\end{align}
	\begin{align}\label{eq_asymptw}
	w(\rho) &= - 2 \, a_0^2 \, e^{2\rho} + \left[  \left(\frac{1}{2} + 4 \, a_2 - 2 \,  c \right) + 4 \, c \, \rho \right]  \nonumber\\
	& \quad + \Big\{  \Big[ -1-352 \, a_2^2 + 192 \, a_4 + 8 \, (2-3 \, c) \, c + 32 \, a_2 \, (5 \, c-1) \nonumber\\
	& \quad + 216 \, \Big( 8 (H_2^2 + H_2 \, K_2 + K_2^2) + 8 (H_2 \, \tilde H + \tilde K \, K_2 ) + 4( \tilde H \, K_2 + H_2 \, \tilde K) + 3 ( \tilde H^2 + \tilde H \, \tilde K + \tilde K^2) \Big)\Big]  \nonumber\\
	& \quad +16 \Big[ 5 \, c \, ( c-12 a_2) + 54  \Big( 6 \, ( H_2 \, \tilde H + \tilde K \, K_2 ) +5 ( \tilde H^2 + \tilde H \, \tilde K + \tilde K^2 ) + 3 ( \tilde H \, K_2 + H_2 \, \tilde K ) \Big)  \Big] \rho \nonumber\\
	& \quad - 96 \Big[  5  \, c^2 - 27 \, \Big( \tilde H^2 + \tilde H \, \tilde K + \tilde K^2  \Big)\Big] \rho^2   \Big\} \frac{e^{-2\rho}}{a_0^2} + \cdots  \ .
	\end{align}
\end{subequations}
}Looking at the expansions above we note that when $\rho \to \infty$ we have the limits $f \to 1$ and $w \to e^{2\rho}$: both these behaviours are consistent with an AlAdS solution.

In the coordinates~\eqref{change_psi_appendix}, the metric and the gauge fields become
\begin{subequations}
\begin{align}
\diff s^2 &= g_{\rho \rho} \diff  \rho^2 + g_{\theta \theta} (\sigma_1^2 + \sigma_2^2) + g_{\psi \psi} \sigma_3^2 + g_{tt} \diff t^2 + 2 g_{t \psi} \, \sigma_3 \, \diff t\ \, ,  \\
A^I &= A^I_t \, \diff t + A^I_\psi \, \sigma_3\ .
\end{align}
\end{subequations}
The various components of the fields assume the following expressions
\begin{subequations}
\begin{align}
g_{\rho \rho} &= f^{-1}\ , \qquad \qquad g_{\theta \theta} = f^{-1} a^2\ , \qquad \qquad g_{\psi \psi} = - f^2  w^2 + f^{-1} (2 a a^\prime)^2\ , \notag \\[1mm]
g_{tt} &= -f^2 (1 + \chi \, w)^2 + \chi^2 f^{-1} (2 a a^\prime)^2 \ ,\ \ \quad g_{t \psi} = -f^2 (1 + \chi \, w) \, w + \chi \, f^{-1} (2 a a^\prime)^2\ , \\
A^I_t  &= \left(\, f + \chi \, f \, w \right) X^I  + \chi \, U^I \ ,\qquad\qquad
A^I_\psi  = f \, w \, X^I + U^I \ .
\end{align}
\end{subequations}
Instead of giving the explicit near-boundary expansions for the various components of the metric, the gauge fields and the scalars in the radial coordinate $\rho$ we used until now, we prefer to switch to the Fefferman-Graham radial coordinate $r$, since it provides a clearer and easier holographic interpretation of the solution.

\subsubsection{Near-boundary solution in Fefferman-Graham coordinates}\label{app_FG}
We begin by presenting the general expected forms the fields of our solution should take in Fefferman-Graham coordinates.
The metric should assume in these coordinates the following form: 
\begin{equation}
\label{FGmetric}
\diff s^2 = \frac{\diff r^2}{r^2} + h_{ij} (x,r) \, \diff x^i \, \diff x^j\ ,
\end{equation}
having denoted the radial coordinate as $r$, while $x^i$ describe the various hypersurfaces at fixed $r$. Each of these hypersufaces has an induced metric $h_{ij}$ that presents the following large $r$ expansion
\begin{equation}
\label{FGmetricomponents}
h_{ij}(x,r) \,=\, r^2 \bigg[ h_{ij}^{(0)} + \frac{h_{ij}^{(2)}}{r^2} + \frac{h_{ij}^{(4)} + \tilde{h}_{ij}^{(4)} \, \log{r^2} + \tilde{\tilde{h}}_{ij}^{(4)} \, \left(\log{r^2}\right)^2}{r^4} + \dots \bigg]\ .
\end{equation}
Obviously all the coefficients in the expansion above depend only on the $x^i$ coordinates and are thus independent on $r$. In a similar way, one can build in the radial gauge $A^I_r = 0$  the following large $r$ expansion for the gauge fields
\begin{equation}
\label{FGgauge}
A^I(x,r) \,=\, A^{I \, (0)} + \frac{A^{I \, (2)} + \tilde{A}^{I \, (2)} \, \log{r^2}}{r^2} + \dots\ .
\end{equation}
For the large $r$ ansatz we shall assume for the scalar fields, the situation is more involved, since this varies according to the mass of the scalar fields and to the conformal dimension of the corresponding dual SCFT operators. The scalar fields $X^I$ involved in the present paper have a mass that fulfils the relation $m^2 \ell^2 = -4$, so the corresponding SCFT operators have dimension $\Delta = 2$. From~\cite{Bianchi:2001kw} we can therefore read that such scalars should present a Fefferman-Graham expansion of the form
\begin{equation}
\label{FGscalars_up}
X^I = \, \bar{X}^I + \frac{{}^{(0)}\phi^{I} + {}^{(0)}\tilde{\phi}^{I} \, \log{r^2}}{r^2} + \frac{{}^{(2)} \phi^{I } + {}^{(2)}\tilde{\phi}^{I} \, \log{r^2} + {}^{(2)}\tilde{\tilde{\phi}}^{I} \, \left(\log{r^2}\right)^2}{r^4} + \dots \ .
\end{equation}
To move in Fefferman-Graham coordinates, we need to switch our radial coordinate $\rho$ to the Fefferman-Graham radial coordinate $r$ so that the $g_{\rho \rho}$ component of the metric becomes $\frac{\diff r^2}{r^2}$. This means that we have to impose that
\be
f^{-1/2}(\rho) \, \diff \rho =  \frac{\diff r}{r} \,,
\ee
which is equivalent to solve the ODE 
\be
\frac{\diff r}{\diff \rho} =  f^{-1/2}(\rho) \, r(\rho) \,.
\ee
Finding an analytical solution of this equation turns out to be very hard, therefore we solve it perturbatively at large $\rho$ obtaining
\be
\begin{split}
	a_0^2 r^2 &= a_0^2 e^{2\rho} + \frac{16 a_2 +12 c +1}{24} + \frac{2c}{3} \, \rho \\
	& \quad + \Bigg[ \Bigg(
	\frac{128 a_2 c+8 c (13-30 c)-768 a_2^2}{2304}  + \frac{1}{768} + \frac{39}{8} \left( \tilde H^2 + \tilde H \tilde K + \tilde K^2 \right) \\
	& \quad \qquad  + 9 \left(  H_2^2      +2 H_2 \tilde H+ H_2 K_2+ H_2 \tilde K+ \tilde H K_2+ 2 K_2 \tilde K     + K_2^2 \right) \!\Bigg) \\
	& \quad  \quad +\left(  \frac{(c-12 a_2 )c}{18} + 3\left( 2 \tilde H^2+ 2 H_2 \tilde H+ H_2 \tilde K+ \tilde H K_2+ 2 \tilde H \tilde K+2 K_2 \tilde K+2 \tilde K^2  \right)  \right) \rho \\
	& \quad\quad +\left( 3 \left(\tilde H^2+\tilde H \tilde K+\tilde K^2\right)-\frac{c^2}{3}  \right) \rho^2  \Bigg] \frac{e^{-2\rho}}{a_0^2} + {\cal O}\left( e^{-3\rho}\right) .
\end{split}
\ee
This change of coordinates is such that with respect to ($t,r,\theta,\phi, \psi$) the metric becomes
\be
\diff s^2 = \frac{\diff r^2}{r^2} + h_{\theta \theta} \, (\sigma_1^2 + \sigma_2^2) + h_{\psi \psi} \, \sigma_3^2 + h_{tt} \, \diff t^2 + 2 \, h_{t\psi} \, \sigma_3 \, \diff t \,,
\ee
with all the components $h_{ij}$ having a large $r$ expansion which is consistent with~\eqref{FGmetricomponents}. The various coefficients of the metric component expansions assume the following expressions:%
{\footnotesize 
\be
\begin{split}
	h_{\theta \theta}^{(0)} &= a_0^2 \,, \quad h_{\theta \theta}^{(2)}  =-\frac{5 c}{6}-\frac{1}{8} \,, \quad \tilde{ \tilde h}_{\theta \theta}^{(4)} = -\frac{3 \left(\tilde H^2+\tilde H \tilde K+\tilde K^2\right)}{2a_0^2} \,, \\
	\tilde h_{\theta \theta}^{(4)} &=\frac{-4 c^2+c+9 \left(\tilde K (-2 H_2+\tilde H-4 K_2)+\tilde H (-4 H_2+\tilde H-2 K_2)+\tilde K^2\right)}{6 a_0^2} \,, \\
	h_{\theta \theta}^{(4)} &=   \frac{1024 a_2^2-384 a_2 c+1536 a_4+8 c (74 c-15)-1}{768 a_0^2}  \\
	& \quad - \frac{3}{8a_0^2}\left( 40 H_2^2+8 H_2 (8 \tilde H+5 K_2+4 \tilde K)+49 \tilde H^2 \right . \\
	& \qquad\qquad\quad  \left. +32 \tilde H K_2+49 \tilde H \tilde K+40 K_2^2+64 K_2 \tilde K+49 \tilde K^2\right) ,
\end{split}
\ee
\be
\begin{split}
	h_{\psi \psi}^{(0)} &=a_0^2 (1-4c)  \,,  \quad h_{\psi \psi}^{(2)} = - \frac{(-1 + 4 c) (-3 + 28 c)}{24} \,, \quad \tilde{ \tilde h}_{\psi \psi}^{(4)} = \frac{3 (4 c-1) \left(\tilde H^2+\tilde H \tilde K+\tilde K^2\right)}{2a_0^2} \,, \\
	\tilde h_{\psi \psi}^{(4)} &= \frac{1-4c}{6a_0^2}\! \left[  2 c (4 c-1) - 9 \left(  4 H_2 \tilde H+2 H_2 \tilde K+5 \tilde H^2+2 \tilde H K_2+5 \tilde H \tilde K+4 K_2 \tilde K+5 \tilde K^2 \right)\right] \, \\
	h_{\psi \psi}^{(4)} &= \frac{1}{62208 a_0^2} \Bigg[ \Big( 3 (4608 a_4-995328 a_6+9604 c-75) + 16 \left(24 c (a_2 (391-208 a_2)+336 a_4) \right.  \\
	& \qquad \quad \left. -144 a_2 (4 a_2 (64 a_2+3)+2064 a_4+3)-9 (2976 a_2+1157) c^2+13420 c^3\right) \!\! \Big) \\
	& \quad + \frac{1}{24} \Big(   -8 H_2^2 (576 a_2+268 c+2592 K_2+2520 \tilde K+9) \\
	& \quad -8 H_2 \left(\tilde H (608 a_2+616 c+5040 K_2+3960 \tilde K-34)  -576 H_4+2592 K_2^2-288 K_4 \right. \\
	& \quad \left. +K_2 (576 a_2+268 c+5040 \tilde K+9)+\tilde K (304 a_2+308 c+1980 \tilde K-17) \right) \\
	& \quad +\tilde K (7296 K_4-64 K_2 (76 a_2+77 c+495 \tilde H)+\tilde H (2944 a_2-3308 c-8712 \tilde H+871)+272 K_2) \\
	& \quad +\tilde K^2 (2944 a_2-3308 c-8712 \tilde H+871)+2944 a_2 \tilde H^2-2432 a_2 \tilde H K_2 \\
	& \quad - 4608 a_2 K_2^2-3308 c \tilde H^2-2464 c \tilde H K_2-2144 c K_2^2  +192 H_4 (38 \tilde H+12 K_2+19 \tilde K) \\
	& \quad - 15840 \tilde H^2 K_2+871 \tilde H^2  -20160 \tilde H K_2^2+136 \tilde H K_2+3648 \tilde H K_4-72 K_2^2+4608 K_2 K_4 \Big) \Bigg] , 
\end{split}
\ee
\be
\begin{split}
	h_{t\psi}^{(0)} &= h_{t\psi}^{(2)}  = \tilde h_{t\psi}^{(4)} =  \tilde{\tilde h}_{t\psi}^{(4)} =0\,, \\
	h_{t\psi}^{(4)} &=- 2 h_{\theta \theta}^{(4)} - 2 \frac{h_{\theta \theta}^{(0)} }{h_{\psi \psi}^{(0)} }  \, h_{\psi \psi}^{(4)}  + \frac{128 a_2 (4 c-1)+8 c (38 c+1)-5}{192 a_0^2} \\
	& \qquad -\frac{6 \left(4 H_2^2+2 H_2 (2 (\tilde H+K_2)+\tilde K)+\tilde H^2+\tilde H (2 K_2+\tilde K)+(2 K_2+\tilde K)^2\right)}{a_0^2} \,,
\end{split}
\ee
\be
\begin{split}
	h_{tt}^{(0)} &= - \frac{4 a_0^2}{1-4c} \,, \quad h_{tt}^{(2)} =-  \frac{4 c+3}{6(1-4 c)} \,, \quad \tilde{\tilde h}_{tt}^{(4)} = \frac{6 \left(\tilde H^2+\tilde H \tilde K+\tilde K^2\right)}{ a_0^2 (1-4 c)}   \,, \\
	\tilde h_{tt}^{(4)} &= \frac{6 \left(\tilde K (2 H_2+5 \tilde H+4 K_2)+\tilde H (4 H_2+5 \tilde H+2 K_2)+5 \tilde K^2\right)}{a_0^2 (1- 4 c)} \,, \\
	h_{tt}^{(0)} &= 8\, \frac{h_{\theta \theta}^{(0)}}{h_{\psi \psi}^{(0)}} \, h_{\theta \theta}^{(4)} + 4 \left( \frac{h_{\theta \theta}^{(0)}}{h_{\psi \psi}^{(0)}} \right)^2 h_{\psi\psi}^{(4)} - \frac{1}{48 h_{\psi \psi}^{(0)}} \left[ 3 + 2 \left( 1 - \frac{g_{\psi \psi}^{(0)}}{g_{\theta \theta}^{(0)}} \right) + 11 \left(  1 - \frac{g_{\psi \psi}^{(0)}}{g_{\theta \theta}^{(0)}}\right)^2  \right] \\
	& \quad + \frac{24 \left(4 H_2^2+2 H_2 (2 \tilde H+2 K_2+\tilde K)+\tilde H^2+\tilde H (2 K_2+\tilde K)+(2 K_2+\tilde K)^2\right)}{a_0^2 (1-4 c)} \,.
\end{split}
\ee}%
Comparing with App.~A of~\cite{Cassani:2014zwa} we may notice that both the leading order terms and the next-to-leading order ones are the same as in minimal gauged supergravity. The effect of the supergravity vector multiplet fields is instead manifest at the following order where coefficients dependent on $\HL, H_2, H_4, \KL, K_2$ and $K_4$ appear. In~\cite{Cassani:2018mlh} the backreaction of the fields appeared at the same order, but it was dependent only on the three parameters of the function $H$ introduced there, playing a role analogous to $\HL, H_2$ and $H_4$. As it is well-known, the free parameters of the metric should appear in the coefficients $h^{(0)}$ and $h^{(4)}$ of the Fefferman-Graham expansion. In particular the parameters controlling $h^{(0)}$ should play the role of the source for the energy-momentum tensor $T_{\mu \nu}$ of the dual CFT, while $h^{(4)}$ should be its expectation value. The analysis of~\cite{Cassani:2014zwa} shows that we should expect five free parameters appearing in the metric and these are found to be $a_0$, $c$, $a_2$, $a_4$ and $a_6$, as it is possible to check looking at the various $h^{(0)}$ and $h^{(4)}$ coefficients above. 

Let us now proceed to examine the scalar fields. They are indeed found to take the form~\eqref{FGscalars_up} and their Fefferman-Graham coefficients are
\be
\begin{split}
	{}^{(0)}\tilde \phi^1 &= \frac{\tilde H}{a_0^2} \,, \\
	{}^{(0)} \phi^1 &= \frac{2 H_2 + \tilde H}{a_0^2} \,, \\
	{}^{(2)}\tilde{\tilde \phi}^1 &= \frac{(\tilde H^2 + \tilde H \tilde K+ \tilde K^2)}{a_0^2} \,, \\
	{}^{(2)}\tilde \phi^1 &= \frac{\tilde H (4 c+96 H_2+48 K_2+144 \tilde K+3)+48 \tilde K (H_2+2 K_2+3 \tilde K)}{24 a_0^4} \,, \\
	{}^{(2)} \phi^1 &=  \frac{1}{12 a_0^4} \Big[ H_2 (4 c+48 K_2+72 \tilde K+3)+48 H_2^2 \\
	& \qquad\quad   +3 \left(\tilde H (-4 c+24 K_2+36 \tilde K+1)  -12 \tilde H^2+4 (2 K_2+3 \tilde K)^2\right)  \Big]\,, 
\end{split}
\ee
\be
\begin{split}
	{}^{(0)}\tilde \phi^2 &= \frac{\tilde K}{a_0^2} \,, \\
	{}^{(0)} \phi^2 &= \frac{2 K_2 + \tilde K}{a_0^2} \,, \\
	{}^{(2)}\tilde{\tilde \phi}^2 &= \frac{(\tilde H^2 + \tilde H \tilde K+ \tilde K^2)}{a_0^2} \,, \\
	{}^{(2)}\tilde \phi^2 &= \frac{(4 c+3) \tilde K+48 (\tilde K (H_2+3 \tilde H+2 K_2)+\tilde H (2 H_2+3 \tilde H+K_2))}{24 a_0^4} \,, \\ 
	{}^{(2)} \phi^2 &=  \frac{1}{12 a_0^4} \Big[  (4 c+3) K_2+3 (1-4 c) \tilde K \\
	& \quad +12 \left(4 H_2^2+2 H_2 (6 \tilde H+2 K_2+3 \tilde K)+9 \tilde H^2+6 \tilde H K_2+9 \tilde H \tilde K+4 K_2^2-3 \tilde K^2\right)   \Big]\,. 
\end{split}
\ee
We do not report the expansion for the third scalar $X^3$ since it is fixed by the constraint~\eqref{Scalars_Up_constraint}. We note furthermore that the expansion for the scalar $X^2$ coincides with the one for $X^1$ upon performing the switching
\begin{equation}
\label{Switching_Coeff}
\HL \leftrightarrow \KL\,, \qquad H_2 \leftrightarrow K_2 \,, \qquad H_4 \leftrightarrow K_4  \,, 
\end{equation}
which is equivalent to switch the coefficients of $H$ with the ones of $K$ and viceversa. 
Both $X^1$ and $X^2$ should have two free parameters in the coefficients ${}^{(0)}\tilde \phi^{1,2}$ and ${}^{(0)} \phi^{1,2}$ which are interpreted respectively as the sources and the expectation values of the dual scalar operators. Looking at the expression of the coefficients above we clearly see that the free parameters are $\HL, \KL$, which are associated with the sources, and $H_2, K_2$, which are associated with the expectation values. 
We do not report the expansions of the lower-index scalars $X_I$ since they can be easily obtained using their definition~\eqref{Def_XI_Down}. 

Finally we look at the gauge fields. We find that they can be expressed in the Fefferman-Graham form~\eqref{FGgauge} and their coefficients are:
\be
\begin{split}
	{}^{(0)}A^1_t &= \frac{4 c+36 \tilde H-3}{12 c-3} \,, \quad  {}^{(2)}\tilde A^1_t = 0 \,, \\
	{}^{(2)}A^1_t &= \frac{1}{72 a_0^2 (4 c-1)} \Big[ -5+384 a_4-32 c+288 H_2-1728 H_4+288 \tilde H \\
	& \quad + 8 \Big( 32 a_2^2+4 a_2 (5 c+36 H_2-18 \tilde H-2)+29 c^2-72 c \tilde H \\
	& \quad -27 \left( 24 H_2^2+4 H_2 (14 \tilde H-6 K_2-5 \tilde K)+41 \tilde H^2-20 \tilde H K_2 \right. \\
	& \qquad \qquad \left. -19 \tilde H \tilde K-24 K_2^2-40 K_2 \tilde K-19 \tilde K^2  \right)\! \Big)\Big] , 
\end{split}
\ee
\be
\begin{split}
	{}^{(0)}A^2_t &= \frac{4 c+36 \tilde K-3}{12 c-3} \,, \quad  {}^{(2)}\tilde A^2_t = 0 \,, \\
	{}^{(2)}A^2_t &= \frac{1}{72 a_0^2 (4 c-1)} \Big[   256 a_2^2+32 a_2 (5 c+36 K_2-18 \tilde K-2)+384 a_4+232 c^2\\
	& \qquad\qquad -32 c (18 \tilde K+1)+288 K_2-1728 K_4+288 \tilde K-5 \\
	& \qquad\qquad  + 216 \big( 24 H_2^2+4 H_2 (10 \tilde H+6 K_2+5 \tilde K)+19 \tilde H^2\\
	& \qquad \qquad+20 \tilde H K_2+19 \tilde H \tilde K-24 K_2^2-56 K_2 \tilde K-41 \tilde K^2 \big)\Big] \, ,
\end{split}
\ee
\be
\begin{split}
	{}^{(0)}A^1_\psi &=\frac{4 c}{3}+6 \tilde H   \,, \quad   {}^{(2)}\tilde A^1_\psi = \frac{(1- 4 c) (2 c-9 \tilde H)}{6 a_0^2}  \,, \\
	{}^{(2)}A^1_\psi &=  \frac{1}{144 a_0^2} \Big[ 256 a_2^2+384 a_4 + 136 c^2-1728 H_4 + 72 \tilde H+1  \\
	& \qquad  +32 c (54 H_2+9 \tilde H-1)-32 a_2 (7 c-36 H_2+18 \tilde H-1) \\
	& \qquad -72 \big( 72 H_2^2-3 \tilde K (20 H_2+19 \tilde H+40 K_2)+168 H_2 \tilde H \\
	& \qquad -72 H_2 K_2+2 H_2+123 \tilde H^2-60 \tilde H K_2-72 K_2^2-57 \tilde K^2\big) \Big] \, ,
\end{split}
\ee
\be
\begin{split}
	{}^{(0)}A^2_\psi &=  \frac{4 c}{3}+6 \tilde K \,, \quad   {}^{(2)} \tilde A^2_\psi =  \frac{(1- 4 c) (2 c-9 \tilde K)}{6 a_0^2}   \,, \\
	{}^{(2)}A^2_\psi &=   \frac{1}{144 a_0^2} \Big[ 256 a_2^2+384 a_4+136 c^2-144 K_2-1728 K_4+72 \tilde K+1 \\
	& \qquad +32 c (54 K_2+9 \tilde K-1)-32 a_2 (7 c-36 K_2+18 \tilde K-1) \\
	& \qquad + 216 \big( 24 H_2^2+19 \tilde H^2+20 \tilde H K_2+19 \tilde H \tilde K-24 K_2^2 \\
	& \qquad  -56 K_2 \tilde K_4 H_2 (10 \tilde H+6 K_2+5 \tilde K)-41 \tilde K^2\big) \Big] .
\end{split}
\ee
As for the scalars, we have not reported the third gauge field $A^3$, since it is not independent from $A^1$ and $A^2$ and can be therefore computed by them. Furthermore we also note that the gauge field $A^2$ can be obtained from $A^1$ switching the parameters as in~\eqref{Switching_Coeff}. Supersymmetry constraints the form of the gauge fields, relating them to the metric and the scalar fields. Indeed, looking at the expressions of the coefficients above, we see that the only free parameters appearing in the gauge fields are $H_4$ and $K_4$; all the other parameters were found before either in the metric or in the scalar fields. Furthermore the equations of motion in Fefferman-Graham coordinates do not determine the coefficients ${}^{(0)}A^{1,2}$ and ${}^{(2)}A^{1,2}$; these are rather fixed by supersymmetry.


\subsection{More on the near-horizon solution} \label{app_more_IR}
In this appendix we report additional information about the near-horizon solution we presented in sec.~\ref{sec:near_horizon}. In particular, here we show the small $\rho$ expansions of all the functions not presented in the main text. We begin by reporting the first terms of the expansions of the functions $a$, $H$ and $K$, which determine the whole near-horizon solution:%
\newpage
{\footnotesize 
\begin{align}
\label{aHK_nearhorizon}
a \,&=\,  \air \, \rho  + \air_3\, \rho^3 + \air_4\, \rho^4 + \frac{3 \, \air_3^2}{10 \, \air}\rho^5+ \mathcal{O}(\rho^6)\ , \nn \\[1mm]
H \,&=\,  \eta \, \air^2 \,  \rho ^2 + 2 \, \air \, \air_3 \, \eta \,  \rho ^4 + \frac{1}{529 \, \air^4 - 92 \, \air^2 - 6912 \left(\hir^2+ \hir \,  \kir + \kir^2 \right)+ 4} \bigg[2 \, \air \, \air_4 \bigg(-768 \, \kir ^2 \left(\air^2+9 \hir \right) \notag \\
& \qquad \qquad  - 768 \, \hir \,  \kir  \left(\air^2+9 \, \hir \right)+ \hir  \left(345 \, \air^4 + 4 \, \air^2 (96 \, \hir -19) - 6912 \, \hir^2 + 4 \right) \bigg) \bigg] \rho^5 + \CO(\rho^6) \notag \ , \nn \\[1mm] \notag \\
K \,&=\,  \kir \, \air^2 \,  \rho ^2 + 2 \, \air \, \air_3 \, \kir \,  \rho ^4 + \frac{1}{529 \, \air^4 - 92 \, \air^2 - 6912 \left(\kir^2+ \kir \,  \hir + \hir^2 \right)+ 4} \bigg[2 \, \air \, \air_4 \bigg(-768 \, \hir ^2 \left(\air^2+9 \kir \right) \notag \\
& \qquad \qquad  - 768 \, \kir \,  \hir  \left(\air^2+9 \, \kir \right)+ \kir  \left(345 \, \air^4 + 4 \, \air^2 (96 \, \kir -19) - 6912 \, \kir^2 + 4 \right) \bigg) \bigg] \rho^5 + \CO(\rho^6) \, . 
\end{align}}%
Recall that $\iota$ is not an independent parameter, but it is fixed as in~\eqref{iota_BH}.
Although at the displayed order the near-horizon expansion of the $a$ function is equal to the one of minimal gauged supergravity~\cite{Cassani:2014zwa}, the presence of non-trivial scalars becomes evident at next orders where terms dependent on $\eta$ appear. 

Using these expansions we can compute the functions $f$ and $w$ via eqs. \eqref{f_from_a_U1} and \eqref{w_from_a_U1} respectively. We obtain the following
\begin{align}
    \label{f_IR}
    f & = \frac{12 \, \air^2 \, \rho^2}{\Delta} +\frac{24 \, \air \, \air_3}{\Delta^4} \bigg\{8 \bigg[-64 \air^6 + 30 \air^4 + \air^2 \left(5184 \left(\eta ^2+\eta  \iota +\iota ^2\right)-3\right) \notag \\
    & \qquad  -648 (72 \eta -1) \iota  (\eta +\iota )\bigg]+5184 \eta ^2-1  \bigg\} \, \rho^4 +  \CO(\rho^6) \, ,
\end{align}%
\begin{align}
    \label{w_IR}
    w &= \frac{-16 \air^4 + 8 \air^2 + 1728 \left(\hir^2 + \hir \, \kir +\kir^2 \right) -1}{48 \air^2 \, \rho^2} \notag \\
    &\qquad + \frac{\air_3 \left[-272 \alpha ^4+64 \alpha ^2-1728 \left(\eta ^2+\eta \, \iota +\iota ^2\right)+1\right]}{24 \alpha ^3} + \CO(\rho) \, ,
\end{align}
where $\Delta$ is defined as in \eqref{Delta_BH}.
Looking at~\eqref{f_IR}, we note that $f$ begins as $\rho^2$, so that it vanishes as $\rho \to 0$. As explained in the main text, this is totally compatible with a black hole solution whose horizon is situated at $\rho = 0$. Note furthermore that terms controlled by $\hir$ and $\kir$ appear already at the leading order in both the functions; therefore it is evident that these expansions differ from the ones of~\cite{Cassani:2018mlh} and from the minimal gauged supergravity ones~\cite{Cassani:2014zwa,Blazquez-Salcedo:2017ghg} providing a generalization of both of them. 

The last functions we explicitly report are the $U^I$, which appear in the gauge fields. The functions $U^1$ and $U^2$ are easily computed by~\eqref{UI_function_U1},~\eqref{UI_function_U2} and we find
\begin{align} \label{UI_functions_IR}
U^1 & = \frac{1}{3} \left(4 \air^2-36 \hir -1 \right) + 12 \, \air \, \air_3 \, \rho ^2 + \CO(\rho^3) \, , \notag \\
U^2 & = \frac{1}{3} \left(4 \air^2-36 \kir -1 \right) + 12 \, \air \, \air_3 \, \rho ^2 + \CO(\rho^3) \, , 
\end{align}
while $U^3$ can be obtained by these using the constraint~\eqref{UI_functions_constraint}.


\subsection{First integrals and equations for the subleading parameters}
\label{app_First_Integrals}
Here we report the explicit expressions for the near-boundary parameters $a_4, \, H_4, \, K_4, \, a_6$ one can find by evaluating the first integrals $\CK_1, \, \CK^{(1)}_2, \, \CK^{(2)}_2, \, \CK_3$ in the near-boundary region and the form the latter assume in the near-horizon one. 

By evaluating \eqref{KappaCharges_U1} in the near-boundary and solving the resulting equations for $a_4$,~$H_4$,~$K_4$,~$a_6$ we find the following relations 

{\small 
\be
\begin{split} \label{First_Integral_UV}
a_4 &= \frac{1}{384}\Big[  5  -256 a_2^2+32 a_2 (2-5 c)+8 c (4-13 c)-144 {\cal K}_1 \\
& \qquad\qquad +216 \Big(8 H_2^2+4 H_2 (6 \tilde H+2 K_2+3 \tilde K)+17 \tilde H^2+12 \tilde H K_2 \\ 
& \qquad \qquad\qquad \qquad +17 \tilde H \tilde K+8 K_2^2+24 K_2 \tilde K+17 \tilde K^2\Big) \! \Big] , \\
H_4 &= \frac{1}{6} \tilde H (1-4 c -2 a_2+24 K_2+36 \tilde K) +H_2 \left(\frac{2 a_2}{3}-4 \tilde H+4 K_2+4 \tilde K+\frac{1}{6}\right) \\
& \qquad -2 H_2^2-3 \tilde H^2+4 K_2^2+8 K_2 \tilde K+6 \tilde K^2 +\frac{{\cal K}_2^{(1)}}{4}  \,, \\
K_4 &= \frac{1}{6} \tilde K (1-4 c -2 a_2+24 H_2+36 \tilde H) +K_2 \left(\frac{2 a_2}{3}-4 \tilde K+4 H_2+4 \tilde H+\frac{1}{6}\right) \\
& \qquad -2 K_2^2-3 \tilde K^2+4 H_2^2+8 H_2 \tilde H+6 \tilde H^2 +\frac{{\cal K}_2^{(2)}}{4} \,, \notag \\ 
\end{split}
\ee
\be
\begin{split}
a_6 &=\frac{1}{93312} \Bigg\{ 80640 a_2^3+288 a_2^2 (197 c-90)+8840 c^3-6000 c^2 \\
& \quad + 27 c \Big[ 25 + 304 {\cal K}_1 - 24 \big(312 H_2^2+4 H_2 (274 \tilde H+78 K_2+137 \tilde K) \\
& \quad +971 \tilde H^2+548 \tilde H K_2+971 \tilde H \tilde K+312 K_2^2+1096 K_2 \tilde K+971 \tilde K^2\big) \Big]	\\
& \quad +3 a_2 \Big[8 c (1913 c-732)-9 \Big(24 \big(1224 H_2^2+12 H_2 (274 \tilde H+102 K_2+137 \tilde K)+2129 \tilde H^2 \\
& \quad +1644 \tilde H K_2+2129 \tilde H \tilde K+1224 K_2^2+3288 K_2 \tilde K+2129 \tilde K^2\big)-2064 {\cal K}_1+29\Big)\!\Big] \\
& \quad + 9 \Big[ 36 \Big(216 H_2^2 (24 K_2+34 \tilde K+1) +24 H_2 \Big(2 \tilde K (507 \tilde H+306 K_2+7) \\
& \quad +612 \tilde H K_2+28 \tilde H+216 K_2^2+9 K_2+6 {\cal K}_2^{(2)}+12 {\cal K}_2^{(1)}+507 \tilde K^2\Big) \\
& \quad +\tilde H^2 (12168 K_2+22122 \tilde K+557)+6 \tilde K (112 K_2+58 {\cal K}_2^{(2)}+29 {\cal K}_2^{(1)}) \\
& \quad +\tilde H \Big(6 \left(4056 K_2 \tilde K+8 K_2 (153 K_2+7)+3687 \tilde K^2\right)+174 ({\cal K}_2^{(2)}+2 {\cal K}_2^{(1)}) \\
& \quad  +557 \tilde K\Big) +72 K_2 (3 K_2+4 {\cal K}_2^{(2)}+2 {\cal K}_2^{(1)})+557 \tilde K^2\Big)-180 {\cal K}_1-27 {\cal K}_3+8 \Big]  \Bigg\} .
\end{split}
\ee}%
As a check of their correctness, we can evaluate these relations in the limit leading to the family of solutions of~\cite{Cassani:2018mlh} and compare the results with the expressions for the same parameters reported in that paper. As explained in the main text, in order to take this limit we have to set $H = K$ and to rescale the two functions by $\frac{1}{6}$. This is equivalent to set $H_2 = K_2 = \frac{H_2^\text{limit}}{6}$ and to impose the same to $H_4, \, K_4, \, \HL, \, \KL$. We have verified that, performing this limit, all the above expressions for the near-boundary free parameters reduce to the ones reported in~\cite{Cassani:2018mlh}. 
Note that equations~\eqref{First_Integral_UV} are valid for every AlAdS solution with the characteristics discussed in sec.~\eqref{sec:near_boundary} and allow to trade the most subleading near-boundary parameters with the four independent first integrals $\CK_1, \, \CK^{(1)}_2, \, \CK^{(2)}_2, \, \CK_3$. Although the number of arbitrary parameters remains the same, this procedure simplifies many expressions.

In the same fashion, we use the near-horizon expansions~\eqref{aHK_nearhorizon} for the functions $a, \, H, \, K$ to explicitly evaluate the first integrals~\eqref{KappaCharges_U1} for the black hole solution we have constructed in this paper. We obtain the following relations
\be
\begin{split} \label{First_Integral_IR}
{\cal K}_1 &=  \frac{5}{144} -\frac{1}{9} \alpha ^2 \left(\alpha ^2+1\right) +12 \left(\eta ^2+\eta \, \iota +\iota ^2\right)  \,, \\
{\cal K}_2^{(1)} &= 8 \eta ^2-16 \iota ^2  -\frac{2}{3} \eta  \left(2 \alpha ^2+24 \iota +1\right)		\,, \\
{\cal K}_2^{(2)} &= 8 \iota ^2-16 \eta ^2  -\frac{2}{3} \iota  \left(2 \alpha ^2+24 \eta +1\right)	\,, \\
{\cal K}_3 &= \frac{7}{108}  -48 \eta ^2 \\
& \quad +  \frac{4}{27} \left[8 \alpha ^6+3 \alpha ^4-3 \alpha ^2 \left(864 \left(\eta ^2+\eta  \, \iota +\iota ^2\right)+1\right)-324 (144 \eta +1) \, \iota  \, (\eta +\iota )\right]	\, ,
\end{split}
\ee
where $\iota$ is fixed as in~\eqref{iota_BH}. As for the near-boundary results, we can check that the relations above reproduce those of~\cite{Cassani:2018mlh} after taking the appropriate limit discussed in sec.~\ref{sec:near_horizon}. 

As we can see from equations~\eqref{First_Integral_IR}, the first integrals are functions of the near-horizon parameters $\air$ and $\hir$ and are fully determined by these. As a consequence, we can determine the parameters $a_4, \, H_4, \, K_4, \, a_6$ in terms of the remaining near-boundary ones and the near-horizon ones $\air, \, \hir$. All the near-boundary parameters should be fully determined in terms of the near-horizon ones; if we had more first integrals at our disposal we could use them to get relations similar to~\eqref{First_Integral_UV} for the other near-boundary parameters and obtain the full dependence of the latter in terms of $\air, \, \hir, \,  \xi$ . However we are not able to find any other first integral (if any exists), therefore we should resort to a numerical procedure to determine the dependence of the remaining near-boundary parameters on the near-horizon ones. This is done in~\cite{Cassani:2018mlh} and also in~\cite{Blazquez-Salcedo:2017ghg,Cassani:2014zwa}.


\section{Holographic renormalization}\label{app_HR}

In this appendix we report some details about how to compute some relevant physical quantities using holographic renormalization and we show their explicit expressions. We divide this appendix into two sections; in the first one we show the procedure to follow in order to obtain the renormalized Lorentzian on-shell action, while in the second one we explicitly compute the energy-momentum tensor, the holographic electric current and the scalar one-point functions. As already stated in the main text, we will always work with the Fefferman-Graham radial coordinate $r$, presented in app.~\ref{app_FG}. We furthermore adopt a minimal subtraction scheme where the only counterterms added are those necessary to cancel the divergences. We will follow closely the notation of~\cite{Cassani:2018mlh}.

\subsection{The renormalized on-shell action}
As we explained in sec.~\ref{sec:Properties}, we can define a Lorentzian renormalized on-shell action as
\be
S_{\rm ren} = \lim_{r_0 \to \infty} S_{\rm reg} \,,
\ee
where the regularized action is
\be
 S_{\rm reg}  = S_{\rm bulk} + S_{\rm GH} + S_{\rm ct} \,.
\ee
The first term coincides with the bulk action~\eqref{Bulk_action}. Using the trace of the Einsten equations~\eqref{Einstein_equations} and combining the result with the Maxwell equation~\eqref{Maxwell_general}, we can rewrite $S_\text{\rm bulk}$ as:
\be
\label{Bulk_action_intermediate}
S_{\rm bulk} = \frac{2}{3 \kappa^2} \int_{M_{r_0}} {\cal V} \star 1 - \frac{1}{3 \kappa^2} \int_{M_{r_0}} \diff \left( Q_{IJ} A^I \wedge \star F^J \right) .
\ee
Recalling the form of the scalar potential~\eqref{scalarpot}, we can show that the following relation holds
\be
\label{Potential_wrt_P}
{\cal V} \star 1 = \half \, \diff \left(  a^2 p \, \diff t \wedge \sigma_1 \wedge \sigma_2 \wedge \sigma_3  \right) \, ,
\ee
where $p$ is the Ricci potential given in~\eqref{P}. Looking at the expression of the latter and recalling the near-horizon expansion of $a$ given in~\eqref{aHK_nearhorizon}, we can immediately conclude that $a^2 p \to 0$ at the horizon; therefore thanks to eq.~\eqref{Potential_wrt_P} it is trivial to integrate the first term of~\eqref{Bulk_action_intermediate} on $M_\rho$ and the only contribution will be given by the upper limit of integration. We set this upper limit to be $r_0$: this is just a cutoff we use to regulate the large-distance divergences of the various pieces of the action. The second term of~\eqref{Bulk_action_intermediate} can also be reduced to a boundary one via a De Rham theorem since $Q_{IJ} A^I \wedge \star F^J $ vanishes at the horizon in the gauge we have chosen and is regular everywhere outside the horizon.
Therefore we rewrite the bulk action as
\be
\label{Bulk_Action_Final}
S_{\rm bulk} = - \frac{16\pi^2}{3 \kappa^2} \, \left. a^2 p \right|_{r_0} \int \diff t + \frac{1}{3 \kappa^2} \int_{\pd M_{r_0}} Q_{IJ} A^I \wedge \star F^J  \,.
\ee
We can thus evaluate the bulk action by plugging the near-boundary expansions we presented in sec.~\ref{sec:near_boundary} into the above expression~\eqref{Bulk_Action_Final}. We obtain for the first term
\be
\label{Bulk_Gravity_Evaluated}
\begin{split}
- \frac{16\pi^2}{3 \kappa^2} \, \left. a^2 p \right|_{r_0} \int \diff t  &= - \frac{8\pi^2 \ell^2}{\kappa^2} \Big[ 4 a_0^4 \left( \frac{r_0}{\ell}\right)^4 - \frac{1}{3} (4c +3) a_0^2 \left( \frac{r_0}{\ell}\right)^2 - \frac{32}{9} \,c^2 \log \frac{r_0}{\ell} \\
& \quad+  \frac{1}{36} c (38 c-128 a_2 +1)+\frac{3}{32} - 2 {\cal K}_1 - 12 (\tilde H^2 + \tilde H \tilde K + \tilde K^2)\Big]  \int \diff t \, ,
\end{split}
\ee
while the second one evaluates to
\be
\label{Bulk_Gauge_Evaluated}
\begin{split}
&\frac{1}{3 \kappa^2} \int_{\pd M_{r_0}} Q_{IJ} A^I \wedge \star F^J = - \frac{8\pi^2 \ell^2}{\kappa^2}\Big\{ - \frac{2}{9} \Big[   8 \left(2 c^2+27 \left(\tilde H^2+\tilde H \tilde K+\tilde K^2\right)\right) \log \frac{r_0}{\ell}  + 16 a_2 c  \\
& \quad   -12 c^2+c+9 \left(6 \left(\tilde K (2 H_2+\tilde H+4 K_2)+\tilde H (4 H_2+\tilde H+2 K_2)+\tilde K^2\right)+{\cal K}_1 \right) \Big]\Big\} \int \diff t  \,.
\end{split}
\ee

We now turn to the second piece of the regularized action, which is the Gibbons-Hawking term. As it is well known, this contribution is needed to make the Dirichlet variational problem for the metric well-defined. It reads
\be
\label{Gibbons_Hawking}
S_{\rm GH} = \frac{1}{\kappa^2} \int_{\pd M_{r_0}} \diff^4 x \, \sqrt{h} \, K \,, 
\ee
where $K = h^{ij} K_{ij} $ is the trace of the extrinsic curvature 
$ K_{ij} = \frac{r}{2\ell} \, \pd_r h_{ij}$. Once we plug the near-boundary expansions of the various quantities in~\eqref{Gibbons_Hawking}, the evaluation of the Gibbons-Hawking term is straightforward, and gives
\be
\begin{split}
\label{Gibbons_Hawking_Evaluated}
S_{\rm GH} &= - \frac{8\pi^2 \ell^2}{\kappa^2}   \Big[ - 16 a_0^4  \left( \frac{r_0}{\ell}\right)^4 +  \frac{1}{3} (4c +3) a_0^2 \left( \frac{r_0}{\ell}\right)^2 + 96  \left(\tilde H^2+\tilde H \tilde K+\tilde K^2\right) \log \frac{r_0}{\ell}  \\
& \quad \qquad \qquad \; + 48 \left(\tilde K (H_2+\tilde H+2 K_2)+\tilde H (2 H_2+\tilde H+K_2)+\tilde K^2\right) \Big] \int \diff t .
\end{split}
\ee
The last piece of the renormalized action, $S_{\rm ct}$, contains all the counterterms which are needed to cancel the divergences appearing in $S_{\text{bulk}} + S_{\text{GH}}$. These local boundary terms for the five-dimensional Fayet-Ilopoulos gauged supergravity have been explicitly constructed in~\cite{Cassani:2018mlh} by generalizing the results presented in~\cite{Bianchi:2001kw} to the setup where the metric, the gauge fields and the scalar fields are all non trivial. The counterterm action reads
\begin{align}
S_{\rm ct} &= - \frac{1}{\kappa^2} \int_{\pd M_{r_0}} \diff^4 x \, \sqrt{h} \Big[ {\cal W} + \Xi \, R - \frac{{\cal W} - 3\ell^{-1}}{\log \frac{r_0^2}{\ell^2}} \nonumber \\
& \qquad \qquad \qquad\qquad\qquad + \frac{\ell^3}{16} \log \frac{r_0^2}{\ell^2} \left( R_{ij} R^{ij} - \frac{1}{3} \, R^2 - 2 \ell^{-2} Q_{IJ} F^I_{ij} F^{J \, ij}   \right) \Big],
\end{align}
where $\CW = 3 \ell^{-1} \bar X_I X^I$ is the superpotential already presented in the main text and $\Xi = \frac{\ell}{4}\, X_I \bar{X}^I $.  We underline that we are using a minimal subtraction scheme where no finite counterterms are added to the divergent ones. Evaluating this part of the action on our supergravity background we obtain
\be
\begin{split}
\label{Counterterms_Evaluated}
S_{\rm ct} &= - \frac{8\pi^2 \ell^2}{\kappa^2}   \Big[   12 a_0^4  \left( \frac{r_0}{\ell}\right)^4 - 144    \left(\tilde H^2+\tilde H \tilde K+\tilde K^2\right) \log \frac{r_0}{\ell}  \\
& \qquad  + \frac{8}{3} \left( c^2-27 \left(\tilde K (H_2+\tilde H+2 K_2)+\tilde H (2 a_2+\tilde H+K_2)+\tilde K^2\right) \right) \Big] \int \diff t \, .
\end{split}
\ee

Adding up all the pieces of the action given by eqs.~\eqref{Bulk_Gravity_Evaluated},~\eqref{Bulk_Gauge_Evaluated},~\eqref{Gibbons_Hawking_Evaluated},~\eqref{Counterterms_Evaluated} we get the final result
\be
S_{\rm ren} = - \frac{\pi^2 \ell^2}{\kappa^2} \Big[ \frac{16}{9} - \frac{14}{9} \, v^2 + \frac{19}{36}\, v^4 - 192 (\tilde H^2 + \tilde H \tilde K + \tilde K^2 )  \Big] \int \diff t\, ,
\ee   
which is the result~\eqref{Renormalized_Action_Computed} reported in the main text. All the power-law and logarithmic divergences of the various pieces of the action cancel non-trivially against themselves when we perform the sum.

\subsection{Holographic one-point functions}
We now briefly show how it is possible to compute holographic one-point functions of the main relevant operators dual to our supergravity fields. 

We start from the holographic stress-energy tensor. It is defined as:
\begin{equation}
\langle T_{ij} \rangle = - \lim_{r_0 \to \infty} \frac{r_0^2}{\ell} \, \frac{2}{\sqrt{h}} \, \frac{\delta S_{\rm reg}}{\delta h^{ij}} \, ,
\end{equation} 
recalling the form of the various pieces of $S_{\rm{reg}}$, we perform the variation with respect to the metric obtaining
\be
\label{Energy_Momentum_Tensor_Appendix}
\begin{split}
\langle T_{ij} \rangle = -\frac{1}{\kappa^2} \lim_{r_0\to \infty} \frac{r_0^2}{\ell^2} & \Bigg[ K_{ij} - (K - {\cal W}) h_{ij} -  \frac{{\cal W} - 3\ell^{-1}}{\log \frac{r_0^2}{\ell^2}}  \, h_{ij}  - \frac{\ell}{2} \, \left( R_{ij} - \half \, R\, h_{ij} \right) \\
& \quad - \frac{\ell^3}{4} \log \frac{r_0^2}{\ell^2} \left( - \half \, B_{ij} - \frac{2}{\ell^2} \, Q_{IJ} F^I_{i k} F^J{}_j{}^k + \frac{1}{2\ell^2} \, h_{ij} Q_{IJ} F^I_{kl} F^{J \, kl}   \right)\!  \Bigg],
\end{split}
\ee
which precisely coincides with eq.~\eqref{Stress_Energy_Tensor} shown in sec.~\ref{sec:Properties}.
We find that the stress-energy tensor can be written as
\begin{equation}
\langle T_{ij} \rangle \, \diff x^i \, \diff x^j = \langle T_{tt} \rangle \, \diff t^2 + \langle T_{\theta \theta} \rangle \, \left(\sigma_1^2 + \sigma_2^2 \right) + \langle T_{\psi \psi} \rangle \, \sigma_3^2 + 2 \,  \langle T_{t \psi} \rangle \, \diff t \, \sigma_3 \, ,
\end{equation}
where the components explicitly read %
{\small 
\be
\begin{split}
\langle T_{tt} \rangle  &= \frac{1}{\kappa^2 a_0^2 v^4 \ell} \Bigg[\frac{2}{27}+\frac{v^2}{9}-\frac{7 v^4}{36}+ \frac{89 v^6}{864} - 4 {\cal K}_1 +{\cal K}_3 \\
& \quad + 24 \left(\tilde H^2 (18 \tilde K-1)+\tilde H (3 {\cal K}_2^{(2)}+6 {\cal K}_2^{(1)}+\tilde K (18 \tilde K-1))+\tilde K (6 {\cal K}_2^{(2)}+3 {\cal K}_2^{(1)}-\tilde K)\right)\\
& \quad -2 v^2 \left(6 \left(\tilde H^2+\tilde H \tilde K+\tilde K^2\right)+{\cal K}_1 \right) \Bigg] , 
\end{split}
\ee
\be
\begin{split}
\langle T_{t \psi} \rangle  &= \frac{1}{\kappa^2 a_0^2 v^2} \Bigg[ \frac{1}{54} \left(-108 {\cal K}_1 \left(v^2-1\right)-27 {\cal K}_3+2 \left(v^2-1\right)^3\right) \\
& \quad -12 \Bigg( \tilde H^2 \left(18 \tilde K+v^2-1\right)+\tilde H \left(3 {\cal K}_2^{(2)}+6 {\cal K}_2^{(1)}+\tilde K \left(18 \tilde K+v^2-1\right)\right) \\
& \quad +\tilde K \left(6 {\cal K}_2^{(2)}+3 {\cal K}_2^{(1)}+\tilde K \left(v^2-1\right)\right)\!\Bigg) \Bigg] , 
\end{split}
\ee
\be
\begin{split}
\langle T_{\psi \psi} \rangle  &= \frac{\ell}{3456 \kappa^2 a_0^2 } \Bigg[  24 (53-192 a_2) v^4+1728 {\cal K}_1 \left(5 v^2-2\right)+864 {\cal K}_3-1117 v^6-480 v^2+64 \\
& \quad  + 10368 \Big(
5 v^2 \left(\tilde H^2+\tilde H \tilde K+\tilde K^2\right)+\tilde H^2 (36 \tilde K-2)\\
& \quad +2 \tilde H (3 {\cal K}_2^{(2)}+6 {\cal K}_2^{(1)}+\tilde K (18 \tilde K-1))+2 \tilde K (6 {\cal K}_2^{(2)}+3 {\cal K}_2^{(1)}-\tilde K)
\! \Big)   \Bigg] , 
\end{split}
\ee
\be
\begin{split}
\langle T_{\theta \theta} \rangle  &= \frac{\ell}{384 \kappa^2 a_0^2 } \Bigg[ 16 (16 a_2-5) v^2-576 {\cal K}_1+67 v^4+32\\
& \quad + 3456 \Big( \tilde K (2 H_2+\tilde H+4 K_2)+\tilde H (4 H_2+\tilde H+2 K_2)+\tilde K^2
\! \Big)   \Bigg] , 
\end{split}
\ee}%
and the trace of the stress-energy tensor is 
\be
\langle T^i{}_i \rangle = \frac{3}{\kappa^2 a_0^4} \, 12 \left(\tilde K (H_2+\tilde H+2 K_2)+\tilde H (2 H_2+\tilde H+K_2)+\tilde K^2\right) .
\ee
As we immediately see by looking at the explicit form of the stress-energy tensor, all the divergences, including the logarithmic ones, are removed, so that $\langle T_{ij} \rangle$ is finite in the limit.

Similarly to the stress-energy tensor, we can define the holographic conserved currents as
\begin{equation}
    \langle j_I^i \rangle = \lim_{r_0 \to \infty} \frac{r_0^4}{\ell^4} \, \frac{1}{\sqrt{h}} \, \frac{\delta S_{\rm reg}}{\delta A_i^I} \, .
\end{equation}
By varying the action with respect the boundary gauge fields, as prescribed by the formula above, we obtain:
\be
\label{Electric_Current_Appendix}
\langle j_I^i \rangle = - \frac{1}{\kappa^2}  \lim_{r_0\to \infty} \frac{r_0^2}{\ell^2} \Bigg[  \ep^{ijkl} \left( Q_{IJ} \star F^J +\frac{1}{6} C_{IJK}A^J \wedge F^K \right)_{jkl} + \ell \, \nabla_j \left( Q_{IJ} F^{J \, ji}\right) \log \frac{r_0}{\ell}  \Bigg] .
\ee
Evaluating eq.~\eqref{Electric_Current_Appendix} on our supergravity background, we find that the two non-vanishing components of the conserved currents are
\be
\begin{split}
\langle j_1^t \rangle &= - \frac{1}{36 \kappa^2 \ell^2 a_0^4} \Bigg[   6 \tilde H \left(18 \tilde K+v^2-1\right)+\frac{1}{3} \left(54 {\cal K}_1 +162 {\cal K}_2^{(1)} +324 \tilde K^2-\left(v^2-1\right)^2\right) \Bigg] , \\
\langle j_1^\psi \rangle &= + \frac{1}{54 \kappa^2 \ell^2 a_0^4 v^2} \Bigg[ v^2 (36 a_2-162 H_2-5)-9 \tilde H \left(36 \tilde K+5 v^2-2\right) 	\\
& \quad \qquad\qquad\qquad \qquad -54 {\cal K}_1 -162 {\cal K}_2^{(1)}-324 \tilde K^2+\frac{25 v^4}{4}+1\Bigg] ,
\end{split}
\ee
\be
\begin{split}
\langle j_2^t \rangle &= - \frac{1}{36 \kappa^2 \ell^2 a_0^4} \Bigg[   6 \tilde K \left(18 \tilde H+v^2-1\right)+\frac{1}{3} \left(54 {\cal K}_1 +162 {\cal K}_2^{(2)} +324 \tilde H^2-\left(v^2-1\right)^2\right) \Bigg] , 		\\
\langle j_2^\psi \rangle &= + \frac{1}{54 \kappa^2 \ell^2 a_0^4 v^2} \Bigg[ v^2 (36 a_2-162 K_2-5)-9 \tilde K \left(36 \tilde H+5 v^2-2\right) 	\\
& \quad \qquad\qquad\qquad \qquad -54 {\cal K}_1 -162 {\cal K}_2^{(2)}-324 \tilde H^2+\frac{25 v^4}{4}+1\Bigg] ,
\end{split}
\ee
\be
\begin{split}
\langle j_3^t \rangle &= -\langle j_1^t \rangle-\langle j_2^t \rangle - \frac{1}{36 \kappa^2 \ell^2 a_0^4} \Bigg[  54 {\cal K}_1 +108 (\tilde H^2 +\tilde H \tilde K + \tilde K^2) -\left(v^2-1\right)^2 \Bigg] , \\
\langle j_3^\psi \rangle &= -\langle j_1^\psi \rangle-\langle j_2^\psi \rangle  \\
& \quad   + \frac{1}{54 \kappa^2 \ell^2 a_0^4 v^2} \Bigg[ 3 - 162 {	\cal K}_1 - 324  (\tilde H^2 +\tilde H \tilde K + \tilde K^2) + 3 (36 a_2 -5) v^2+\frac{75 v^4}{4} \Bigg] .
\end{split}
\ee
We notice that $\langle j_1^t \rangle \leftrightarrow \langle j_2^t \rangle $ if $\tilde H \leftrightarrow \tilde K$ and ${\cal K}_2^{(1)} \leftrightarrow {\cal K}_2^{(2)}$.

Finally we compute the one-point function of the scalar operators, which is defined as
\begin{equation}
    \langle {\cal O}_I \rangle = \lim_{r_0 \to \infty} \left(\frac{r_0^2}{\ell^2} \, \log \frac{r_0^2}{\ell^2} \, \frac{1}{\sqrt{h}} \, \frac{\delta S_{\rm reg}}{\delta X^I} \right) \, .
\end{equation}
By varying our regularized action with respect the scalar fields, we obtain\footnote{While performing the variation, one has to keep in mind that the scalar fields fulfill the constraint~\eqref{constraint}. This implies that $\bar{X_I} \, \delta \, {}^{(0)} \phi^I = 0$}
\be
\langle {\cal O}_I \rangle = \frac{2}{\kappa^2} \, \bar{Q}_{IJ} \, \, \, {}^{(0)} \phi^J \, ,
\ee
where ${}^{(0)} \phi^I$ is the $\CO(r^{-2})$ term in the Fefferman-Graham expansion of the scalar fields defined in app.~\ref{app_FG}.
Evaluating explicitly the scalar one-point functions we obtain
\be
\begin{split}
\langle O_1 \rangle &= -  \frac{3}{\kappa^2 a_0^2} \left( 2 H_2 + \tilde H\right) , \\
\langle O_2 \rangle &= -  \frac{3}{\kappa^2 a_0^2} \left( 2 K_2 + \tilde K \right) , \\
\langle O_3 \rangle &= -\langle O_1 \rangle -\langle O_2 \rangle \,.\\
\end{split}
\ee
We conclude by remarking that all the one-point functions we evaluated in this section fulfill the Ward identities reported in eqs.~(4.18)-(4.20) of~\cite{Cassani:2018mlh}, where the Weyl and chiral anomalies are defined as reported in eqs.~(4.21),(4.22) of the same paper.


\newpage 
\bibliographystyle{nb}
\bibliography{U1cubebiblio}

\end{document}